\documentclass[en,vawiessen]{udesoftec} 
\usepackage{udesoftec-extra}
\usepackage{amssymb}

\DeclareFixedFont{\ttb}{T1}{txtt}{bx}{n}{12} 
\DeclareFixedFont{\ttm}{T1}{txtt}{m}{n}{12}  

\usepackage{color}
\definecolor{deepblue}{rgb}{0,0,0.5}
\definecolor{deepred}{rgb}{0.6,0,0}
\definecolor{deepgreen}{rgb}{0,0.5,0}

\usepackage{listings}

\newcommand\pythonstyle{\lstset{
		language=Python,
		basicstyle=\ttm,
		morekeywords={self},              
		keywordstyle=\ttb\color{deepblue},
		emph={MyClass,__init__},          
		emphstyle=\ttb\color{deepred},    
		stringstyle=\color{deepgreen},
		frame=tb,                         
		showstringspaces=false
}}

\lstnewenvironment{python}[1][]
{
	\pythonstyle
	\lstset{#1}
}
{}

\newcommand\pythonexternal[2][]{{
		\pythonstyle
		\lstinputlisting[#1]{#2}}}

\newcommand\pythoninline[1]{{\pythonstyle\lstinline!#1!}}

\usetikzlibrary{shadows}

\date{}  

\title{Proof-of-Turn:\\
	Blockchain consensus using\\
	a round-robin procedure	as one\\
	possible solution for cutting costs\\
	in mobile games
	}
\author{Dominik Braun}
\typeofdoc{Master Thesis}
\institution{Presented to the Faculty of Economics and Business Administration of\par University of Duisburg-Essen}
\academicfield{Business Information Systems}
\semester{Summer semester 2021}

\authorbox{
	\begin{tabularx}{.7\linewidth}{ll}
		By:&Braun, Dominik\\
		&Friedenstr. 105\\
		&47053, Duisburg\\
		&Mat. No.: 3086262 \\
		\\
		Examiner:&Prof. Dr. Stefan Eicker\\
		&Prof. Dr. Tobias Kollmann\\
		\\
		Academic field:&Wirtschaftsinformatik\\
		\\
	\end{tabularx}
}

\abstract{
	This master thesis deals with Blockchain Technology in mobile turn based peer to peer games.
	First, it investigates the capabilities of Blockchain Technology to be used for gaming applications.
	In this regard, among others, \textit{Proof-of-Mechanisms}, \textit{Vote-based Consensus}
	and several \textit{Performance Improvements} are described.
	Second, several smart contracts are introduced to show the general
	feasibility of turn based games hosted on Blockchain Technology.
	More specific, \textit{Hidden transactions}, \textit{Randomization},
	\textit{Piles of Cards}, \textit{Fog of War} elements, \textit{Data allocation improvements}
	and other smart contracts are specified.	
	Third, a special Proof-of-Turn consensus mechanism, based on the Blockchain Technology,
	is defined to enable game publishers to cut costs in the means of their provided game servers.
	Herein, \textit{Byzantine Fault Tolerance}, \textit{Peering}, the \textit{CAP Theorem},
	\textit{Interoperability} among other characteristics are covered.	
	Last, these measures shall additionally raise the trust level among the players in mobile turn based games.
}

\renewcommand{\udesoftecoverride}{
    \renewcaptionname{ngerman}{\labelabstracttitle}{Zusammenfassung}
    \renewcaptionname{british}{\labelabstracttitle}{Abstract}
}

\gdef\udesoftec@usedlistofitems@acronyms{}

\label{udesoftec:needsloa}

\renewcaptionname{british}{\udesoftecAcronymsTitle}{Abbreviations}%

\begin{document}
	
	\newacronym{A}{A}{Availability}
	\newacronym{BT}{BT}{Bloat Transaction}
	\newacronym{BC}{BC}{Blockchain}
	\newacronym{BCT}{BCT}{Blockchain Technology}
	\newacronym{BFT}{BFT}{Byzantine Fault Tolerance}
	\newacronym{BPMN}{BPMN}{Business Process Modeling Notation}
	\newacronym{C}{C}{Consistency}
	\newacronym{CF}{CF}{Consensus Finality}
	\newacronym{CM}{CM}{Consensus Mechanism}
	\newacronym{DDoS}{DDoS}{Distributed Denial of Service}
	\newacronym{DNs}{DNs}{Distribution Nodes}
	\newacronym{ETH}{ETH}{Ethereum}
	\newacronym{GCN}{GCN}{Garbage Collecting Node}
	\newacronym{LN}{LN}{Leading Node}
	\newacronym{MMORPG}{MMORPG}{massively multiplayer online role-playing game}
	\newacronym{NFTs}{NFTs}{Non-fungible tokens}
	\newacronym{PBFT}{PBFT}{Practical Byzantine Fault Tolerance Consensus}
	\newacronym{P}{P}{Partition Tolerance}
	\newacronym{PoA}{PoA}{Proof-Of-Authority}
	\newacronym{PoET}{PoET}{Proof-Of-Elapsed-Time}
	\newacronym{PoP}{PoP}{Proof-Of-Play}
	\newacronym{PoT}{PoT}{Proof-Of-Turn}
	\newacronym{PoS}{PoS}{Proof-Of-Stake}
	\newacronym{PoW}{PoW}{Proof-Of-Work}
	\newacronym{Ripple}{Ripple}{Ripple Consensus}
	\newacronym{rrTs}{rrTs}{relevant (revealed) Transactions}
	\newacronym{SC}{SC}{Smart Contract}
	\newacronym{TCM}{TCM}{Tendermind Consensus Model}



\chapter{Introduction}
\label{chap:Introduction}

Since the beginning of peer to peer games, the game industry emerged towards online games in order to serve a growing number of simultaneous players \cite[192]{Nagygyorgy.2013}.
After general networking moved from small local token ring- and Ethernet-based networks \cite[1]{Smythe.1999}
towards huge online communities, gaming became a \textit{massive multiplayer online}-phenomenon (\citet[1]{Williams.2008}; \citet[1]{Wang.2012})
and publishers needed to offer servers to coordinate the high amount of players \cite[41]{Lee.2008} and to prevent cheating.
Nevertheless, all peers need to trust the publisher's server, whilst the server has to carry the costs of availability and network computation.
Consequently, power consumption, computation workload and likewise attached costs for \textit{networking and peering} are mainly carried by 'the central participant' of the network.
But sometimes the central element - e.g. a publisher - is not able to run the central server without the need to monetize its offerings.
One consequence was the rise of private dedicated servers, hosted by private persons (e.g. using a peer-to-peer architecture \cite[41]{Lee.2008}), which were 'detached from the publisher's direct influence'.
Additionally, the technical/mental hurdle to set up a dedicated server, whilst other more convenient games exist,
as well as the single point of failure 'out of the publisher's reach'
are considered reasons against a rise and large distribution of private, peer-owned dedicated servers.
Nowadays, along with reduced server costs\footnote{\hspace{0.1cm}Reduction of running costs due to Moore's Law (\cite{Mann.2000}, 2000). Here: less power consumption or increased sessions per hardware unit.},
many game publishers rely on in-game purchases, advertisements and data mining to target those players who are willing to pay more (\cite{Davis.2012}).
Consequently, consumers "[...] are regularly exposed to predatory behaviour from the games publishers, with microtransactions and loot boxes [...]" \cite[14]{Laneve.2019}.
Other distributed server architecture as proposed in the documents "A Distributed Server Architecture for Massively Multiplayer Online Games" from \citet{Khan.2006} and
"Matrix: adaptive middleware for distributed multiplayer games" from \citet{Besancon.2019}, to the writer's knowledge, never settled in the consumer world. \\
On the contrary, \gls{BCT} offers distributed computing in 'trustless environments' and
offers many of those by the (game) publishers needed characteristics.
Generally speaking, a \gls{BC} "[...] is an innovative technology, which can have a high impact in numerous industries,
such as healthcare [1], supply-chain [2], finance [3] and video games [4]. BC are append-only ledgers shared across a network of clients" \cite[84]{Besancon.2019}. \\
This thesis shall bring \gls{BCT} closer to practical use cases in the gaming industry.
Therefore, this study may especially be of interest to the following readers:
\begin{enumerate}
	\item Researchers interested in enhancing \gls{BCT} and affiliated mechanisms.
	\item Practitioners willing to understand how blockchain may affect the software (gaming) industry.
	\item Managers and administrative staff from the software (gaming) industry looking for potential savings of server operation costs.
\end{enumerate}
The remainder of the document is structured as follows:\\
This section will provide the problem statement, research design as well as guiding research questions in detail.
In section 2, a review of \gls{BCT} and included mechanisms to achieve consensus are presented.
Section 3 outlines recent applications of \gls{BCT} in the gaming industry as well as mechanisms to map known tabletop game mechanisms into digital contracts.
In section 4, a novel \gls{CM} for \gls{BCT} is presented and discussed in detail.
Finally, in section 5 the work is concluded and directions are provided for future research.

\FloatBarrier

\section{Current Situation}
\label{sec:CurrentSituation}
First of all, data and scientific research targeting the gaming industry is scarce.
Still it is known that  most "[...] game projects fail to meet their financial expectations [...]" \cite[17]{Bethke.2003}.
Nevertheless, the number of games is growing constantly (\cite{statista.com.2021}) and providers are forced
to minimize their (running) costs (\cite{Koster.2018}).
"Furthermore, small developers are finding it hard to stand out in a market dominated by giant companies that dwarf them through marketing or polish,
with the average indie developer earning only around \$10 thousand a year" (\citet[14]{Laneve.2019} using data from \citet[6]{Gamesetwatch.com.2014}. \\
Although there are several ways to cut costs, one of them is the use of dedicated servers which are paid by single peers or clans of the game network (\cite{Wikipedia.2021c}).
Especially in the long run, overall costs \cite[13]{Weilbacher.2012} may be reduced by these servers and help publishers to keep the game alive and establish a network of players and enthusiasts.
Some publishers offer dedicated servers (\cite{VALVe.2021}) as one possible solution for externalizing expensive hardware and maintenance.
Nevertheless, there are some reasons against dedicated gaming servers:
\begin{enumerate}
	\item \textbf{Giving away the source code} \\
	Behind this phrase, the publisher fears its game's code to be analyzed and cloned to provide any type of copycat product \cite[2]{Li.2014}.
	Two prominent examples for this phenomenon in the game industry is the massive imitation of the mobile game 'flappy bird' (\cite{Li.2014})
	as well as the co-evolution of the 'battle royale mode' (\cite{KooistraJ..2018}).
	Hence, these game mechanic ideas skyrocketed and (many) late adopters followed, but not because the code was given 'out of hands' in the first place.
	Therefore the argument can be considered obsolete as each game's general mechanics can be obtained by simply playing.
	Nevertheless, the fear is not negligible in general as sometimes source code is stolen (\cite{vice.com.2021}).
	
	\item \textbf{Update laziness} \\
	Games progress, therefore both 'security'- as well as 'game-mechanics'-updates
	shall comply with publishers' and gamers' expectations.
	But dedicated server providers are detached from the publishers' direct influence.
	The providers act by intrinsic motivation, ranging from \textit{control} over \textit{fame} up to \textit{monetization}.
	Hence, publishers as principals and providers as agents follow different incentives - which presents a \textit{Principal-agent problem} as generally described by \citet{STIGLITZ.1989}.
	
	\item \textbf{Nerdy peers} \\
	As a server needs to work 24/7 to serve whenever needed, there need to be maintainers \cite[211]{Lowell.2004}.
	These maintainers are not supposed to be casual gamers, but peers which offer expertise with web-servers, scripting and so forth (\cite{teamfortress.com.2021}).
	On the one hand, without these type of persons, the offer of dedicated servers is worthless and
	consequently cheap 'scalability and accessibility' can not be achieved.
	On the other hand, a game publisher has to consider an additional group of (silent) stakeholders.
	
	\item \textbf{Shadow server} \\
	A \textit{shadow server} is considered a server which runs on one players device, without the player's knowledge/consent.
	This is not recommended as players could 'force quit' the game or shut down their system (unintended).
	Consequently the (dedicated) shadow server would end to serve and remaining players are kicked out of their games which presumably leads to a bad consumer experience.
	
	\item \textbf{Lost peer management \& data collection} \\
	In the assumption that dedicated servers are generally detached from the publisher's reach,
	metrics and other data get lost although gamers stay within the game's ecosystem.
	Still, publishers have a "Need for Metrics" \cite[152]{Palmer.2002} to find flaws in their games.
	Thus, dedicated servers are seen as a misleading approach.
	
	\item \textbf{Performance} \\
	Bad performance of dedicated servers leads to bad consumer experience \cite[10]{Weilbacher.2012}.
	Nevertheless, the reasons for bad performance are manifold, from slow server devices not suitable to host their
	(large scale) games up to slow broadband connection \cite[10]{Weilbacher.2012}.
	
	\item \textbf{Integrity \& Cheating} \\
	The combination of nerdy peers and the possibility to modify game content (e.g. within a 'modding community')
	could lead to cheating behavior and consequent changed game characteristics (\cite{Morris.2003}).
	Although there are basic possibilities to check the integrity of server instances (\cite{Agosta.2003}; \cite{Deswarte.2004}),
	the publishers or other players can't rely on the system to be somehow silently manipulated.
	
\end{enumerate}
Still, the number of persons 'willing and able' to provide dedicated servers is seen low compared to the number of gamers playing in the manifold games.
To the writers knowledge, despite the enormous amount of mobile (Android) apps offering server-based network multiplayer
modes (\cite{itch.io.2021}),
there are only some mobile games, such as \textit{'Among Us'} and \textit{'Minecraft'} offering dedicated servers.
Those games have been designed for desktop primarily and have only thereafter been ported to mobile operating systems (Here: Android \& iOS).
Compared to the 'Hard-core' gamers, mobile and 'casual' gamers can be considered less likely to set up additional systems
to boost their game sessions due to reduced general involvement \cite[388]{Prugsamatz.2010}.
As small, independent developers might not be able to afford additional running servers,
there is a (growing) demand for distributed computing in the mobile gaming ecosystem.
At this point the interest in the present work originates from the insights into game development and knowledge about \gls{BCT}.
During the last years, \gls{BCT} emerged and offers the possibility to establish distributed networks to share data without a central authority.
For now, "[...] an analysis of the different projects, [...]" has shown "[...] that the main problems blockchain games face are related to scalability and transaction speed" \cite[47]{Laneve.2019}.
Nevertheless, if there is a possibility to tweak a Proof-/Voting-Based Consensus Model \cite[3]{Khan.2020} to fit the need of the gaming industry,
it might keep cooperative niche games alive or kick start some community driven game development projects \cite[14]{Laneve.2019}.

\FloatBarrier

\section{Problem Statement}
\label{sec:ProblemStatement}

The biggest issue seen for game publishers is the shortage of money, both during production and operation (\cite{Koster.2018}).
Sometimes, when games do not settle in the market, publisher servers need to be shut down (\cite{rockpapershotgun.com.2021}).
This beholds true for all genres as network- and cross selling effects play a key role for the survival of online (multiplayer) games \cite[45]{Rong.2018}. \\
Once game servers are shut down, players can only play offline, if at all.
Additionally, games are established in the perception of long term revenue.
Hence, once an investment decision for a game idea is made, long availability is a key feature.
Only an accessible game can continue to yield revenue.
Last, due to tight budget plans during a game's creation process, additional effort for dedicated servers is mostly not met as an active community has not yet existed.
Consequently, many game publishers are in a problematic situation between \textit{bankruptcy} and \textit{squeezing their consumers} for income generation \cite{Sotamaa.2021}.
Once a game is established, besides fixing bugs and implementing new features, running servers are a key driver of 'keep alive costs'.
The question arises whether there is any backend technology to further minimize some of the games running server costs.
The solution has to be:
\begin{enumerate}
	\item \textbf{Distributed}, to relieve the central running server from some/many workloads.
	\item \textbf{Trusted}, preventing malicious behavior and cheating.
	\item \textbf{Cross-plattform} applicable to take advantage of gamers using different operating systems.
	\item \textbf{Scalable \& Fast}, at least for some designated game niche use cases.
	\item \textbf{Leightweight} in terms of preventing heavy computation on single devices.
\end{enumerate}
On the first look, \gls{BCT} offers some of these characteristics.
Nevertheless, as "[...] it stands, blockchain technology does not seem applicable for the design of the most popular game genres such as first-person shooters or real-time strategy[...]" \cite[3]{Serada.2020}.
Still, "[...] several attempts have been made in this direction (e.g., EOS Knights, HyperDragons, and Epic Dragons), and many more are likely to follow [...]" \cite[3]{Serada.2020}. \\
Therefore this document will examine \gls{BCT} thoroughly to find a suitable application area.

\FloatBarrier

\section{Research Design \& Questions}
\label{sec:ResearchDesignQuestions}

The main research question which shall be answered with this thesis is:
\begin{center}
	\textit{"Can \gls{BCT} be used to reduce publisher's server costs whilst \\
		providing (mobile) players a suitable gaming experience?"}
\end{center}
To guide the further procedure, no explicit framework is used.
As a starting point the recent literature has been reviewed to reflect the state of research about \gls{BCT} and related \gls{CM}s.
The results of the grounding research and the \gls{CM}s performance characteristics are given in the chapter \hyperref[chap:BCT]{Blockchain Technology}.
Additionally, as \gls{BCT} is still pretty young regarding the field of information systems - starting from \cite{Nakamoto.2009} in 2009 -
there is only scarce literature to be found in the intersection with the gaming industry, yet.
Still, some applications and use cases of \gls{BCT} in games could be found.
"The industry has started its exploration on this topic by integrating traditional games with blockchain systems" \cite[1]{Min.2019b}.
Further more, \citet[1]{Min.2019b} state that blockchain games have already become an important component of decentralized applications and have held a considerable market capitalization.
Nevertheless, these applications are barely suitable for game play related issues.\\
As \gls{BCT} builds on trust and agreed rules, the mechanism called \gls{SC} has to be mentioned.
\gls{SC}s are rulesets which can be used to establish agreements between participants of \gls{BCT} networks.
As there seem to be reasons, which prevent even slow paced games to be conducted on \gls{BCT}, the following guiding question has been raised:
\begin{center}
	\textit{"Which kind of \gls{SC}s need to be established to cover typical in-game mechanics?"}
\end{center}
Both, existing games using \gls{BCT} as well as game specific \gls{SC}s are given in chapter \hyperref[chap:BlockchainInGames]{Blockchain in Games}.
Moreover, during the search for suitable \gls{CM}s for games hosted on \gls{BCT},
it appeared that games with realtime-mechanics, such as first person shooters and \gls{MMORPG},
can be considered as not suitable for \gls{BCT} backbones \cite[19]{Serada.2020}.
Therefore, fast paced games became out of scope as distributed computation lacks performance regarding speed most of the time \cite[3]{Serada.2020}.
Thus this present work concentrates on slow paced games, such as asynchronous 'turn based (strategy)'-games \cite[1]{Bergsma.2008} and raises a second guiding question:
\begin{center}
	\textit{"What is the best fitting \gls{BCT} \gls{CM} to cover an asynchronous game play scenario?"}
\end{center}

To address this research gap, the different Proof-/Voting-Based Consensus Models are evaluated and
a novel \gls{CM} (Chapter: \hyperref[chap:PoT]{Proof-of-Turn}) is proposed.
\gls{PoT} is a special merge of existing protocols to fit the need of turn based games and
remains a theoretical proof of concept for a (new) Proof-/Voting-Based \gls{CM}.
Consequently the research hypothesis is:
\begin{center}
	\textit{"The \gls{PoT} \gls{CM} leads to a reduction of server running costs for game publishers."}
\end{center}
Additionally the following working assumptions are used:
\begin{enumerate}
	\item \textit{Throughout a game a majority of peers is constantly online.}
	\item \textit{The network offers enough transmission performance (throughput/bandwidth).}
	\item \textit{Every peer has the needed storage available.}
	\item \textit{Depending on the 'real world'-scenario, the publisher's servers help on special \\
		occasions (e.g. a trusted timestamp).}
	\item \textit{Symmetric and asymmetric encryption is secure.}
\end{enumerate}
Due to the intersection of this research between \textit{business administration} and \textit{computer science},
the document falls into the research area of \textit{economics of information systems}. \\
Last, in other contexts gamers, players, peers and nodes are sometimes used as synonyms,
but in the following they need to be distinguished:
\begin{enumerate}
	\item \textbf{Gamers} are persons, who play a game.
	\item \textbf{Players}, which are considered digital in-game characters, are controlled by \textit{gamers}.
	\item \textbf{Nodes} are hardware systems, which is part of a \gls {BCT}-(gaming-)network.
	\item \textbf{Peers} are persons, who can be \textit{gamers} but predominantly own or maintain a node.
\end{enumerate}
Upcoming the core principles of \gls{BCT} are described.


\chapter{Blockchain Technology}
\label{chap:BCT}

Although the general pattern of linked data,
primarily using the terms '\textit{(One-Way) Hash Chain}' or '\textit{Merkle (Hash) Tree}' (\citet{Hu.2005}),
was known before the therm \gls{BCT}, the rise of Bitcoin,
enabled by \citeauthor{Nakamoto.2009}'s core paper in 2009 presenting the fundamental \gls{PoW} protocol, pushed \gls{BCT} into public attention.
In this thesis, \gls{BC} is from now on considered to be a (special) instance of the abstract term \gls{BCT} \label{sec:BCI}.
Additionally, \gls{BCT} is classified as a database technology.
More specific, \gls{BCT} "[...] is a form of distributed ledger technology, deployed on a peer-to-peer network where all data are replicated, shared, and synchronously spread across multiple peers" \cite[13]{Butijn.2020}.
Nowadays many different Blockchain network approaches exist, serving different (special) needs.
Before further details of the different \gls{BCT} approaches are presented, a short recapitulation is given, why all this effort is important. \\
Especially for cryptocurrencies, but also within all the other use cases (e.g.: \cite{Serada.2020}),
fraud in distributed trust-less environments shall be prevented \cite[1]{Xu.2019}.
Within cryptocurrencies, which are build upon \gls{BCT}, fraud is primarily known as \textbf{double spending} of money \cite[1]{Nakamoto.2009}.
This is why \citet{Nakamoto.2009}'s core paper was a breakthrough, as it offers "[...] a solution to the double-spending problem using a peer-to-peer network" \cite[1]{Nakamoto.2009}.
Generally, the "[...] blockchain protocol was designed to maintain a permanent traceable record between two parties that is transparent and
open to public scrutiny without the need of a middleman to authenticate transactions" \cite[483]{Gainsbury.2017}.
Hence \gls{BCT} is not only capable of dealing with money - it can also administer any kind of digital assets. \\
Consequently, in the context of online games, fraud 
can be equated with cheating - the illegal transactions which raise the probabilities to win a game.
Later on the space between providing needed information to keep the network alive and the need for hiding information,
not yet allowed to be revealed, will become small (see section \hyperref[sec:HiddenTransactionsPlusRandomization]{Hidden transactions \& Randomization}).
Especially under theses given circumstances, the prevention of double spending has to be ensured.
To understand the mechanics and techniques to prevent fraud better, this chapter offers core characteristics and an overview on recent mechanisms.
Six of them are briefly described:

\begin{enumerate}
	\item \textbf{Chain Structure} \\
	Data is stored in a linked chain, consisting of 'blocks of data'.
	The blocks are linked by fingerprints (hashes) of previous blocks and the write/update-process is cryptographically secured
	as shown in figure \ref{fig:BlockchainStructure} (alike \citet[654]{Lin.2017}).
	"The exception to this rule is the \textbf{Genesis Block}, which is the first block in the chain and has a predetermined hash"
	\cite[181]{Oliveira.2019}.
	The \textit{genesis block} will be of further importance in section \hyperref[sec:DataAllocationImprovements]{Data allocation improvements}
	of chapter \hyperref[chap:PoT]{Proof-of-Turn},
	\begin{figure}
		\centering{
			\includegraphics[width=.95\linewidth,keepaspectratio=true]{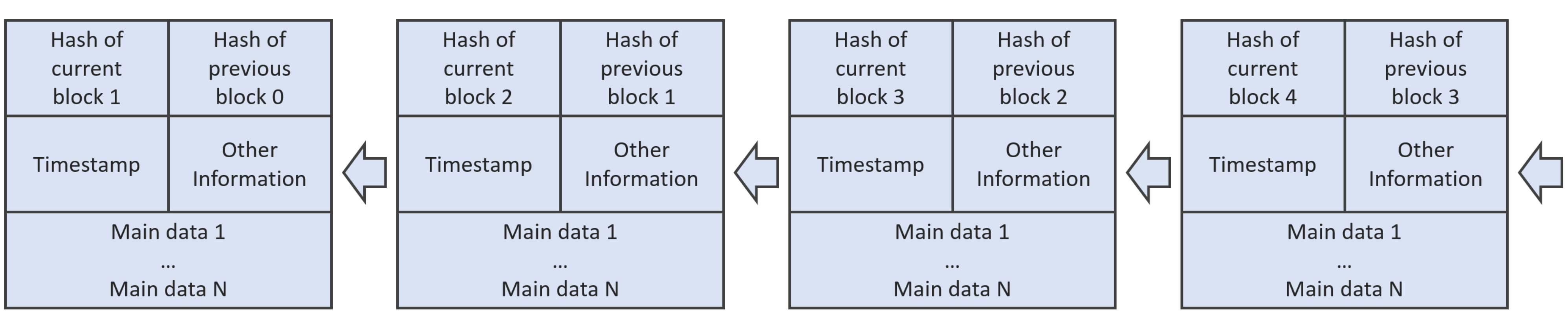}
			\caption{Blockchain Structure (Adapted from \citet{Lin.2017})}
			\label{fig:BlockchainStructure}
		}
	\end{figure}

	\item \textbf{Decentralization, Distribution \& Disintermediation} \\
	"The blockchain is designed for distributing and synchronizing the data across multiple networks"
	\cite[6]{Sharma.2020}.
	"In traditional centralized transaction systems each transaction needs to be validated by a (trusted) third party (e.g., a bank). The decentralized workings of \gls{BCT} enables the direct transfers of digital assets between two counter parties without this third party leading to direct disintermediation" \cite[17]{Butijn.2020}.
	Although \gls{BC}s may be computed by only one peer, the benefit of \gls{BCT} stems from the network and the distribution of data itself,
	reducing the single point of failure \cite[79764]{Bodkhe.2020}.
	Finally all nodes strive to have the whole \gls{BC} stored to prevent fraud and to keep persistence with the other nodes \cite[79766]{Bodkhe.2020}.
	
	\item \textbf{Persistence \& Immutability} \\
	A \gls{BC} "[...] is a permanent record of transactions. Once a block is added, it cannot be altered. This creates trust in the transaction record" \cite[53]{SultanK..2018}.
	Generally data can only be invalidated by reference.
	Nevertheless, data can not be be deleted entirely \cite[88]{Sharma.2020}.
	Hence published data is immutable and distributed transparently throughout the network \cite[60-61]{Butijn.2020}.
	
	\item \textbf{Transparency} \\
	Moreover, until the last node is shut down, the \gls{BC}'s data is immutable published and all peers have the full data set \cite[88]{Sharma.2020}.
	As long as data in blocks is not encrypted to some peers, the BC is transparent and all peers possess the same collective knowledge \cite[483]{Gainsbury.2017}
	Additionally, as (new) data is always signed by a peer, information is traceable and transactions can be verified by all peers \cite[88]{Sharma.2020}.
	
	\item \textbf{Consensus Driven} \\
	Although there are special types, generally all nodes have the right to write blocks, if they follow the given ruleset \cite[54]{Dib.2018}.
	The given ruleset for adding data is called the \gls{CM} and differs regarding the \gls{BC}'s use case \cite[88]{Sharma.2020}.
	
	\item \textbf{Mining} \\
	The action of adding a new block to a \gls{BC} "is called mining
	and the nodes executing the calculations are referred to as miners in the Bitcoin nomenclature" \cite[4]{Butijn.2020}.
	The name stems from the reward - any type of tokens - granted by many \gls{CM}s.
	
\end{enumerate}
\noindent Before the usage of \hyperref[sec:BlockchainInGames]{Blockchain in Games} is introduced in the next chapter,
this chapter presents different \textit{network types and characteristics of \gls{BC}s} as well as commonly used \gls{CM}s.

\FloatBarrier

\section{Network types \& characteristics of Blockchains}
\label{sec:TypesOfBC}

In this section, \textit{accessibility of \gls{BC} networks} will be described before the \textit{layer structure} for pushing data is explained. 
Last, characteristics of \textit{Blocks, Transactions and Smart Contracts} are given.

\begin{enumerate}
	\item \textbf{Accessibility of BC networks} \\
	The following definitions deal with the accessibility of \gls{BC}s.
	Note that the following characteristics regarding accessibility do not influence
	the allocated storage space on the network's nodes - every node stores the full \gls{BC}.
	Generally it is distinguished between an \textbf{access policy} in the \textit{private/hybrid/public}-scheme
	and a \textbf{validation policy} in in the \textit{permissionless/permissioned}-scheme \cite[2]{Daniel.2019}.	
	The networks can be categorized as follows (Table \ref{tbl:BlockchainNetworkTypes}):
	
	\begin{enumerate}
		\item "In \textbf{public permissionless networks}, consensus mechanisms are required to be very strict due to the lack of trust between the participants.
		This is justified by the main feature of a public network, the equality between nodes.
		Once a network participant, the user receives a pair of cryptographic keys to sign and perform transactions.
		Any node can be a miner and participate in the consensus mechanism.
		The problems associated with public blockchains are the fees that must be paid to encourage users to participate in the network and to mine blocks;
		the concern with the network scalability; and the time interval to confirm the transactions.
		Such problems are related to the fact that public permissionless networks are collaborative environments and,
		therefore, depend on the benign behavior of nodes. Besides, for sensitive data networks,
		the fact that all information is available to everyone represents a challenge to data privacy.
		Due to the characteristics of public permissionless networks, costly consensus mechanisms are required" \cite[182]{Oliveira.2019}.
		\label{def:PublicPermissionlessBCNetworks}
		In this manner, \gls{BC} "[...] systems like Bitcoin and Ethereum are called permissionless, i.e. any node on the Internet can join and become a miner" \cite[1]{Angelis.2018}.
		
		\item To cure the needed computation power of \gls{BC} systems in \textit{public permissionless networks},
		"[...] \textbf{public permissioned networks} were developed to implement less costly consensus mechanisms on public networks.
		The difference between the public permissionless networks and the permissioned ones is the different roles that can be played by nodes in the network.
		In public permissioned networks, the node only participates of the network after proper verification of its identity" \cite[182]{Oliveira.2019}.
		\label{def:PublicPermissionedBCNetworks}
		
		\item "The \textbf{private permissionless networks} differ from public networks because they restrict the entry of participants.
		The private network is usually governed by a single institution or a set of institutions, which determine who are the nodes allowed to participate in the network.
		Participant nodes have equal functions and carry the same importance.
		Once authorized to participate in the network, the node can generate transactions and blocks, and participate in consensus" \cite[182]{Oliveira.2019}.
		\label{def:PrivatePermissionlessBCNetworks}
		
		\item "The \textbf{private permissioned networks} only allow some nodes to participate in the consensus process,
		and only a subset of these nodes can generate the next block" \cite[182]{Oliveira.2019}.
		\label{def:PrivatePermissionedBCNetworks}
	\end{enumerate}
	\begin{table}
		\centering
		\begin{tabularx}{0.48\textwidth}{ l | c | c }
			& Permissionless & Permissioned \\ \hline
			Public & \hyperref[def:PublicPermissionlessBCNetworks]{a)} & \hyperref[def:PublicPermissionedBCNetworks]{b)} \\ \hline
			Private & \hyperref[def:PrivatePermissionlessBCNetworks]{c)} & \hyperref[def:PrivatePermissionedBCNetworks]{d)} \\ \hline
			Hybrid & \hyperref[def:HybridPermissionlessBCNetworks]{e)} & \hyperref[def:HybridPermissionedBCNetworks]{f)} \\
			\hline
		\end{tabularx}
		\caption{Blockchain Network Types}
		\label{tbl:BlockchainNetworkTypes}
	\end{table}
	\noindent If \citet{SultanK..2018}'s definitions are considered as well,
	accessibility behaves within a grey scale between \textit{public}, \textit{private} and \textit{hybrid}. \\	
	\textbf{Public} \gls{BC}s ".. have no single owner; are visible by anyone;
	their consensus process is open to all to participate in;
	and they are full decentralized. Bitcoin is an example of a public blockchain" \cite[53]{SultanK..2018}. \\
	\textbf{Private} \gls{BC}s ".. use privileges to control who can read from and write to the blockchain.
	Consensus algorithms and mining usually aren’t required as a single entity
	has ownership and controls block creation" \cite[53]{SultanK..2018}. \\
	\textbf{Hybrid} \gls{BC}s ".. also known as consortium, these blockchains are public only to a privileged group.
	The consensus process is controlled by known, privileged servers using a set of rules agreed to by all parties.
	Copies of the blockchain are only distributed among entitled participants; the network is therefore only partly decentralized" \cite[53]{SultanK..2018}.
	\noindent Concluding, hybrid \gls{BC}s are operating in a public network, controlled by all nodes and hence do not fit into the bipolar private, public scheme.
	Hybrid entries are added as well:	
	\begin{enumerate}[start=5]
		\item \textbf{Hybrid permissionless} \gls{BC}s operate in a public network, but only an associated node can grant access \cite[53]{SultanK..2018}.
		No additional consensus has to take place \cite[53]{SultanK..2018}.
		This does not differ much from a \textit{public permissionless} \gls{BC}
		except the first entry barrier. \label{def:HybridPermissionlessBCNetworks}
		
		\item \textbf{Hybrid permissioned} \gls{BC}s, consequently, operate in a public network as well,
		but access needs to be granted by a majority of nodes (regarding a consensus protocol) \cite[53]{SultanK..2018}.	\label{def:HybridPermissionedBCNetworks}
	\end{enumerate}
	Within both hybrid types, again, the nodes are distributed and not controlled by any higher entity. \\
	
	From here on it is important to keep in mind that a round-robin consensus model is aimed for.
	The round-robin "[...] consensus model is usually used in \textit{permissioned} Blockchain. It allows nodes to take turn one by one for generating blocks" \cite[3]{Khan.2020}.
	Although the following \gls{PoT} mechanism (Section: \hyperref[chap:PoT]{Proof-Of-Turn}) could also be used for \textit{public permissionless} \gls{BC} networks,
	subsequent the focus will be on private as well as hybrid permissioned \gls{BC}s.
	
	\item \textbf{Layer structure} \\
	\gls{BCT} generally consists of four layers \cite[181]{Oliveira.2019},
	which pass data from the publishing (pushing) source node towards the global distribution throughout the network (Figure \ref{fig:LayersOfBlockchains}).
	\begin{figure}[!b]
		\centering{
			\includegraphics[width=.55\linewidth,keepaspectratio=true]{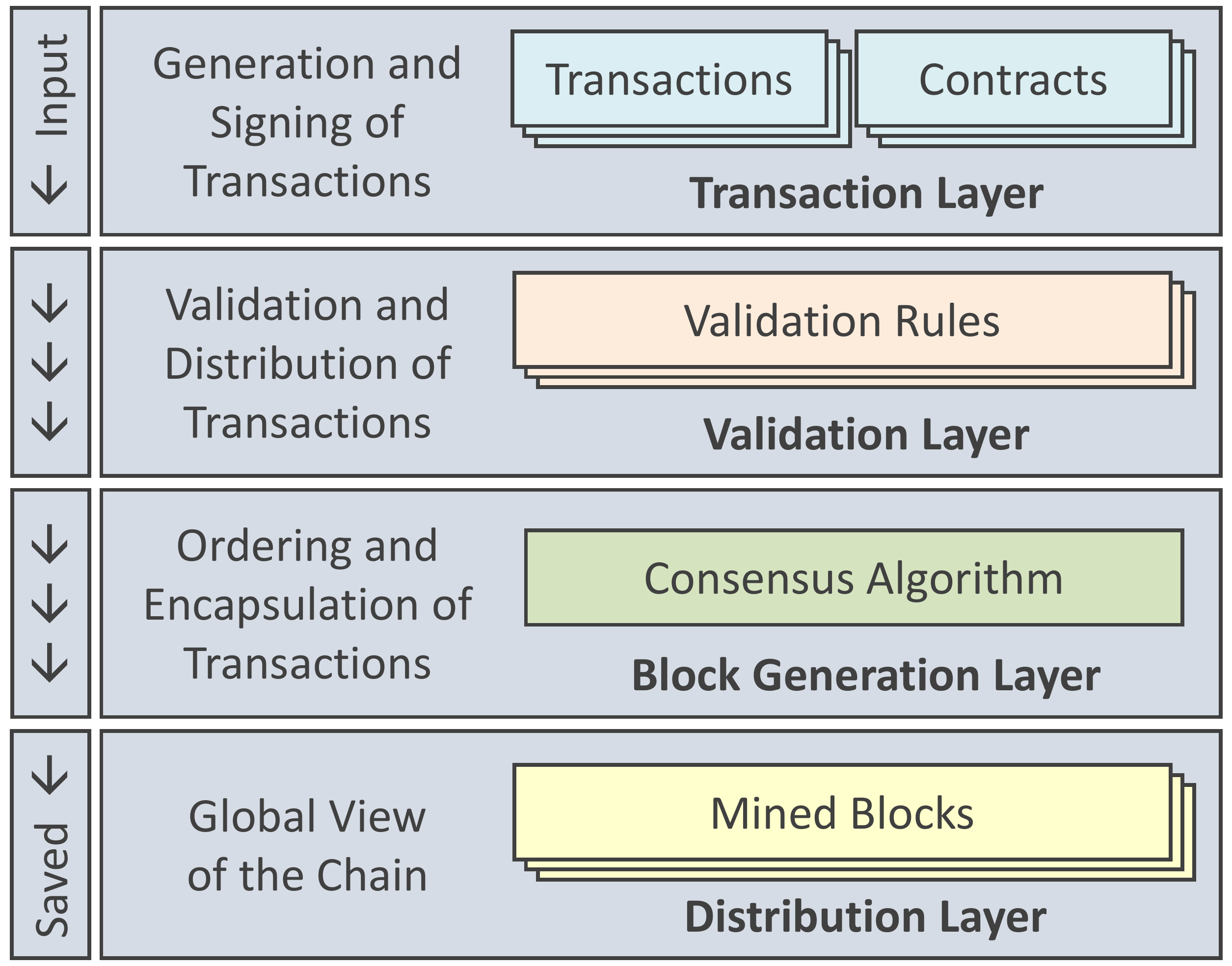}
				\caption{\citeauthor{Oliveira.2019}'s layers of \gls{BC} consensus}
			\label{fig:LayersOfBlockchains}
		}
	\end{figure}
	First, in the \textit{Transaction Layer}, new, local data from \hyperref[def_TransactionsAndSC]{transactions} is signed and given to the validation layer \cite[181]{Oliveira.2019}.
	Second, in the \textit{Validation Layer},
	each of the transactions are validated to the respective predefined rules \cite[181]{Oliveira.2019} - for example, defined by a \gls{SC}.
	As many/all nodes need to comply with the given data, the \hyperref[def_TransactionsAndSC]{transactions} are here (already) distributed among the nodes \cite[181]{Oliveira.2019}.
	\label{sec:SmartContract}
	Third, in the \textit{Block Generation Layer}, many \hyperref[def_TransactionsAndSC]{transactions} are combined and summed up in one block \cite[181]{Oliveira.2019}.
	This block is then passed into any \gls{CM} \cite[181]{Oliveira.2019}.
	This mechanism will distinguish the next added block and has to prevent hostile behavior of nodes (e.g. publishing a block against the consensus rules) \cite[54]{Dib.2018}.
	It can be seen as the gatekeeper for the data to be distributed.
	Please note that the wrapping of transactions into one block is primarily used to reduce computational effort \cite[1204]{Kim.2018},
	as some \gls{CM}s, like  Bitcoin's \gls{PoW}, are computation hungry \cite[11008]{Liu.2018}.
	Still, every transaction may also be appended individually to the \gls{BC}, which is only recommended along computationally lightweight \gls{CM}s.
	Different emendations regarding this manner are given in the following subsection \hyperref[sec:PerformanceImprovements]{Performance improvements}.
	Last, in the \textit{Distribution Layer}, all added blocks are 'finally' stored and the \gls{CM} 'ensures' that data,
	once written here, cannot be deleted \cite[182]{Oliveira.2019}. \\
	Finally, the "[...] data structure of a blockchain, whether public or consortium,
	corresponds to a linked list of blocks containing transactions also referred to as the 'ledger'" \cite[52]{Dib.2018}.

	\item \textbf{Blocks, Transactions and Smart Contracts} \\
	Next to the metadata the linked blocks, building the backbone structure of \gls{BCT}, consist of data, as shown in figure \ref{fig:BlockchainStructure}.
	More specifically all blocks can be seen as new, updated states of a \gls{BC} database \cite[1]{Danzi.52018}.
	Hence they are transactions, sometimes without direct impact or constraints to other nodes, sometimes even with mandatory consultation \cite[1]{Danzi.52018}.
	Moreover, a transaction can be "[...] an exchange of assets that is managed under the entity service’s rules" \cite[51]{SultanK..2018}.
	"These rules also form the basis for smart contracts.
	A smart contract is a set of logic rules in the form of a coded script which can be embedded into the blockchain to govern a transaction" \cite[51]{SultanK..2018}.
	The specific rules of \gls{SC}s and their validation are, understandably, implementation dependent and out of scope.
	Nevertheless, some game related meta-types of \gls{SC}s will be covered
	in the subsection \hyperref[sec:GSSCs]{Game specific smart contracts}
	of the upcoming	chapter \hyperref[chap:BlockchainInGames]{Blockchain in Games}.
\end{enumerate}
For now the skeleton of \gls{BCT} is known.
Thus, \gls{CM}s are now presented in further detail.

\FloatBarrier

\section{Consensus Mechanisms}
\label{sec:ConsensusMechanisms}

A major driver of progress in \gls{BCT} research lies within new \gls{CM}s, which unlock new fields of application.
As stated before, the \gls{CM}s are the gatekeepers for new data within each \gls{BC}.
A \gls{CM} can be grounded on some sort of \textbf{proof}, which has to be delivered by the pushing node \cite[106]{NguyenG.T..2018}.
Further, a \gls{CM} can be grounded on \textbf{votes} as well \cite[106]{NguyenG.T..2018}.
Generally, nodes who participate in the \gls{CM} need any incentive to meet the effort \cite[2]{Catalini.2016}.
Before some popular \gls{CM}s and their benefit system are described and discussed, two characteristics, \gls{BFT} and \gls{CF} need to be addressed: \label{sec:ByzantineFaultTolerance} \\
\textbf{\gls{BFT}} describes the share of nodes within a system, which are allowed to become hostile/unavailable
\textit{before the system is vulnerable} to break down, freeze or becomes prone for fraud \cite[1]{Gramoli.2017}. \label{sec:FinalStateSecurity} \\
\textbf{\gls{CF}} is a phrase to describe the \textit{indelible finality} of written data \cite[3]{Angelis.2018}.
Depending on the chosen \gls{CM}, data in the distribution layer (Figure \ref{fig:LayersOfBlockchains}, Layer 4) are 'only supposed to be final' \cite[3]{Angelis.2018}.
The low probability of \textit{non-finality-data} (e.g. in \gls{PoW}) makes data only implicitly 'consensus final' \cite[3]{Angelis.2018}.
Nevertheless, other \gls{CM}s - e.g. based on voting - provide data with (directly) secured final states \cite[3]{Angelis.2018}.
From a more formal perspective, \citet{Angelis.2018} uses \citet{Vukolic.2016}'s comparison on \gls{PoW} with \gls{BFT}-like approaches
"[...] introducing the distinguishing property of \textbf{consensus finality}: the impossibility of reaching consensus without fully distributed agreement.
In blockchain’s jargon, it amounts to the impossibility of having forks.
As expected, \gls{PoW} does not enjoy \textit{consensus finality} (as forks can happen), while all \gls{BFT}-like approaches does (all parties reach agreement before consensus)" \cite[3]{Angelis.2018}. \\
With this knowledge, \gls{CM}s by \textbf{proof} and \gls{CM}s by \textbf{votes} are given.
The reader has to be reminded that the list is limited on purpose due to manifold \gls{CM}s.

\subsection{Proof-of-Mechanisms}
There are several Proof-of-Mechanisms which differ in their characteristics. \label{sec:PbC}
To get an overview important/general mechanisms are introduced:

\begin{enumerate}
	\item \textbf{Proof-of-Work} \label{sec:PoW} \\
	The mechanism, which is supposed to be the most popular one, is the \gls{PoW} mechanism.
	It is used in the Bitcoin network and was introduced by Satoshi \citet{Nakamoto.2009}.
	Here, nodes have to solve mathematical problems to generate new blocks \cite[4]{Butijn.2020}.
	Pieces of the underlying \hyperref[sec:def_Cryptocurrency]{cryptocurrency} are used as
	incentives for finding the next suitable block (\citet[8]{Nakamoto.2009}; \citet[2]{Catalini.2016}).
	Hence the predominant motivation for the mining nodes to keep the network alive is supposed to be extrinsic.
	Additionally, the Bitcoin implementation adapts its energy consumption along the rising computation power
	within the network to generate a steady output of one block every ten minutes \cite[2254]{Ma.2020}.
	Generally an "[...] attacker would need the majority of the network’s computational power to rewrite history and calculate new blocks" \cite[3-4]{Demi.2021}. \\
	Still, "[...] each node needs to keep the history of all the transactions made in the network, so the storage space keeps increasing,
	and the number of transactions that can be processed by the network is quite limited, around 7 transactions per second [...]" \cite[81]{Besancon.2019}.
	Nevertheless Bitcoin scales along its stakeholders, who are not mining, pretty well - there is no limit in participants \cite[8]{Nakamoto.2009}.
	However, \gls{PoW} fails to reduce power consumption along an increasing mining network \cite[1-2]{Saleh.2020}.
	
	\item \textbf{Proof-of-Stake} \label{sec:DefPoS} \\
	\gls{PoS} was introduced to overcome the energy waste issue of \gls{PoW} (\citet[56]{Chaudhry.2018}; \citet{Narayanan.2018}).
	Whilst everyone is allowed to mine in the \gls{PoW} approach, the \gls{PoS} protects the network's
	stakeholders - those nodes with intrinsic motivation to keep the chain's integrity - from external nodes \cite[56]{Chaudhry.2018}.
	The \gls{PoS} \gls{CM}"[...] saves more energy as compared to the \gls{PoW} model.
	It is an energy efficient variant of \gls{PoW}.
	In \gls{PoS}, miners have to showcase and declare the ownership on a certain amount of currency.
	It is presumed that people with more cryptocurrency or coins would not attack the blockchain network" \cite[3]{Khan.2020}.
	Nevertheless, as nodes always seek to gain rewards, \gls{PoS} is running the risk of a \textit{Nothing-at-Stake problem},
	wherein "a validator will always update the ledger whenever given the opportunity even if the update necessarily perpetuates disagreement" \cite[2]{Saleh.2020}.
	The reduced power consumption stems from the reduced computation power (less mining nodes) within the network \cite[56]{Chaudhry.2018}.
	
	\item \textbf{Proof-of-Authority} \label{sec:DefPoA} \\
	Third, the \gls{PoA} approach, also known as \textit{Proof-of-Identity}, is presented:
	If a node wants to publish transactions, the node has to prove its "[...] identity and should be verifiable in the blockchain network.
	Basically, the publishing nodes are putting their \textbf{identity and reputation} for being a publishing node.
	Publishing node's reputation is directly linked to the publishing node's behavior.
	Any malicious activity by publishing [...] can \textbf{damage the reputation} of the node in the Blockchain network.
	Node reputation would increase if it acts in a manner that Blockchain users agreed with.
	Nodes with less reputation are \textbf{less likely to get the chance to publish a new block}. [...] 
	This model is preferred in permissioned blockchain for it requires a lot of trust on the nodes"	\cite[3]{Khan.2020}.
	
	\item \textbf{Proof-of-Elapsed-Time} \label{sec:DefPoET} \\
	For this mechanism special hardware (Software Guard Extensions - Intel SGX) is needed
	to create a Trusted Execution Environment called TEE, which works on the basis of exact timing \cite[55]{Dib.2018}.
	The \gls{PoET} Consensus Model "[...] selects random leader via election protocol. [...]
	A leader is selected to add the next block to the Blockchain in this model. [...]
	Every miner asks for the running code within the TEE for a waiting time and the miner with the lowest waittime becomes a leader node.
	TEE function can prevent tampering by any internal or external source.
	The only drawback in this consensus model is that it requires special hardware implementation" \cite[3-4]{Khan.2020}.
	
	\item \textbf{Proof-of-Play} \label{sec:DefPoP} \\
	Fifth, in the \gls{PoP} approach the interaction of a player behind a node is evaluated.
	"In the current \gls{PoP} design, the evaluation is adjusted according to the player’s ability.
	This provides a fair chance for everyone to pass through the evaluation stage with enough effort of their ability" \cite[23]{Yuen.2019}.
	Finally, the chosen node writes all parallel published data into one block \cite[23]{Yuen.2019}.
	
\end{enumerate}
\noindent Many of the presented Proof-of-Mechanisms were enhanced using several names, such as: \\
\textbf{Delegated Proof-of-Stake}, \textbf{Proof-of-Activity} (\gls{PoP}/\gls{PoS}-hybrid), \textbf{Proof-of-Achievement} (\citet{Komiya.2019}),
\textbf{Proof-of-Burn}, \textbf{Proof-of-Capacity} (or Proof-of-Storage), \textbf{Proof-of-Excellence} \cite[5]{King.2012}, \textbf{Proof-of-Existence}, \textbf{Proof-of-Importance},
\textbf{Proof-of-SequentialWork}, \textbf{Proof-of-Validation} (PoV), \textbf{Proof-of-Vote} as well as \textbf{Proof-of-WorkOrKnowledge} and \textbf{Proof-of-ZeroKnowledge}. \\
To dig deeper into most of these algorithms, \citet[5]{Butijn.2020} and \citet[3]{Angelis.2018} are recommended as starting points.
Nevertheless, these enhancements are considered to be out of scope due to their specificity. \\
Until now the permission to write a block was either dependent on available computation power (\gls{PoW}),
share of stake (\gls{PoS}), reputation within the network (\gls{PoA}),
exact timing (\gls{PoET}) or some kind of special effort (\gls{PoP}). \\
In Proof-of-Mechanisms much independence among the writing nodes is given - nodes are never asked to consent (explicitly) \cite[8]{Nakamoto.2009}.
Therefore competing versions of the \gls{BC}, known as \textbf{forks}, occur \cite[60]{Butijn.2020}.
"A user only needs to keep a copy of the block headers of the longest proof-of-work chain, which he can get by querying network nodes until he's convinced he has the longest chain, and obtain the Merkle branch linking the transaction to the block it's timestamped in" \cite[5]{Nakamoto.2009}.
A miner "[...] follows the \textbf{longest-chain rule} if she always chooses the last block of one of the longest chains" \cite[2]{Ewerhart.2020}.
This behavior, formally known as the \textit{Longest Chain Rule}, was first described by \citet{Courtois.2014}.
"Note that the \textit{longest-chain rule} is a class of strategies, rather than a single strategy" \cite[2]{Ewerhart.2020}.
The \textit{Longest Chain Rule} is the reason why blocks in some Proof-of-Mechanism \gls{BC}s may never reach $100$\% \gls{CF} \cite[3]{Angelis.2018}. \\
"Because of the possibility of forks, there is no such thing as absolute reliability of the data retrieved from the blockchain.
It is decreasingly high toward the most recent blocks data, as one only get the version of the ledger stored on a node at a given time,
so that a blockchain-specific time-dependent reliability weight has to be determined" \cite[62]{Dib.2018}.
Some Proof-of-Mechanisms have found their way around this flaw, such as \gls{PoET} and \gls{PoP}.
Nevertheless, on vote based (/consortium) \gls{BC}s, "[...] the use of adapted
consensus algorithm allows for “block finality”: once a block has been validated,
it remains on the main chain and forks are not allowed" \cite[54]{Dib.2018}.

\subsection{Vote-based Consensus}
\label{sec:VbC}
In contrast to \textit{Proof-of-Mechanisms}, \textbf{Vote-based-Mechanisms} - also
known as \textbf{consortium} mechanisms - only accept blocks which already reached consensus \cite[3]{Angelis.2018}. 
The accepted blocks reach a $100$\% \gls{CF} directly after approval \cite[3]{Angelis.2018}.
"The consensus is coordinated by the distributed nodes controlled by consortium partners which will come to a decentralized arbitration by voting. The key idea is to establish different security identity for network participants, so that the submission and verification of the blocks are decided by the agencies’ voting in the league without the depending on a third-party intermediary or uncontrollable public awareness. Compared with the fully decentralized consensus–Proof of Work (\gls{PoW}), vote-based consensus has controllable security, convergence reliability, only one block confirmation to achieve the transaction finality, and low-delay transaction verification time" \cite[466]{Kejiao.2017}. \\

Again there are multiple solutions:
\begin{enumerate}
	\item \textbf{Practical Byzantine Fault Tolerance Consensus Model} \label{sec:PracticalBFTCM} \\
	Malicious "[...] attacks, operator mistakes, and software errors are common causes of failure and
	they can cause faulty nodes to exhibit arbitrary behavior, that is, Byzantine faults" \cite[399]{Castro.2002}.
	The \gls{PBFT} is designed to prevent Byzantine faults.
	"It is mostly used in permissioned Blockchain such as in Hyperledger as it can manage up to 1/3 malicious byzantine replicas.
	In practical \gls{PBFT}, the next block is accepted in a round process.
	There are certain processes to follow in every round for choosing a primary node.
	The practical \gls{PBFT} model's process is classified into 3 well-defined phases: preprepared, prepared, and commit.
	For changing the state, the node should have at least 2/3 votes from all nodes.
	It will ensure that all nodes are recognizable and know with each and every node.
	[...] In \gls{PBFT} each node needs to inquire about other nodes"
	\cite[4]{Khan.2020}.
	\bigbreak
	
	\item \textbf{Ripple Consensus Model} \label{sec:RippleCM} \\
	This model will further be called \textbf{Ripple}.
	Every "[...] node needs to create a unique node list (UNL). All \gls{Ripple} nodes are part of UNL and are considered as reliable nodes for all other nodes to rely on them.
	No node are to go against UNL.
	\gls{Ripple} network encourages all nodes to communicate with various nodes available in that UNL to achieve consensus in the network.
	Every node in the UNL should make 40\% of overlap with all other nodes.
	In the \gls{Ripple} network, consensus could be achieved in a couple of rounds where all nodes understand the responsibility for assembling the transaction
	with the state of the candidate set in a "a well-known data structure" and transmit its candidate sets to nodes in the current UNL.
	Nodes are responsible for validating the transaction, vote for a specific transaction and transmit the votes to the network based on the results of the collective votes.
	Every node filters out its candidate set and transaction receiving the highest votes are transferred to coming round of consensus.
	Right after achieving the 80\% votes of candidates set from every node available in the UNL then that particular candidate set becomes the ledger in \gls{Ripple} term.
	Another round of consensus would start with the new and pending transaction which could not get accepted.
	Consensus in the whole network can only be achieved after every sub-network achieves consensus"
	\cite[4]{Khan.2020}.
	
	\item \textbf{Tendermind Consensus Model} \label{sec:TendermindCM} \\
	\citet[4]{Khan.2020} summarizes the \gls{TCM} from \cite{Kwon.2014} as follows:
	\gls{TCM} "[...] is another variant of \gls{PBFT} model which mostly works with permissionless Blockchain.
	In this model, clients have the privilege of directly creating a new transaction to the nodes.
	The clients in this model utilize the gossip protocol to broadcast the transaction for the validating nodes.
	Within the Brach broadcast pattern, the progression would be done with the external validating situation so
	validating nodes are required to collect the transaction by gossip right before it gives a right to include the transaction in the block. \\
	The significant difference between Tendermind and \gls{PBFT} is that it continuously rotates the leader node right.
	Tendermind takes \gls{PBFT} view change mechanism into the common case pattern.
	Here this shows as it is waiting to timeout, and validating nodes wait for the leader node to convey the first message in the Bracha broadcast pattern.
	which is more relevant to the view change the timer in \gls{PBFT}.
	As the timer expires validating nodes vote to a nil block and validating nodes take part n the Bracha Broadcast message pattern on a continuous basis.
	The Tendermind was suffering from many issues and the obvious one was livelock, presenting locking and unlocking votes by validating nodes."
	
\end{enumerate}

\noindent The last two approaches which are covered are enhancements of above given \gls{CM}s: \textit{MultiChain} and \textit{Child-chains}. \\
\textbf{MultiChain} "[...] enforces a round robin schedule, in which the permitted miners must create blocks in rotation
in order to generate a valid blockchain" \cite[7]{Greenspan.2015}. \label{def:MultiChain}
Hereby the round robin is implicitly enforced regarding a value called \textit{mining diversity} \cite[7]{Greenspan.2015}.
"A value of 1 ensures that every permitted miner is included in the rotation, whereas 0
represents no restriction at all. In general, higher values are safer, but a value too close to 1 can
cause the blockchain to freeze up [...]"
\cite[7]{Greenspan.2015} due to inactive miners - the network stalls.
If a network split occurs, "[...] the fork with the longer chain will be adopted as the global consensus.
The diversity threshold ensures that the longer blockchain will belong to the island
containing the majority of permitted miners, since the other island’s chain will quickly freeze up"
\cite[8]{Greenspan.2015}.
\textit{MultiChain} is only an example for a general round robin approach. \\
\textbf{Child-chains} can be used with every \gls{BCT} and presuppose parallel \gls{BC} threads.
Further, \textit{side-chains} assume that there are transactions which necessarily belong on the main
tread (primary chain) and others which do not need to be stored on the primary chain \cite[1206]{Kim.2018}.
Therefore the ladder can reduce the load on the \textit{primary chain}.
Here transactions which are written to the \textit{primary chain} are called \textit{on-chain} and those written to any \textit{sidechains}, \textit{off-chain}.
"In a traditional on-chain transaction, each transfer would have to be validated by all peers in the network before the transfer is labelled as complete, which keeps it very slow.
In contrast, in an off-chain transfer, not all peers need to wait for the transfer to be verified before it is labelled as successful or complete" 
\cite[15]{Sharma.2020}. \\
Concluding, \textit{MultiChain} seeks a reduction of needed overall computation power and \textit{child-chains} reduce the reconciliation effort.
Last, as \textit{child-chains} are not tied to a special \gls{CM}, they are covered in the following subsection \hyperref[sec:PerformanceImprovements]{Performance improvements}.
A comparison of the given \gls{CM}s follows thereafter.

\subsection{Performance Improvements}
\label{sec:PerformanceImprovements}
Next to the general functionalities to achieve consensus, there are some scalability solutions, which can be applied to multiple upper level \gls{CM}s.
On this topic, \citet[1204]{Kim.2018} offer five categories of improvement-methods they found in their survey.
Whilst On-chain, Off-chain and Child-chain improvements operate within a main-chain's network (only),
Side-chain and Inter-chain try to connect \gls{BC}s out of different networks.
The key facts are given briefly:

\begin{enumerate}
	\item "The \textbf{on-chain} solution refers to methods that increase scalability by modifying only elements within a blockchain" \cite[1204]{Kim.2018}.
	One prominent example would be to increase the number of transactions within one block until it is considered a 'Big block' \cite[1204]{Kim.2018}. \\
	In the following specific case, the "[...] Bitcoin Core protocol limits blocks to 1 MB in size.
	Each block contains at most some 4,000 transactions.
	Blocks are added to the blockchain	on average every 10 minutes,
	therefore the transaction rate is limited to some 7 transactions per second" \cite[1]{Goebel.2017}.
	\begin{table}[!b]
		\centering
		\begin{tabularx}{0.80\textwidth}{ l | c | c | c }
			& Transaction / second & Transactions / block & Block size \\ \hline
			Bitcoin Core & $7$ & $4.000$ & $1$ MB \\ \hline
			Big block (1) & $70$ & $40.000$ & $10$ MB \\ \hline
			Big block (2) & $700$ & $400.000$ & $100$ MB \\ \hline
		\end{tabularx}
		\caption{Blockchain Blocksize}
		\label{tbl:BlockSize}
	\end{table}
	If a block was allowed to be bigger than the traditional 1 MB of Bitcoin Core, it would enable more
	transactions to be processed in the same amount of time (Table \ref{tbl:BlockSize}: 'Big block' 1 \& 2),
	as solving the \gls{PoW}'s mathematical mining problem is barely constraint to the actual block size (\cite{Zhang.2018}).
		
	\item "The \textbf{off-chain} solution is to improve the scalability by processing the transactions at outside the blockchain.
	This is also called a \textbf{state-channel} solution, because it maintains the state of the main-chain and applies the last state that has been processed in the other channel" \cite[1205]{Kim.2018}.
	In other words, in "[...] a traditional \textit{on-chain} transaction, each transfer would have to be validated by all peers
	in the network before the transfer is labelled as complete, which keeps it very slow.
	In contrast, in an \textit{off-chain} transfer, not all peers need to wait for the transfer to be verified before it is labelled as successful or complete" \cite[15]{Sharma.2020}.
	'The Bitcoin Lightning Network' (\citet{Poon.2016}), Sprites (\citet{Miller.2017}) and game channels (\citet{Kraft.2016}) fit into this category.
	All three papers split transactions from the main chain, starting with an introducing interaction just to merge the results of the sidechain branch back into the main chain later on.
	In figure \ref{fig:ChainImprovements} temporary off-chains can be seen. 
	Whilst the main-chain is managed by all nodes of the network,
	the \textit{off-chain state channels} are only governed by the (two) affected nodes.
	Generally, \citet{Poon.2016} can be seen as the founders of
	this idea and \citet{Miller.2017}'s research aims to further improve the speed.
	As those two are located in the cryptocurrency business only, they are out of scope.
	Regarding \gls{BCT} in games, \citet{Laneve.2019} (p. 27) state that this "[...] approach gives the developers all the flexibility they need with proven infrastructures while decentralising only the parts that need it".
	Although \citet{Kraft.2016} is using \gls{BCT} based on a cryptocurrency in both on-chain as well as off-chain transactions, he is located within online gaming business.
	Therefore \citet{Kraft.2016}'s research will become important later on.
	
	\begin{figure}[!b]
		\centering{
			\includegraphics[width=.95\linewidth,keepaspectratio=true]{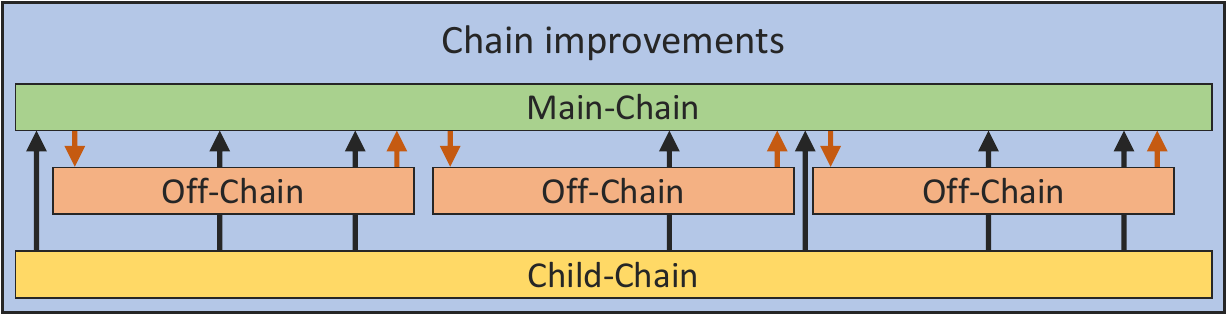}
			\caption{Off-Chain \& Childchain}
			\label{fig:ChainImprovements}
		}
	\end{figure}
	
	\item "The \textbf{child-chain} solution has a parent-child structure, processes the transactions in the child-chain, and records the results in the parent-chain" \cite[1206]{Kim.2018}.
	On the one hand, this can be used for further improvements on cryptocurrency transactions (\citet[1206]{Kim.2018}),
	on the other hand, special transactions can be stored on the child-chain,
	which are only transferred to the parent-chain on fulfilled
	smart contracts (see \hyperref[sec:DataAllocationImprovements]{Data allocation improvements} in chapter \hyperref[chap:BlockchainInGames]{Blockchain in Games}).
	Consequently, in figure \ref{fig:ChainImprovements} a sample permanent \textit{child-chain} 
	is 'every once in a while' used to push certain data into the \textit{main-chain} (dark arrows).
	
	\item "The \textbf{side-chain} approach is to exchange assets of different blockchains with each other.
	And their goal is to bring the function of another blockchain into the current blockchain" \cite[1205]{Kim.2018}.
	In the first place, two \gls{BC}s with different \gls{CM}s are combined to obtain the advantages of both approaches, such as using smart contracts of Ethereum together with Bitcoin's cryptocurrency (\citet{Kim.2018}, p. 1205-1206).
	Here, prominent example which is mentioned more often in the literature is \citet{Back.2014}'s paper about 'Pegged Sidechains'. \\
	In figure \ref{fig:ChainImprovements} the side-chain approach is used,
	if one of the layers needs a divergent \gls{CM} to serve special service characteristics.
	In section \hyperref[sec:Interoperability]{Interoperability} in chapter \hyperref[chap:PoT]{Proof-of-Turn}
	the \gls{PoT} \gls{CM} serves as as an off-chain/side-chain solution for a \gls{PoW} \gls{BC}
	(e.g. for an ecosystem's in-game currency), whilst the
	\gls{PoT} \gls{CM} is served by an off-chain/side-chain \gls{PBFT} \gls{BC} to enable straight forward and lightweight voting (Figure \ref{fig:MultiLayerBC}).
	
	\item "The \textbf{inter-chain} method is a way to enable communication among the various blockchain" \cite[1206]{Kim.2018}.
	Inter-chain is very similar to the side-chain approach and tries to make e.g. Litecoin and Bitcoin interoperable \cite[1206]{Kim.2018}.
	Hence, the inter-chain method can be seen as the infrastructure technology for implementing the side-chain \cite[1206]{Kim.2018}.
\end{enumerate}
Further advantages and disadvantages, here out of scope, can be taken from \citet{Kim.2018} (p. 1206, Table 1., A comparative analysis table of the scalability issues).
Concluding, if possible, a novel \gls{CM} shall offer the possibility to be interoperable, as \citet[81]{Besancon.2019} states that the need for interoperability
"[...] can be found at multiple levels: a) between different BC, b) between different projects running on the same BC, and c) between BC and other technologies used to create decentralized applications." \\
Later on, during the design of \gls{PoT}, \textbf{on-chain} improvements in regards of number of blocks written within a given time slot,
the \textbf{off-chain} method regarding interoperability,
\textbf{child-chains} to counterweight in-game mechanics as well as 
\textbf{side-chain} solutions for mixing \gls{CM} characteristics will be of interest.

\FloatBarrier

\subsection{Comparison of Consensus Mechanisms}

\begin{table}
	\centering
	\begin{tabularx}{0.575\textwidth}{ c | c | c }
		\textbf{Proof based} & \textbf{Write} & \textbf{\gls{BFT}} \\
		\textbf{\gls{CM}s} & \textbf{permission} & \\ \hline \hline
		\gls{PoW} & Heavy computation & \textless $25$ \% \\ \hline
		\gls{PoS} & Stakeholder & \textless $50$ \%  \\ \hline
		\gls{PoA} & Authority & \textless $51$ \%  \\ \hline
		\hyperref[sec:MultiChain]{MultiChain} & Round Robin & \textminus  \\ \hline
		\gls{PoET} & Time & \textminus \\ \hline \hline
	\end{tabularx}
	\caption{Proof based consensus mechanisms \cite[5]{Khan.2020}}
	\label{tbl:SumConsensusMechanisms_1}
\end{table}

\noindent To sum up the given \gls{CM} overview, Table \ref{tbl:SumConsensusMechanisms_1} (Proof based \gls{CM}s) and Table \ref{tbl:SumConsensusMechanisms_2} (Vote based \gls{CM}s)
show the mechanism's characteristics regarding \textit{write permissions} and \textit{\gls{BFT}}.
Values are taken from \citet[5]{Khan.2020}, which match
values from \citet[55]{Dib.2018}.

\begin{table}
	\centering
	\begin{tabularx}{0.55\textwidth}{ c | c | c }
		\textbf{Vote based} & \textbf{Write} & \textbf{\gls{BFT}} \\
		\textbf{\gls{CM}s} & \textbf{permission} & \\ \hline \hline
		\gls{PBFT} & Vote rounds  & \textless $33.3$ \% \\ \hline
		\gls{Ripple} & Leader rotation & \textless $20$ \%  \\ \hline
		\gls{TCM} & Leader rotation & \textless $33.3$ \%  \\ \hline
		\hline
	\end{tabularx}
	\caption{Vote based consensus mechanisms \cite[5]{Khan.2020}}
	\label{tbl:SumConsensusMechanisms_2}
\end{table}

\noindent Although values in Table \ref{tbl:SumConsensusMechanisms_3} (Additional \gls{CM}s) are only anticipated, they are given to complete the list.
\begin{table}
	\centering
	\begin{tabularx}{0.75\textwidth}{ c | c | c }
		\textbf{Proof based} & \textbf{Write} & \textbf{\gls{BFT}} \\
		\textbf{\gls{CM}s} & \textbf{permission} & \\ \hline \hline
		\gls{PoP} & Effort in game & \textminus  \\ \hline
		\gls{PoT} & Round Robin, turn (time)  & Implementation \\
		 & and special rules & dependent \\ \hline
	\end{tabularx}
	\caption{Additional consensus mechanisms}
	\label{tbl:SumConsensusMechanisms_3}
\end{table}
\noindent Table \ref{tbl:SumConsensusMechanisms_1}, \ref{tbl:SumConsensusMechanisms_2} and
\ref{tbl:SumConsensusMechanisms_3} show the different approaches, which can be used to model a game specific \gls{CM}.
All values, to the author's knowledge, are given for plain \gls{CM}s without enhancements from the previous section \hyperref[sec:PerformanceImprovements]{performance improvements}. \\
Nevertheless, the "[...] existing proof-of-based models seem to be supporting open participation
but they are not suitable for the real-time applications where immediate
transaction finality and high transaction rates are the main requirement" \cite[6]{Khan.2020}.
Finally, as all theses mechanisms are solving storage related issues, the '\textit{CAP theorem}' as described by \cite{Brewer.2012}, has to be mentioned as well.
In a distributed data store, such as \gls{BCT}, out of the three following properties \gls{C}, \gls{A} and \gls{P} only two can be ensured \cite[6]{Angelis.2018}.
Thus either \gls{C}$+$\gls{A}, \gls{C}$+$\gls{P} or \gls{A}$+$\gls{P} can fully be met \cite[6]{Angelis.2018}.
Additionally, "[...] an Internet-deployed permissioned blockchain has to tolerate these adverse situations:
(i) periods where the network behaves asynchronously;
(ii) a (bounded) number of Byzantine authorities aiming at hampering availability and consistency" \cite[7]{Angelis.2018}.
Consequently from (i), there is no way around \gls{P} and a trade off between \gls{C} and \gls{A} has to be found \cite[7]{Angelis.2018}.
The CAP theorem will further be covered in chapter \hyperref[chap:PoT]{Proof-of-Turn}, section \hyperref[sec:CAPtheorem]{CAP Theorem measurement}. \\

\noindent Additionally, from the given information it can be concluded that \textit{forks are expensive}.
Consequently, designing a \gls{CM} without forks or \textit{diminished computation in vain} along forks is strived for to increase energy efficiency and to reduce network traffic as well as inconsistencies.
Still, to meet the needs of speed/throughput the interoperability with \textbf{off-chain}, \textbf{side-chain} and \textbf{child-chain} improvements is beneficial. \\
Revising the research question, \textit{"What is the best fitting \gls{BCT} \gls{CM} to cover an asynchronous game play scenario?"}, the answer is not clear, yet.
If there was a novel \gls{CM} design, the lowest possible consultation in regards of the given scenario is aimed for.
This solution does not necessarily need an attached cryptocurrency. \\
In chapter \hyperref[sec:PoT]{Proof-of-Turn} the following characteristics will become important:
\begin{enumerate}
	\item \textit{Round Robin}, in regards of reducing the \textit{permissioned writing node} down to one
	using a fixed succession.
	\item \textit{Time}, not as strict as in \gls{PoET}, but not negligible.
	\item \textit{\gls{BFT}}, to an (at least) reasonable extend of \textless $33.3$ \%, or more (implementation dependent).
	\item \textit{\gls{CF}}, both in the fashion of short-termed forks as well as possible long-termed rewrites of the blockchain.
	\item \textit{\gls{PoP}}, in regards of the final scope of the \gls{PoT} \gls{CM} for games.
	\item \textit{Off-chain}, as stated in \citet{Kraft.2016}'s paper using the phrase \textit{Game Channel}.
	\item \textit{Child-chain(s)/Side-chain(s)}, in the manner of organizing different types of transactions.
	\item \textit{Interoperability}, as an additional desirable characteristic.
\end{enumerate}

\noindent Still, before getting into the details of the \gls{PoT} \gls{CM} in chapter \hyperref[chap:PoT]{Proof-of-Turn},
the next chapter \hyperref[sec:BlockchainInGames]{Blockchain in Games} will list characteristics \gls{BCT} needs to fulfill in general to serve turn based games.


\chapter{Blockchain Technology in Games}
\label{chap:BlockchainInGames}
Since \citet{Nakamoto.2009}'s paper, many different \gls{CM}s were invented,
as given in section \hyperref[sec:ConsensusMechanisms]{Consensus Mechanisms}.
Currently the use cases of \gls{BCT} in games can be grouped by concepts of 'real-world ownership',
'digital tokens which represent game assets', the 'reduction of upkeep costs' as well as 'enhanced engagement from the user-base' \cite[15]{Laneve.2019}.
In other words, \citet[1]{Min.2019} state that \gls{BCT} in games is used for \textit{Rule Transparency},
\textit{Asset Ownership}, \textit{Assets Reusability}, \textit{User-Generated Content} as well as  \textit{Current scalability}.
Therefore, using ".. games to grasp the functionalities of cryptocurrencies and the blockchain technologies that make them possible seems like an obvious choice"
\cite[2]{Serada.2020}, especially as beneficial synergies between the needs of gaming networks and promises of \gls{BCT} can be found.
This intersection seems to be situated around slow paced asynchronous games, as \gls{BCT} is basically not suited for fast applications \cite[19]{Serada.2020}.
Therefore the following chapter wants to answer the first guiding question:
\textit{"Which kind of \gls{SC}s need to be established to cover typical in-game mechanics?"}. \\
Hence, \textbf{first} recent solutions are analyzed and grouped by \textit{ownership}, \textit{gameplay} and \textit{network costs}.
\textbf{Second}, the gaming market is examined briefly and some sample games are presented.
This shows blind spots of \gls{BCT} game play and respective (asynchronous) in-game mechanics.
\textbf{Third}, a problem area is defined to increase tangibility of problems in distributed games.
\textbf{Last} some solutions to these problems are shown, custom tailored for distributed game play. \\
Afterwards, assuming that \gls{BCT} is suitable for (asynchronous) games, the following chapter,
\hyperref[chap:PoT]{Proof-of-Turn}, will provide a \gls{CM} which aims to deliver
the features described at the end of the last chapter, \hyperref[chap:BCT]{Blockchain Technology}.

\section{Ownership,  network costs, gameplay \& scalability}
\label{sec:OwnershipGameplayNetwork}
\begin{figure}
	\centering{
		\includegraphics[width=.95\linewidth,keepaspectratio=true]{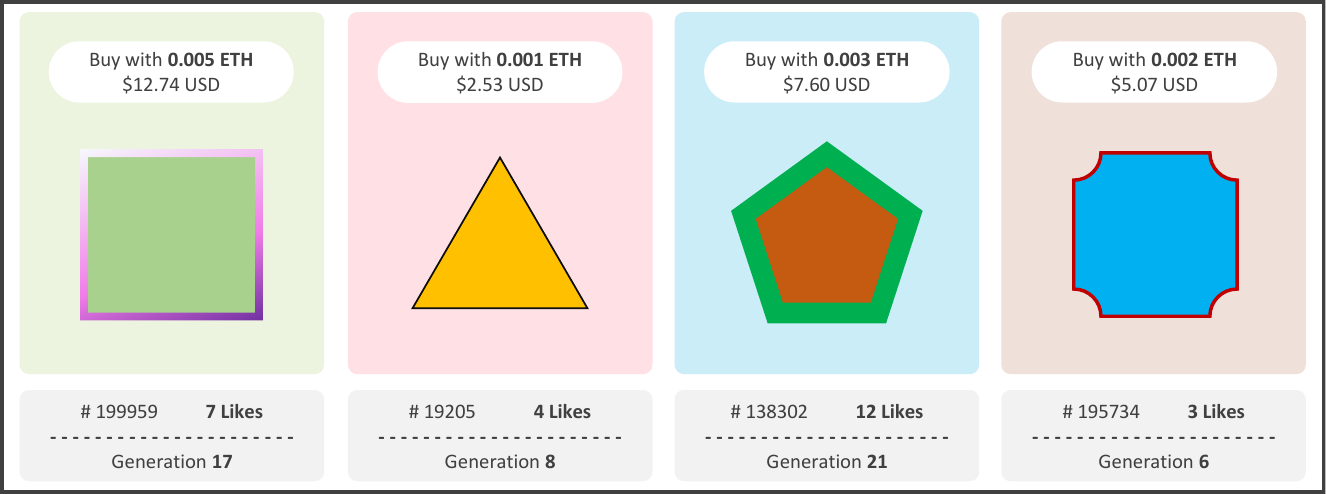}
		\caption{Collectibles store 'CryptoShapes' (Adapted from \citet{CryptoKitties.co.2021})}
		\label{fig:Collectibles}
	}
\end{figure}

For \textbf{ownership} models, the general idea is derived from Bitcoin which administers a cryptocurrency.
Besides or instead of a cryptocurrency, digital assets are regulated by the game's \gls{BC} \cite[2]{Pfeiffer.2020}.
This idea is called \textit{Collectibles (Collection Games)} and, as it is not too circuitous,
it was one of the first mentioned type of games hosted on \gls{BCT}.
"The ownership enables the game assets to be independent of specific game operators, which allow the players to retain their digital properties and in-game relationships, even after the game stops its operation" \cite[1]{Min.2019}.
A prominent example is \textit{CryptoKitties} \cite[2]{Pfeiffer.2020}.
Next to CryptoKitties many other cryptogames were developed (e.g., Cryptopunks, Decentraland, MyCryptoHeroes, HyperDragons, Gods Unchained, Etheremon, Blockchain Cuties, NeoWorld, and Axie Infinity) \cite[2]{Serada.2020}, which fall into this category.
A sample store to buy assets from these \textit{Collectibles} might look like figure \ref{fig:Collectibles}.
Digital assets can be bought using some sort of payment channel (here the cryptocurrency: \gls{ETH}).
Sometimes \textit{Collectibles} offer the opportunity to breed new assets out of the existing ones \cite[2]{Serada.2020}.
During breeding properties (here: border thickness, colors, edges, size etc.) of two shapes are recombined.
To increase asset diversity, sometimes mutations happen and new characteristics (e.g. color gradients, figure \ref{fig:Collectibles}: rectangle) occur, which makes the single asset special.
As a bread shape can be considered a child-generation of the two parent shapes, breeding generations are given (compare \cite{CryptoKitties.co.2021}).
The breeding may be rather seen as part of the \textit{Collectibles}-game than part of the asset ownership itself \cite[2]{Pfeiffer.2020}.
Further, one advantage of \textit{Collectibles} is, that the "[...] possibility of asset trades across different games and blockchain platforms stimulates the players to better engage in the game economy" \cite[1]{Min.2019}.
Regarding \textit{reusability}, game "[...] developers can leverage blockchain to design an ecosystem that allows players to reuse their characters and virtual items across different games.
To this end, newly launched games can directly inherit game assets from the existing ones" \cite[1]{Min.2019}. \\
Moreover, User-Generated Contents "[...] in traditional games are restricted in the specific game, thus, belongs to the game operator.
In contrast, these contents can be preserved by the players, thus, has the potential to be shared among multiple games.
This benefit will, in turn, encourage the players to participate in the construction of new content"
\cite[1]{Min.2019}. \\
These connected games are called 'Layer two games' as they use \gls{NFTs} "[...] from other games to fuel their gameplay.
Players could only access these games by using items they had acquired from other sources" \cite[26]{Laneve.2019}.
Depending on the used cryptocurrencies or tokens, "[...] game operators may benefit from the value increase of the tokens they issued" \cite[1]{Min.2019}. \\

\noindent \textbf{Network costs} \textemdash \hspace{0.075cm} No game has been found which uses \gls{BCT}
only for the sake of reducing upkeep costs of servers, yet.
In this regard, \citet[3]{Serada.2020} states:
"Oftenmost cryptogames are neither developed by game companies nor is their
design shaped by typical business models or assumptions such as profit making." \\

\noindent \textbf{Gameplay} \textemdash \hspace{0.075cm}Here,
the literature emphasizes the value of \textit{rule transparency} as an anti-cheating mechanism and \gls{BCT}'s potential in games (\citet[57]{Dib.2018}; \citet[1]{Min.2019}).
\gls{BCT} "[...] addresses one of the biggest issues with direct peer-to-peer transactions online; that is, a lack of trust.
Historically, a lack of trust has been cited as one of the greatest disadvantages of online gambling [...]" \cite[483]{Gainsbury.2017}.
"Due to the transparent characteristic of blockchain data, players or third-party organizations can audit
the smart contract based games rules, which was hidden in the centralized server in traditional games.
The transparent game rules will enhance the trustworthy of the game operation" \cite[1]{Min.2019}. \\
Yet, except the theoretical \gls{PoP} paper from \citet{Yuen.2019}, the literature only shows one single example of \gls{BCT} in games to effectively replace the \textit{central server network approach} (\citet{Wu.2020}) especially if cryptocurrency ownership models are \textit{excluded}.
\citet[4559-4560]{Wu.2020} came to the same conclusion:
"To the best of our knowledge, this is the first blockchain-related game demostration that achieves a real-time serverless gaming system with an anti-cheating mechanism."
Still no professionally produced game could be found, which already left its BETA-status or
can be considered more than a proof of concept, such as \textit{Huntercoin} (\citet{Kraft.2016}). \\

\noindent \textbf{Current scalability} \textemdash \hspace{0.075cm}Additionally, scalability of BCT has to be put in contrast:
"As it stands, blockchain technology does not seem applicable for the design of the most popular game genres such as first-person shooters or real-time strategy, although several attempts have been made in this direction [...]" \cite[3]{Serada.2020}.
\begin{figure}[!b]
	\centering{
		\includegraphics[width=.95\linewidth,keepaspectratio=true]{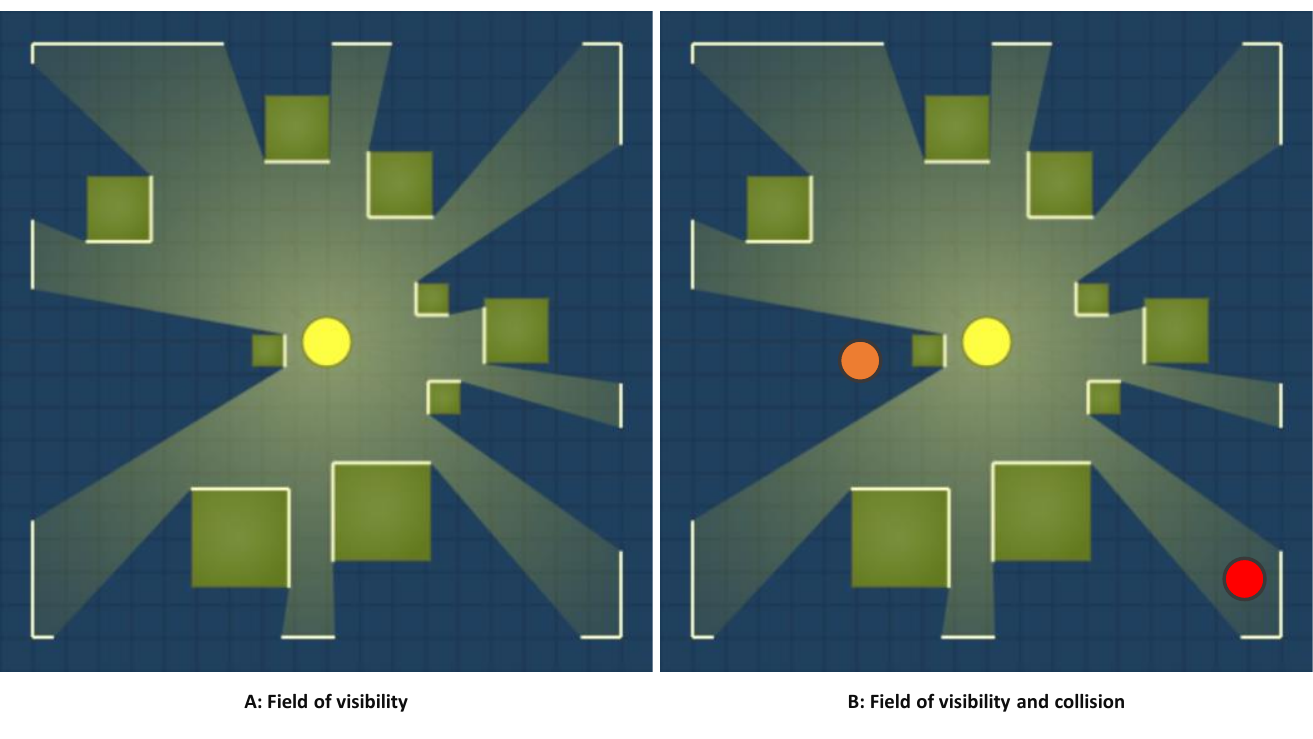}
		\caption{Visibility and Scalability}
		\label{fig:VisibilityScalability}
	}
\end{figure}
This statement derives from \gls{BCT} focusing on consistency and partition tolerance regarding the CAP-Theorem instead of high availability (\citet{Angelis.2018}).
Additionally cryptocurrencies like Ethereum and Bitcoin show that "read availability of blockchains is typically high [...]"
whilst "[...] write availability — for transaction management — is actually low" \cite[64]{Weber.2017}. \\
Furthermore, in a centralized server context, all players ($p$) send their information to the
server - 'one on one' relationships for the players and 'one to $p$' for the server, sum: $2p$.
In \gls{CM}s wherein all nodes are allowed to write, the number of relationships ($r$) becomes squared ($r = p^2$).
Consequently, consultation and throughput increases tremendously with growing networks.
\gls{CM}s which reduce the writing nodes, help to cure this scalability issue. \\
Additionally, a major constraint is the visibility of every player.
It has to be determined who receives necessary (movement) information and who does not.
This could either be done to hide information or to reduce network throughput.
This information is especially restricted by objects blocking the visibility as shown in the 2D-figure \ref{fig:VisibilityScalability} (Playground of \citet{RedBlobGames.2020} was used).
In figure \ref{fig:VisibilityScalability}, a player (circle, middle) is shown with its range of visibility (lighted area).
The visibility is blocked by rectangular objects, which cast shadows.
Yet alone in figure \ref{fig:VisibilityScalability} (A) there are no constraints.
But assuming other players around (Figure \ref{fig:VisibilityScalability}, B) one can tell that distance is not the only measure for other players to be displayed - a 'Line of Sight' has to exist.
Leaving the display to honest nodes is not applicable here as fraud is a major concern in this document.
Adding cryptography and signing information only to specific players to prevent fraud, slows down the network even more.
Consequently, whilst a fast paced four player game might be possible using the right \gls{CM},
40 simultaneous players as in \textit{Star wars Battlefront} (\cite{Wikipedia.2021e})
or even 64 as in \textit{Battlefield} (\cite{Wikipedia.2021}) are far out of reach \cite[3]{Serada.2020}. \\
Last, in distributed environments each player wants to keep its position secret as long as possible to prevent cheating.
Still the position of both players has to be published to calculate the possibility of a 'line of sight'.
This brings the anti-cheat mechanism at odds.
Hence, \gls{BCT} is clearly not a solution for high traffic (shooter) games and
due \gls{BCT}s characteristic to be "barely scalable" \citet[19]{Serada.2020},
\gls{MMORPG}s with multiple simultaneous actions are out of scope as well.

\FloatBarrier

\section{Gaming market}
\label{sec:GamingMarket}

\begin{figure}
	\centering{
		\includegraphics[width=.95\linewidth,keepaspectratio=true]{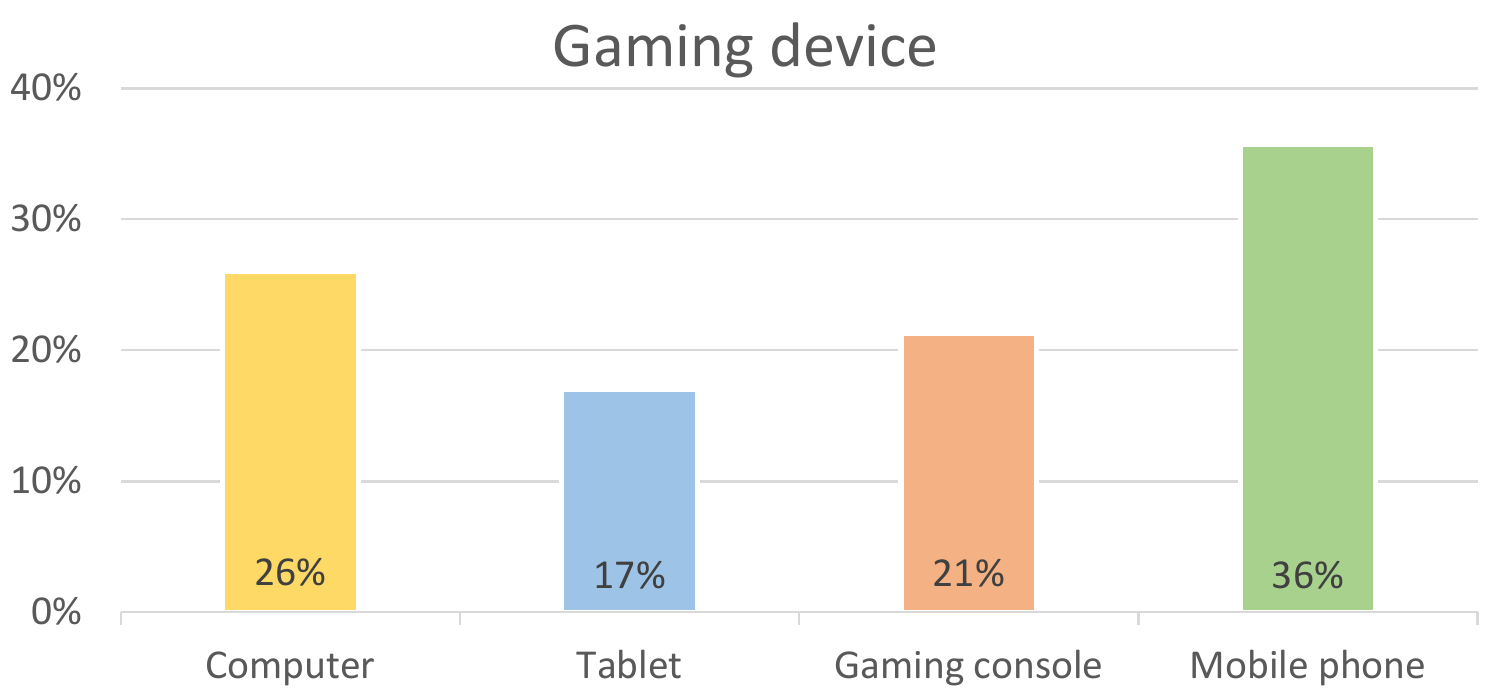}
		\caption{Global share of devices by gaming time (Data from \citet{LimelightNetworks.2020})}
		\label{fig:GameingDeviceData}
	}
\end{figure}

Until now the focus of the document was primarily on the technical side.
This section sheds the light on the recent gaming market - especially on the gamers.
The following data is taken from a market report of \citet{LimelightNetworks.2020}. \\
\textbf{First} it has to be determined which kind of device is suitable for games based on \gls{BCT}.
The given data \cite[7]{LimelightNetworks.2020} attest mobile phones a superior position for gaming globally (Figure \ref{fig:GameingDeviceData}).
This claim is backed by the data of each observed country (Appendix - Figure \ref{fig:GameingDeviceDataTable}).
If a publisher wants to tackle a big market, mobile phones shall be considered before targeting personal computers and gaming consoles.
Additionally, tablets are also part of the mobile market as they are run with the same operating systems (\citet{statista.com.2021b}) and offer the same distribution channels.
Consequently, slow paced (turn based) games which can be played on mobile phones
or cross platform offer a market to be hosted on \gls{BCT}.
Although \hyperref[sec:DataAllocationImprovements]{data allocation improvements} may be established in a game,
\gls{BCT} still needs storage space \cite[81]{Besancon.2019} as data can not just be pulled on demand from a central server.
Still, as a \textit{Counterpoint Research}-article from \cite{Wang.2021} stresses:
"Average smartphone nand flash capacity crossed the 100GB threshold in 2020"
(Appendix: Figure \ref{fig:GameingDeviceStorageSpace}).
This trend enables gaming with smartphones using \gls{BCT}.
The other devices, \textit{computers} and \textit{gaming consoles}, from figure \ref{fig:GameingDeviceData} are supposed to offer more or at least similar amounts of storage capacity. \\
\textbf{Second}, game characteristics shall be mentioned which are preferred by gamers.
Data out of the report \cite[18]{LimelightNetworks.2020} offers five categories of preferred games (Appendix:
Figure \ref{fig:GameCharacteristicsDataTable}).
\begin{figure}[!b]
	\centering{
		\includegraphics[width=.95\linewidth,keepaspectratio=true]{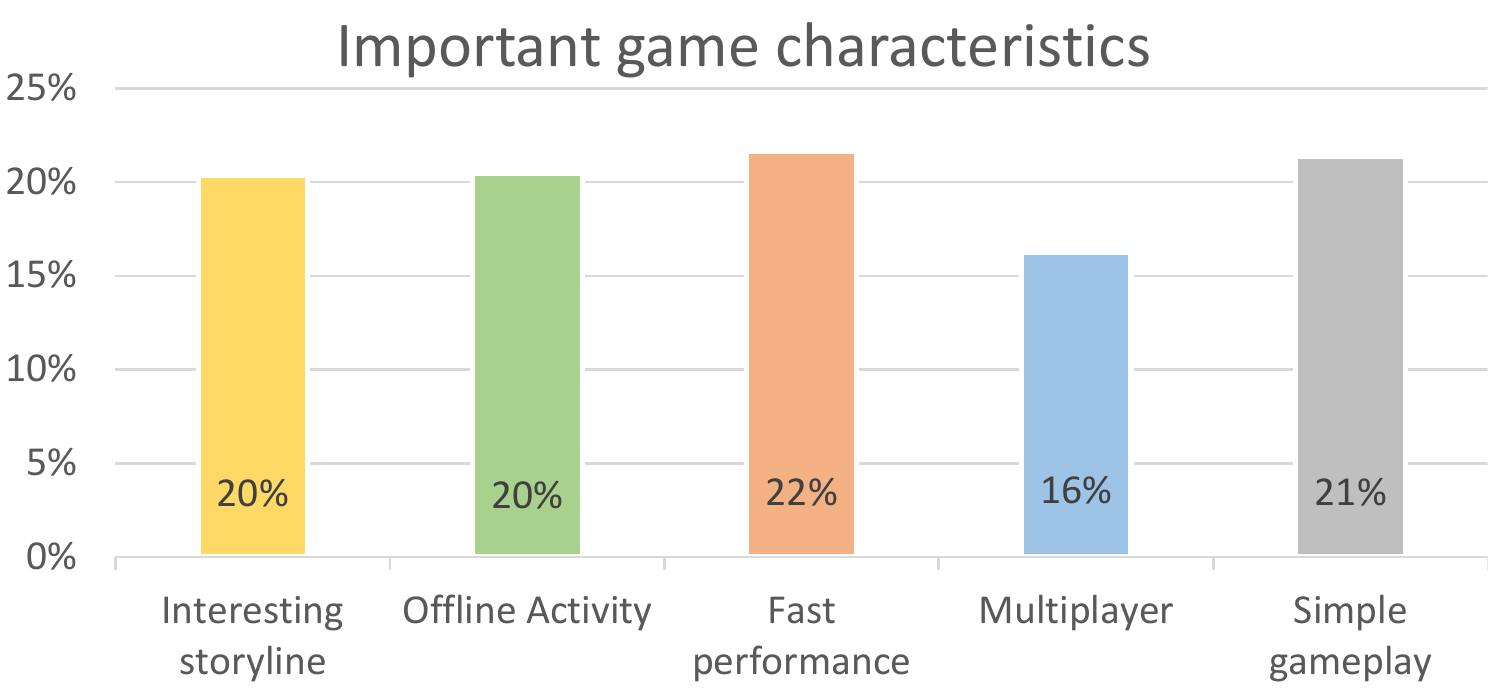}
		\caption{Global preferred game characteristics (Data from \citet{LimelightNetworks.2020})}
		\label{fig:GameCharacteristics}
	}
\end{figure}
The average of each category is shown in Figure \ref{fig:GameCharacteristics}.
The categories '\textit{Interesting storyline}' as well as '\textit{simple gameplay}' are out of scope as 
they have no influence on the game's backbone (e.g. \gls{BCT}).
On the contrary '\textit{Fast performance}', quickly loading and speedy interactions, '\textit{offline activity}',
the ability "to play the game while disconnected from the internet") \cite[18]{LimelightNetworks.2020}
as well as the '\textit{mulitplayer}' option are of importance regarding \gls{BCT}.
Here, \gls{BCT} enables offline time slots \cite[8]{Nakamoto.2009} in multiplayer games without the need for a central server.
Depending on the application - as stated before - fast performance is dependent on the chosen \gls{CM}. \\
\textbf{Third}, "[...] fair play is essential to any game [...]" \cite[44]{Yan.2009} and thus cheating (playing against agreed rules) harms gameplay.
Hence \cite[2]{HAEBERLEN.2010} cites from \cite{McGraw.2010}: "Cheating in online games is an important problem
that affects game players and game operators alike".
Many centralized games suffer in their game experience from this type of hostile behavior.
Therefore some "games try to prevent modding (e.g., multiplayer games to avoid cheating)" \cite[2493]{Lee.2020}.
Others install anti-cheating systems like PunkBuster, Warden or Valve Anti-Cheat \cite[8]{HAEBERLEN.2010}.
"In addition to privacy concerns, this approach has led to an arms race between cheaters and game maintainers,
in which the former constantly release new cheats or variations of existing ones,
and the latter must struggle to keep their databases up to date" \cite[8]{HAEBERLEN.2010}.
In the end anti-cheat measures commonly annoy gamers and reduce gaming system's performance.
With these problems in mind, \gls{BCT} promotes itself, despite its lack of speed
and the multiple data allocation, to be used in slow paced online (multiplayer) games.

\FloatBarrier

\section{Existing games using Blockchain Technology}
\label{sec:GamesOnBCT}

With the knowledge of the previously given four
categories \hyperref[sec:OwnershipGameplayNetwork]{Ownership,  network costs, gameplay \& scalability} , some already existing games, using \gls{BCT} are described:
\begin{enumerate}
	\item As described before \textit{CryptoKitties} is a game based on collectibles wherein digital assets (here: virtual cats) can be collected, bred, bought and sold \cite[2]{Serada.2020}.
	\textit{CryptoKitties} itself was born "[...] from a Hackathon idea and released officially in November 2017" \cite[25]{Laneve.2019}.
	According to \citet[15]{Laneve.2019}'s as well as \citet[18]{Serada.2020}'s research the Ethereum network was slowed down by \textit{CryptoKitties} significantly during that time. 
	"One of the pioneering features of Cyptokitties was the introduction of the ERC-721 protocol for \gls{NFTs},
	which rendered each kitten unique instead of being another form of cryptocurrency" \cite[25]{Laneve.2019}.
	The general case of reusability as described by \cite{Min.2019b} before is reflected by \citet[16]{Pfeiffer.2020}:
	"Applications of The KittyVerse are not only developed by the Cryptokitties producer but also by third developer studies and this shows the special potential of virtual objects/assets on Blockchain basis.
	The possession is with the player and the asset can be used for other games or applications."
	Thus \textit{CryptoKitties} faced many imitators such as \textit{Cryptopunks, Decentraland, MyCryptoHeroes, HyperDragons, Gods Unchained, Etheremon, Blockchain Cuties, NeoWorld, and Axie Infinity} \cite[3]{Serada.2020}.
	Further information about \textit{CryptoKitties} can be found in \citet{Laneve.2019}'s paper (p. 25-26).
	Staying at ownership models, \textit{Forgotten Artifacts} (\citet{lostrelics.io.2021}) is mentioned because it uses blockchain to prove ownership of assets,
	but stays a \textit{role play game} in the first place \cite[29]{Laneve.2019}.
	\textit{Forgotten Artifacts} is described in more detail by \citet{Laneve.2019} (p. 30-34).

	\item \citet{decentraland.org.2021} is a game which combines collectible traits
	with crafting mechanics (alike '\textit{Minecraft}').
	It "proves ownership of virtual land hosted on a decentralised content distribution system" \cite[29]{Laneve.2019}.
	More information on \textit{Decentraland} can be found in \citet{Laneve.2019}'s paper (p. 38-41).
	
	\item Although it can only barely be listed as part of a games list, \citet{Socios.com.2021} is mentioned
	because it connects realworld assets (e.g. voting rights) with digital assets (tokens) stored on a \gls{BC}.
	The digital tokens provide voting rights, which makes \textit{Socios} a voting platform \cite[29]{Laneve.2019}.
	Again, further details can be found in \citet{Laneve.2019}'s paper (p. 35-37). \\
	Generally, as "[...] of April 2019, based on open data sources, the number of cryptogames was estimated to be over 650, excluding gambling games" \cite[3]{Serada.2020}.
	Within this pile of games there are even games containing terms of useage,
	which "[...] are completely contrary to the spirit of the blockchain. Players’ ownership of virtual properties cannot be guaranteed." \cite[2]{Min.2019}.
	
	\item A completely different approach was taken by \textit{Huntercoin} using the cryptocurrency CHI from the XAYA blockchain framework (\citet{xaya.io.2021}).
	"\textit{Huntercoin} started as a 1-year experiment in 2014 to test how well a blockchain
	network could handle thousands of transactions happening in real-time, but
	due to an explosion in popularity development continued [...]" \cite[24]{Laneve.2019}.
	\textit{Huntercoin} uses the \gls{PoW} algorithm \cite[85]{Kraft.2016}.
	"When a miner finishes a block, he gets 10\% of the block reward in CHI,
	the other 90\% gets added to a pot that is then distributed to developers so they can reward to players" \cite[44]{Laneve.2019}.
	Additionally, for "[...] a fee paid in huntercoins, users can create hunters (corresponding roughly to player accounts) in this game world.
	This allows for \textit{human mining}: Parts of the block rewards are not paid to the proof-of-work miner
	but instead placed inside the game world, where hunters can pick them up and bank them to their on-chain address.
	This is not straight-forward, however, and requires skill since other hunters can fight for and steal the coins until they are secured.
	This is intended to give humans a chance to “mine” huntercoins by playing the game" \cite[85]{Kraft.2016}. \\
	A downside of this approach, used by Huntercoin and Motocoin, is given in the \gls{PoP} core paper:
	The "[...] act of play in both Huntercoin and Motocoin [...] becomes incentive-driven due to the blockchain, making the game progress lack entertainment" \cite[21]{Yuen.2019}.
	Moreover, \citet[84]{Kraft.2016} being the developer of huntercoin states that this approach leads to "[...] large growth of the blockchain and heavy resource requirements".
	Last, as playing the game offers the opportunity to earn CHI, Huntercoin sufferd from an army of bots mining ingame until certain countermeasures were introduced \cite[88-89]{Kraft.2016}.
	
	\item Throughout this list, \textit{Taurion} seems like the most interesting game regarding the scope of this thesis.
	Although still in development, it focuses more on a throughout gameplay using \gls{BCT} instead of asset monetization as it "uses a custom blockchain to host the game world and its interactions" \cite[29]{Laneve.2019}.
	Taurion, as Huntercoin, claims to use the XAYA blockchain framework.
	"To handle the speed of transactions and scalability needed to host massively
	multiplayer online games, the XAYA team introduced three mechanisms:
	Atomic Transactions, Game Channels and Ephemeral Timestamps" \cite[42]{Laneve.2019}.
	"Game Channels for Turn-Based Interactions" are described by \cite[90-92]{Kraft.2016} in more details.
	Game channels can be seen as off-chain games, branching off the main game to reenter later on \cite[91]{Kraft.2016}.
	They "[...] interact with the blockchain only for part of their functionality" \cite[27]{Laneve.2019}.
	Recently, \textit{Taurion} is on hold as the developers concentrate on \textit{Soccer Manager Elite}, which primarily fulfills the ownership case (yet again).
\end{enumerate}
Finally, revising all the mentioned use cases, \hyperref[chap:BCT]{BCT} is not (yet) used for cutting costs on infrastructure for the game publishers,
which is one main goal of this document.
Still, \gls{BCT} "[...] could be cost-effective, removing the centralized authority’s need to monitor and regulate transactions and interactions between different members"
\cite[5]{Sharma.2020}.

\FloatBarrier

\section{Problem area}
\label{sec:ProblemSpace}

Once \gls{BCT} is chosen for cost cutting and used as a trust building backend, games in scope need to be chosen and \gls{SC}s need to be defined. \\
The PoP-mechanism already provides a good basis for games and is considered as a basic benchmark throughout this chapter.
Hence it is assumed that the \gls{BCT}, which is used, meets speed criteria in non critical scenarios ($\leq 1$ second delay). \\
Consequently, in-game mechanics which rely on fast reactions, such as first person
shooter games or based on real-time strategy are not supposed to be in the scope as updates might not reach the opponents in time and thus prevent good game play experience. \\
Nevertheless, there are game genres, which are in the scope such as card-, quiz-, collaborative- as well as (slow) strategy games.
First, \textit{Hearthstone} (\citet{playhearthstone.com.2021}) and \textit{Magic: The Gathering Arena} (\citet{magic.wizards.com.2021}) are prominent examples for (online) card games.
\begin{figure}
	\centering{
		\includegraphics[width=.66\linewidth,keepaspectratio=true]{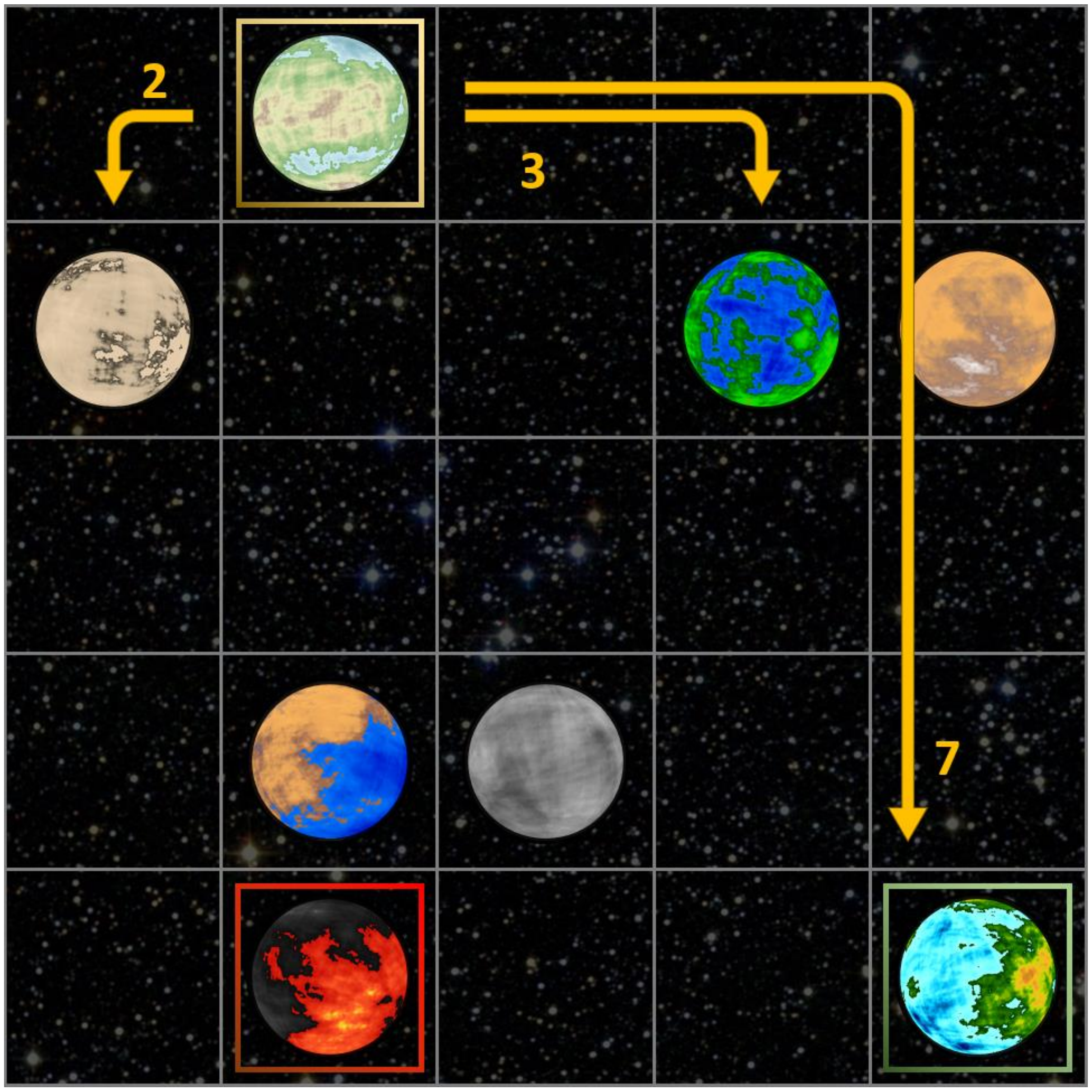}
		\caption{Sample game}
		\label{fig:ReachForTheStars}
	}
\end{figure}
Second, the German mobile app
\textit{Quizduell} (\citet{Quizduell.2021}) can be seen as an example for simple question-answer-games.
Last, \textit{Pen- \& Paper} role-play games and strategy games which allow more than 1 second response time e.g. through separated turns such as \textit{play-by-email} video
games (\citet{Wikipedia.2021b}) are beneficial. \\
Next, a hypothetical game is introduced, \textit{Reach for the Stars} (\hyperref[def:RftS]{RftS}) (see figure \ref{fig:ReachForTheStars}), to become less generic and abstract in the subsequent chapters.
\label{def:RftS}
After a short introduction, significant game mechanics are extracted and described theoretically.
\hyperref[def:RftS]{RftS} is a 2D, turn-based multiplayer game, where players try to colonize all planets in a sample galaxy - alike the game \textit{Risk} (\citet{Hasbro.2021}).
At the beginning, each player starts with only one planet.
Each planet has both, a fixed and a variable amount of production of space ships.
For the variable amount, \textit{randomization} is needed.
From a controlled planet, fleets can be sent to colonize/conquer other planets.
Fleets need fixed game-rounds, based on grid distance to reach planets (Figure \ref{fig:ReachForTheStars}, arrows with grid distance).
Fleet movements are not visible for the other players.
Hence the game lives from the nature of \textit{hidden transactions} and consequently originating \textit{fog of war} (\citet{By.2011}; \citet{Hagelback.2008}; \citet{Setear.1989}).
Fleets can be called back to the origin planet (once), but can not change the direction otherwise (\textit{Follow-Up hidden transactions}).
For the sake of reduced complexity colonization as well as battles are computed by linear algorithms and here not further of importance.
Last, there are two piles for each player to \textit{draw cards} from.
Both decks offer equal cards, but one is a \textit{shared deck} for all players and the other is a \textit{private deck}.
The information given on the cards is not important. \\
From this \hyperref[def:RftS]{RftS}-scenario, we can extract the need for \textit{randomization}, \textit{hidden transactions} including \textit{Follow-Up hidden transactions}, \textit{fog of war} as well as mechanics for \textit{private} and \textit{shared decks} of cards.
In this regard, time constraints on local machines
(e.g. game: M.U.L.E. from 1983, \citet{Wikipedia.2021d})
, as they are not seen to be confirmable, are out of scope.
Last, as storage space on low end devices endagers gamers to keep the game,
possible considerations to reduce storage space shall be taken into account.
For these transactions and movements \hyperref[sec:SmartContract]{SCs} have to be established which can be used in a \hyperref[chap:BCT]{BCT} context. \\

\pagebreak

\FloatBarrier

\section{Game specific smart contracts}
\label{sec:GSSCs}

Although there are innumerable designs of \gls{SC}s, games offer some common patterns.
Some of these patterns like \textit{hidden transactions}, \textit{fog of war}, \textit{pile of cards} or \textit{randomization} were just mentioned and are explained together with \textit{reveal claims triggers} and \textit{disputes} subsequently:

\subsection{Hidden transactions \& Randomization}
\label{sec:HiddenTransactionsPlusRandomization}
\begin{figure}
	\centering{
		\includegraphics[width=.72\linewidth,keepaspectratio=true]{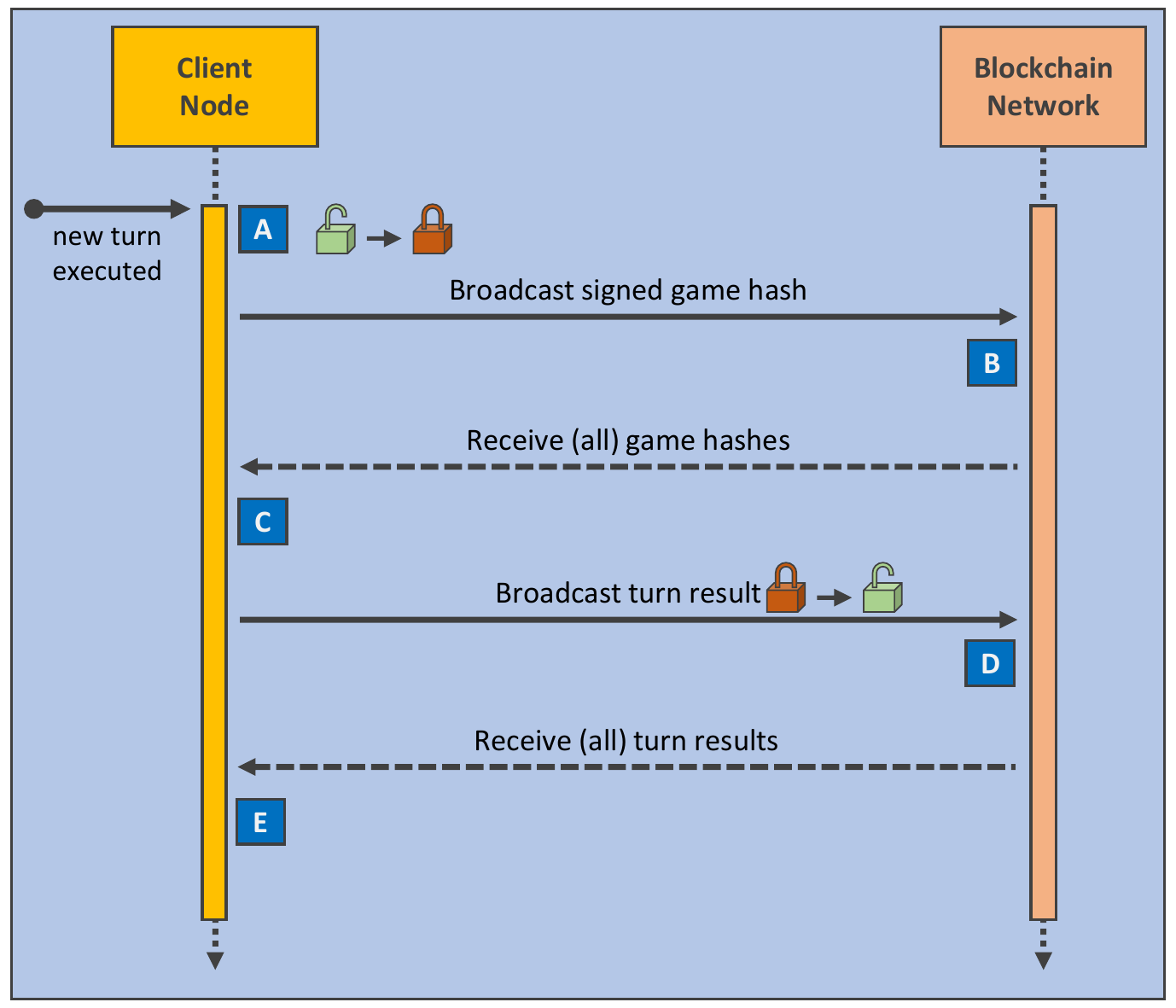}
		\caption{Offset revealing with game hash (From \citet{Yuen.2019})}
		\label{fig:OffsetRevealingGameHash}
	}
\end{figure}

If a game incorporates transactions which shall not be visible to other players immediately, a suitable \gls{SC} is needed. \label{lbl:HiddenTransactions}
The \textit{receiving players} have a justifiable desire to obtain a proof for the \textit{publishing player's} interaction.
Still the \textit{publishing player} wants to sustain the information, contained in the transactions, unpublished as long as possible/needed.
Hence there is a need for the information to be published 'veiled', which becomes 'unveiled' later on.
There are two approaches for this course of action, which from now on will be called \textit{offset revealing}. \\
\textbf{First}, the \textit{publishing player} uses \textbf{encryption} to veil the information and pushes it into the \gls{BC} network.
By the time the needed information has to be revealed, the password of the encryption is published.
As all nodes do not trust each other, they recalculate the transaction with the newly published key.
\textbf{Second}, only the hash of the transaction data \textbf{game hash} could be published \cite[94]{Kraft.2016}.
Both solutions offer benefits and downsides. \label{def:GameHash} \\
On the one hand, a \textit{game hash} is lightweight ($\sim0.032$ KB)\footnote{\hspace{0.1cm}The SHA256-algorithm generates fixed size 256-bit (32-byte) hashes \cite[7]{Rachmawati.2017}.}
and abstracts from the actual storage size of the transaction.
On the other hand, encryption allows to pass the decryption key to single elected players instead of the whole network.
If this measure is needed, \textit{game hashes} might be the inferior solution.
Hence, decryption to a part of the network becomes relatively easy/cheap in the case of encryption.
Still, encrypted data might offer implications on the content (e.g. transaction size) or the game requires data to be written unencrypted to the \gls{BC} later on.
Without using storage restoring procedures (see chapter: \hyperref[sec:DataAllocationImprovements]{Data allocation improvements}),
encryption might lead to unnecessary allocated storage.
Additionally, if there are only a few movement options, which can be iterated using 'brute force',
\textit{salt data} \cite[597]{Morris.1979} should be considered.
\begin{figure}
	\centering{
		\includegraphics[width=.95\linewidth,keepaspectratio=true]{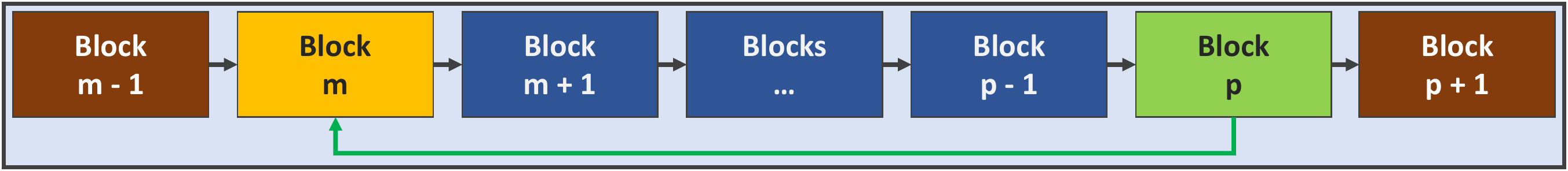}
		\caption{Offset revealing of a single block (As described by \citet{Kraft.2016})}
		\label{fig:OffsetRevealingSingle}
	}
\end{figure}
\noindent However, the \gls{BC} itself would look like figure \ref{fig:OffsetRevealingSingle}
which aligns the blocks according to their publication time from \textit{old} (left) to \textit{new} (right).
In figure \ref{fig:OffsetRevealingSingle} there are blocks 'before \textbf{m-1}', 'after \textbf{p+1}' and 'in between the revelation \textbf{m+1} to \textbf{p-1}'.
Block \textbf{m} equals either the encrypted or hashed data,
whilst block \textbf{p} equals the encryption key or the data described by the hash.
Block \textbf{p} reveals the information contained in block \textbf{m} to the network. \\
This procedure can be used for semi-simultaneous turns as well \cite[94]{Kraft.2016}.
Here 'simultaneous' does only refer to the game play - depending on the \hyperref[sec:ConsensusMechanisms]{CM},
the network can both be \textit{sequential} (e.g.: \gls{PoP}) or \textit{semi-parallel} (e.g.: \gls{PoW}). \\
In \citet[22]{Yuen.2019}'s paper, which introduces \gls{PoP},
a procedure for \textit{simultaneous hidden turns} is given alike figure \ref{fig:OffsetRevealingGameHash}.
\noindent In this case, multiple players have to publish any kind of information \textit{simultaneously}.
The recent transaction's \textit{game hash} will be encrypted (Figure \ref{fig:OffsetRevealingGameHash}, A) and sent (Figure \ref{fig:OffsetRevealingGameHash}, B).
Comprehensibly, none of the players wants to reveal the information before the other has not published his information in the \gls{BC}.
Hence it is waited until all involved nodes have published their \textit{game hash} (Figure \ref{fig:OffsetRevealingGameHash}, C).
As soon as all information is gathered, the information is released (Figure \ref{fig:OffsetRevealingGameHash}, D).
Finally, all players could make their moves simultaneously and check the other peer's results (Figure \ref{fig:OffsetRevealingGameHash}, E). \\
Consequently, synchronous and asynchronous as well as solo, paired and grouped hidden transactions can be served with this type of \gls{SC}s. \\

\begin{figure}
	\centering{
		\includegraphics[width=.95\linewidth,keepaspectratio=true]{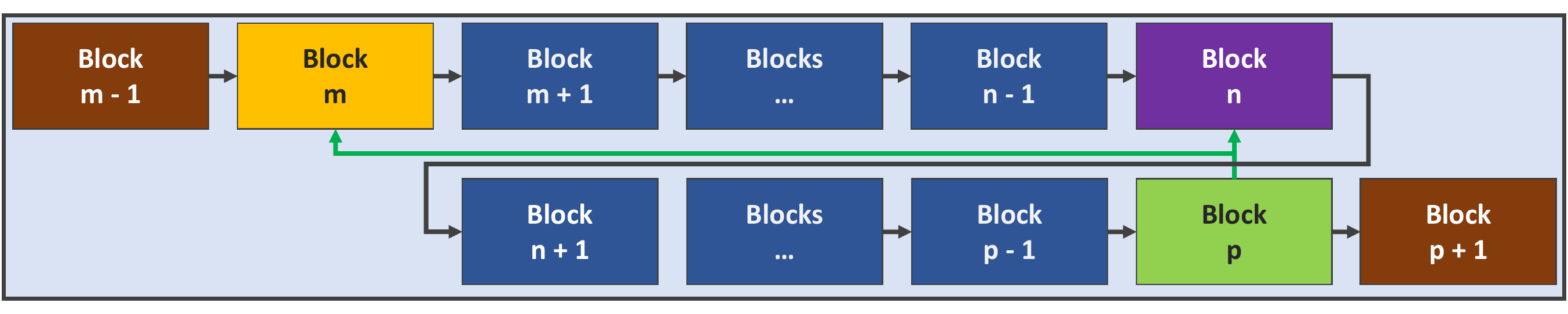}
		\caption{Offset revealing with follow up}
		\label{fig:OffsetRevealingFollowUp}
	}
\end{figure}

\noindent \textbf{Hidden follow-up movements} \\
Likewise to figure \ref{fig:OffsetRevealingSingle}, in a game a following transaction might occur, which implies the existence of the base transaction (Figure \ref{fig:OffsetRevealingFollowUp}). \label{lbl:FollowUpMoves}
This type of transactions for \gls{BCT} has not yet been described in the literature. \\
To stay within the context of \hyperref[def:RftS]{RftS} a fleet, sent to any planet, is called back before its arrival at the intended location.
This movement is only allowed given the side constraint that it has to return to its planet of departure.
Additionally, such a maneuver can only be performed once within a fleet-sending procedure. \\
As long as the implementation dependent \textit{maximum time to hide information} is not exceeded, \textit{hidden follow-up movements} are possible.
On a more abstract level, in figure \ref{fig:OffsetRevealingFollowUp},
\textbf{m} is the base transaction enhanced by the follow-up transaction \textbf{n}.
Both \textbf{m} and \textbf{n} are revealed by block \textbf{p}.
In figure \ref{fig:OffsetRevealingFollowUp} the hidden follow-up movement is consequently finished before block \textbf{p+1} is published.
\noindent If not intentionally, in a poor design of upper level software, the node may be enforced to publish block \textbf{m}
before the fleet has returned to the planet it has originated from.
Hence block \textbf{m} needs to be published by e.g. any \textit{timeout rule}.
The timeout implies (assumption) that the fleet would have had to reach the destination planet already.
Consequently block \textbf{m} represents an inconsistency.
Therefore, at least a statement has to be published that the
fleet's callback has occurred somewhere in between.
Now: 
\begin{enumerate}
	\item The range of rounds for the revisit of the fleet at its origin country to occur becomes limited.
	\item If there is only one block left which could represent the callback, the hidden follow-up movement is revealed by implication.
	The number of rounds between sending and calling back offers all players to calculate the round for the fleet to return.
	\item If the \textit{timeout rule} is set very low, block \textbf{n} has to be published as well before the fleet has finally returned.
\end{enumerate}
As the implications of \textit{hidden follow-up movements} are known, further details remain implementation dependent. \\

\noindent \textbf{Randomization} \\
Randomization in a distributed network needs \textit{hidden transactions} as well as \textit{trigger events} (section \hyperref[sec:RcTeDisputes]{Reveal claims, trigger events \& Disputes}). \label{lbl:Randomization}
To compute (pseudo) random values a common procedure is to take a \textit{random number generator} and give it some sort of \textit{seed}.
Mostly something alike the \textit{system clock time} or requested \textit{random mouse movement} etc. is used.
Nevertheless, if one node \textbf{P} needs a random number, other players might not comply with \textbf{P}'s local generated number - a lack of trust into \textbf{P}'s seed exists.
\citet{Chatterjee.2019} address this phenomenon in detail and describes four approaches to solve the problem.
Herein \citet{Chatterjee.2019} discourage from three of these approaches:
\begin{enumerate}
	\item The last block's hash can be seen as a random value,
	therefore it can generally be used as a seed \cite[403]{Chatterjee.2019}.
	Still, using a value from "[...] the current block as a seed [...]" \cite[403]{Chatterjee.2019} is seen not desired
	as the previous block may be tempered by a party which has interest into a 'fitting' random value \cite[406]{Chatterjee.2019}.
	\citet{Harkanson.2020}'s approach targeting 'Texas Hold’em Poker' can be put in this category.
	
	\item Alternatively an external provider could be used who acts as an oracle for the network \cite[403]{Chatterjee.2019}.
	This solution is out of scope as it requires a trusted external source.

	\item \citet{Chatterjee.2019}'s own algorithm may be used.
	In this case a \gls{BC} environment has to be used 
	which enforces to pay fees \cite[403]{Chatterjee.2019}.
	This solution is encouraged by \citet[403]{Chatterjee.2019}, but the fees in
	form of cryptocurrency tokens are seen as a flaw is this document's context.

	\item Last, the "anyone can submit randomly-generated numbers" \cite[403]{Chatterjee.2019} approach.
	Although \citet[406]{Chatterjee.2019} discourage from conducting this approach, it is given in more detail
	as one of \citet[406]{Chatterjee.2019}'s grounding assumptions can be changed. \\
	But first, the process:
	For the sake of simplicity a simultaneous hidden turn is assumed after a trigger call from \textbf{P} (Figure \ref{fig:OffsetRevealingGameHash}).
	Node \textbf{P} offers a \textit{game hash} which represents
	a certain number for the seed (Figure \ref{fig:OffsetRevealingGameHash} \& table \ref{tbl:RandomizationValues}: \textbf{B}).
	Subsequent \textbf{P} awaits other nodes to add their desired values as well (Figure \ref{fig:OffsetRevealingGameHash} \& table \ref{tbl:RandomizationValues}: \textbf{C}). \\
	Depending on the implementation, \textbf{P} waits until at least one answer arrives or 'a certain time is passed' \cite[406
	]{Chatterjee.2019}.
	On the one hand, the '\textit{at least one answer}'-solution may stall the network - if no answer is given.
	On the other hand, the '\textit{time based}'-solution may lead to attempts to choose a time slot
	which gives no answer and node \textbf{P} picks the most desired seed alone,
	which breaks the intention of randomization. \\
	Once the values are given or the time slot has exceeded, the values are revealed
	(Figure \ref{fig:OffsetRevealingGameHash} \& table \ref{tbl:RandomizationValues}: \textbf{D}) \cite[406]{Chatterjee.2019}.
	After all given values are revealed (Table \ref{tbl:RandomizationValues}: X, Y \& Z in \textbf{D}) or time has exceeded
	(Table \ref{tbl:RandomizationValues}: $\alpha$ in \textbf{D}), \textbf{P} calculates the seed
	out of the given values (Table \ref{tbl:RandomizationValues}: \textbf{E}).
	For the sake of tangibility a simple \textit{sum} is used here.
	Last, \textbf{P} receives a random number out of a predefined \textit{static random generator} (Here: Random of ($1,522$)).
	During the randomization procedure, once a (hidden) seed is given, no adjustment is possible.
	Additionally, none of the participating nodes (\textbf{P, Q, R, S \& T}) can adjust their value to that
	of the others as the one-way hash functions prevent altering in between \cite[406]{Chatterjee.2019}.
	\begin{table}[!b]
		\centering
		\begin{tabularx}{0.83\textwidth}{ c | c | c | c | c }
			\textbf{Node\textbackslash State} & \textbf{Start (B)} & \textbf{Answers (C)} & \textbf{Reveal (D)} & \textbf{Calculation (E)} \\ \hline
			\textbf{P} & X & $-$ & $X = 412$ & $-$ \\ \hline
			\textbf{Q} & $-$ & Y & $Y = 369$ & $-$ \\ \hline
			\textbf{R} & $-$ & $-$ & $-$ & $-$ \\ \hline
			\textbf{S} & $-$ & $\alpha$ & $\alpha = ???$ & $-$ \\ \hline
			\textbf{T} & $-$ & Z & $Z = 741$ & $-$ \\ \hline
			\textit{Seed} & $-$ & $-$ & $-$ & $\sum_{\substack{X,Y,Z}} = 1,522$ \\ \hline
		\end{tabularx}
		\caption{Sample given randomization values}
		\label{tbl:RandomizationValues}
	\end{table}
	Finally, the randomization's process outcome is dependent on every involved node and
	'\textit{non participating}'-nodes cannot veto (afterwards) because they are guilty in the case of '\textit{not participating}'.

	\citet[406]{Chatterjee.2019} sees a flaw in this approach as the time slot of revealing
	each seed (Table \ref{tbl:RandomizationValues}: \textbf{D}) offers the possibility to alter the sum.
	If node \textbf{S} had calculated the sum and came to the conclusion that revealing $\alpha$ would lead to a less desirable random value, \textbf{S} would gain an advantage from revealing late.
	This results in two possible shots for each node \citet[406]{Chatterjee.2019} except the demanding node (this one has to be revealed anyways) and a race to become the last emitting node.
	This scenario only beholds true if each node reveals directly to the whole network.
	On the contrary any other node (e.g.: \textbf{Z}) can add a transaction containing its encrypted value.
	The transaction containing \textbf{Z}'s encrypted value is now part of the \gls{BC}.
	Assuming that there is one node \textbf{P} who asked for a random number,
	\textbf{Z} sends the key only to \textbf{P}.
	Only \textbf{P} can therefore obtain \textbf{Z}'s value during the revelation time slot.
	After the revelation time slot is over, either \textbf{P} or \textbf{Z} reveal
	the key which offers the seed to the whole network.
	Of course this additional step slows down the answer. \\	
	Against \citet[406]{Chatterjee.2019}'s claim that the process "[...] provides no incentives to the participants to submit random numbers [...]", incentives are seen here in the upper level game itself.
	Therefore same results from constantly given static values is inapplicable here as well \cite[406]{Chatterjee.2019}.
\end{enumerate}
It has to be kept in mind that random number generators need to fit the purpose and are likewise encryption, a key attack vector.
Although the thoughts are backed by \citet{Chatterjee.2019}'s paper, the solution is considered intuitive
and supposed for short time frames to cover recalculation risks etc.
It is not error prone by definition.
Although further details regarding random number generators are not in the scope of this document.
	
\subsection{Piles of Cards}
\label{sec:PileOfCards}
Dealing with cards, such as on a player's hand, in a pile or currently being drawn
can be difficult regarding security in a \hyperref[chap:BCT]{BCT} context. \label{lbl:CardDraw}
Any type of the subsequent \textit{card draw scenarios in \gls{BCT}} 
cannot be found in the literature at the present time. \\
The easiest part is distinguishing between a secret hand, just visible to the owner and an open hand, visible to all players.
Moreover decks/piles need to be shuffled or drawing has to be grounded on randomization.
Consequently, drawing cards from decks introduces specific obstacles.
The different cases are shown in table \ref{tbl:Card draw scenarios}. \\
On the one hand it is distinguished here between \textbf{private} and \textbf{public} piles in the means of accessibility.
On the other hand draws differ in regards of \textbf{open}, \textbf{private} and \textbf{hidden}
transactions in the means of visibility/openness to other players. \\
In \textbf{open} transactions a card is shown directly to all players - no restrictions in regards of accessibility exists.
In contrast, \textbf{hidden} transactions are only known to the acting player.
Other players have no indication that a card has been drawn at all - still the acting player receives a card with the attached information.
In between \textit{open} and \textit{hidden} there are \textbf{private} transactions,
which show other players that a card has been taken from the pile, but the card's information is only revealed to the drawing player.

\subsubsection{General draws from piles}
\label{sec:GDfP}
The different types of draws work as follows:
\begin{enumerate}	
	\item A \textbf{private pile} with \textbf{public draw} is supposed to be the least obstacle.
	The cards in the pile are supposed to be known by everyone and only the card to be drawn has to be determined.
	A random number helps to choose the card to be drawn (by index).
	Finally, the card is shown to all players.
	
	\item More complicated is a \textbf{private pile} with \textbf{private draw}.
	Still it is manageable as the drawn card can be proven by \textit{offset revelation}.
	Here the effective pile is only known to the drawing player in regards of order (shuffled cards).
	Again, a random number helps to choose the card to be drawn.
	The draw has to be execute as a hidden transaction for later verification.
	Still the validity of every played card may only be verifiable at the end of the game.
	The latter is e.g. depending on the fact whether the cards in the pile are generally known to other players or not.
	
	\item A \textbf{private pile} with \textbf{hidden draws} needs another mechanism.
	Already a call for randomization, the process to receive random values, implies to draw a card.
	Still, without this call a draw is not supposed to be possible.
	Contrariwise the absence of a \textit{call for randomization} implies that no card can be drawn.
	Therefore faked draw transactions are the consequence.
	But even faked draw transactions have to be documented on the \hyperref[sec:BCI]{BC} to fulfill the promised transparency and trust.
	This aspect will be covered in more detail in the section \hyperref[sec:DataAllocationImprovements]{Data allocation improvements}.
	
	\begin{table}
		\centering
		\begin{tabularx}{0.525\textwidth}{ l | c | c | c }
			\textbf{Pile\textbackslash Draw} & \textbf{Open} & \textbf{Private} & \textbf{Hidden} \\ \hline
			\textbf{Private} & \textbf{1.} & \textbf{2.} & \textbf{3.} \\ \hline
			\textbf{Public} & \textbf{4.} & \textbf{5.} & \textbf{6.} \\
		\end{tabularx}
		\caption{Extended Blockchain Network Types}
		\label{tbl:Card draw scenarios}
	\end{table}
	
	\item A \textbf{public pile} with \textbf{public draw} is supposed to be as easy as a \textit{private pile} with \textit{public draw} (\textit{1.}).
	The only difference is that all players can access the pile to draw from it - the pile is not explicitly reserved for one specific player.
			
	\item A \textbf{public pile} with \textbf{private draw} is supposed to be the most complex situation
	as it requires the highest level of collaboration. \\
	Therefore, it is moved to the following section (\hyperref[sec:PPwPd]{Public pile with private draw}).
	
	\item Last, a \textbf{public pile} with \textbf{hidden draw} offers a special mutually exclusive case.
	As this type prevents to inform other nodes that a card was drawn,
	not even the drawn index from a multiple times shuffled and encrypted pile is allowed to be published.
	Consequently, a node in a subsequent turn may draw an already taken card from the pile.
	Hence a race condition \cite[75]{NetzerR.H.B..1992} in the form of \citet{Nakamoto.2009}'s \textbf{double spending problem} occurs.
	To the current knowledge, this problem connot be solved without external help,
	such as \hyperref[sec:HelperNodes]{Helper Nodes}. 
	These are discussed in further detail in the section
	\hyperref[sec:FurtherCharacteristics]{Further Characteristics} of the \hyperref[chap:PoT]{Proof-Of-Turn} approach.
\end{enumerate}

\subsubsection{Public pile with private draw}
\label{sec:PPwPd}
\begin{figure}[!b]
	\centering{
		\includegraphics[width=.95\linewidth,keepaspectratio=true]{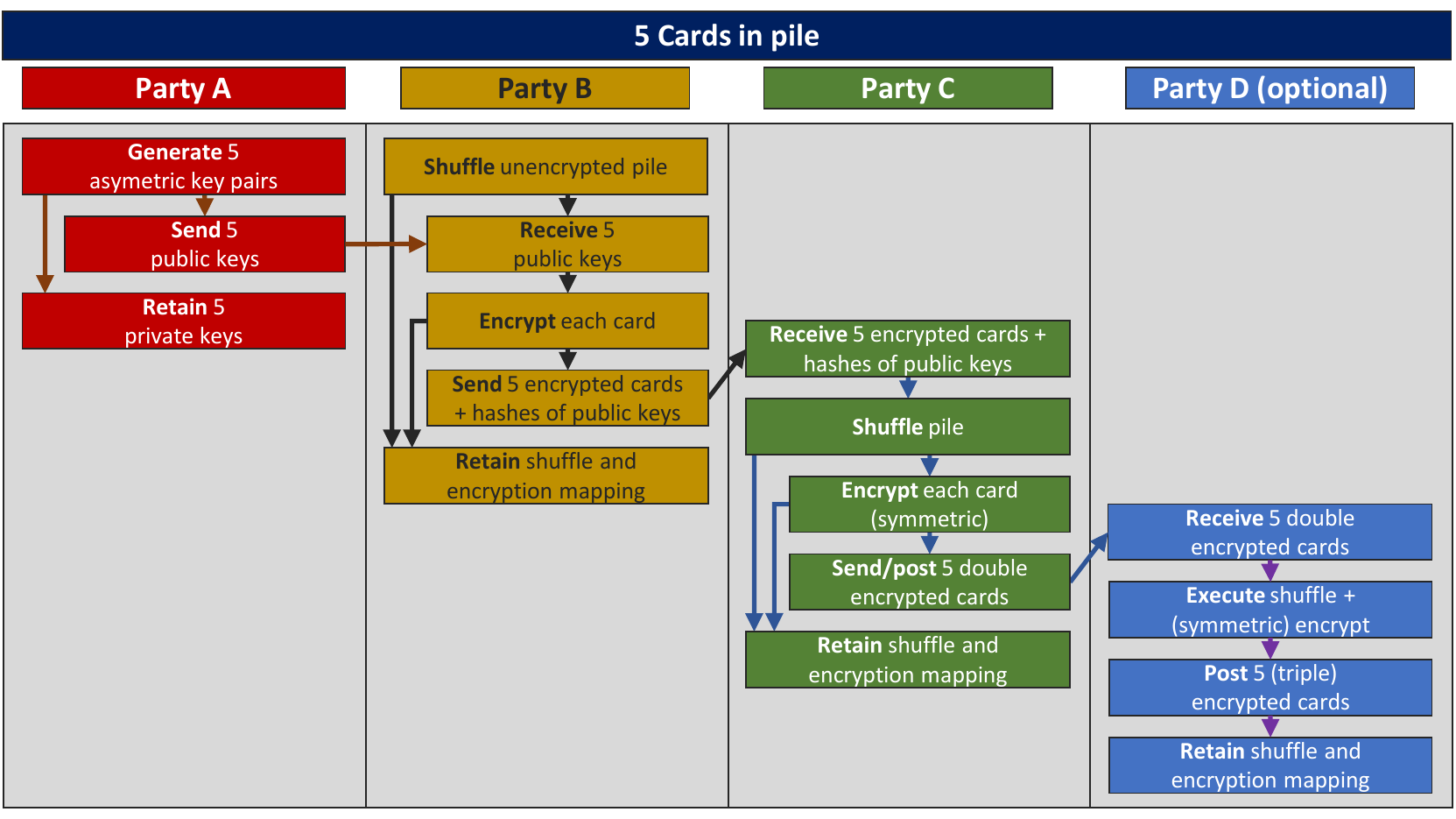}
		\caption{Parties shuffling cards}
		\label{fig:NodesShufflingCards}
	}%
\end{figure}

\begin{figure}
	\centering{
		\includegraphics[width=.95\linewidth,keepaspectratio=true]{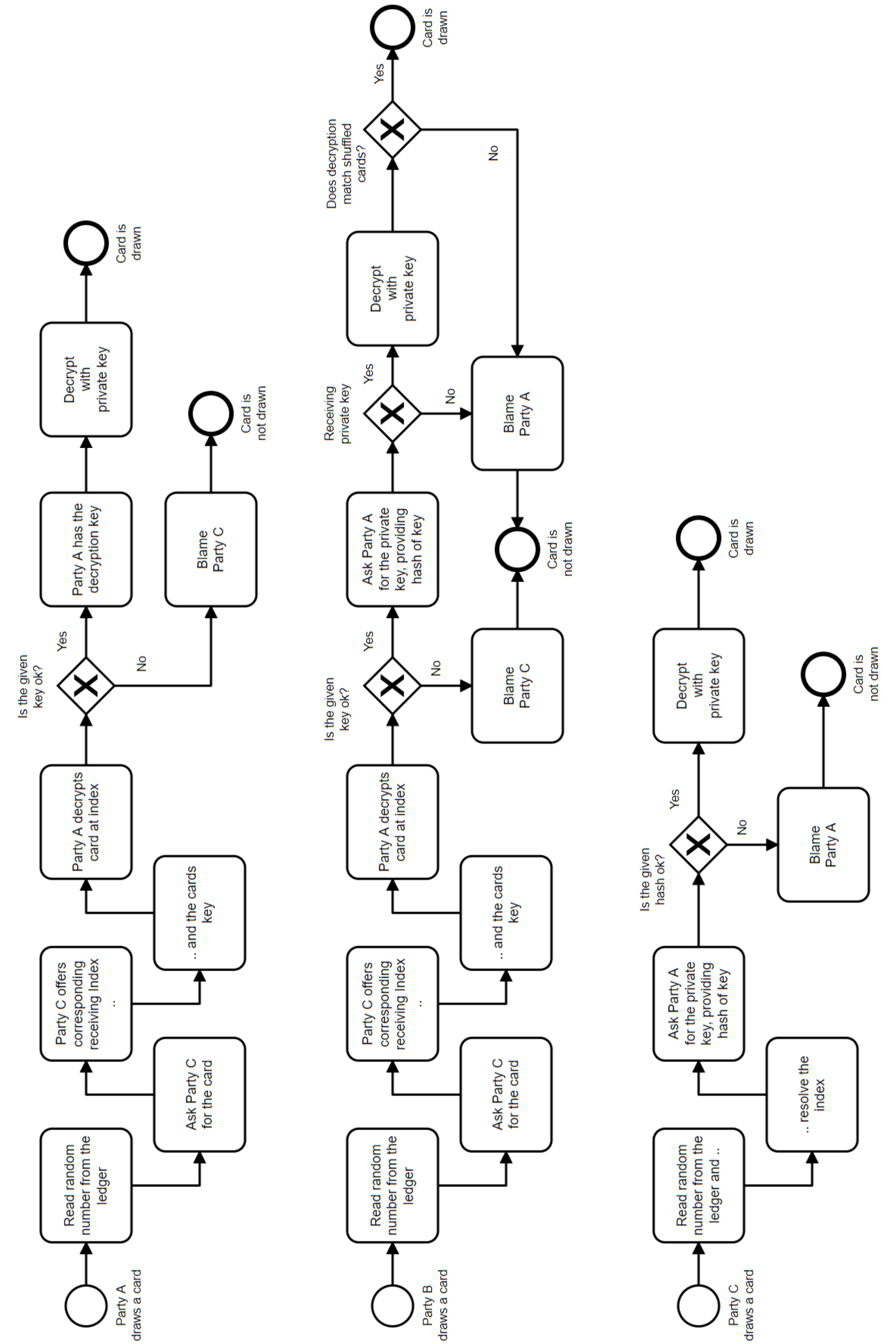}
		\caption{Decryption processes (Nodes A, B and C)}
		\label{fig:ShuffledCardsDecryptionAtoC}
	}
\end{figure}

\begin{figure}
	\centering{
		\includegraphics[width=.87\linewidth,keepaspectratio=true]{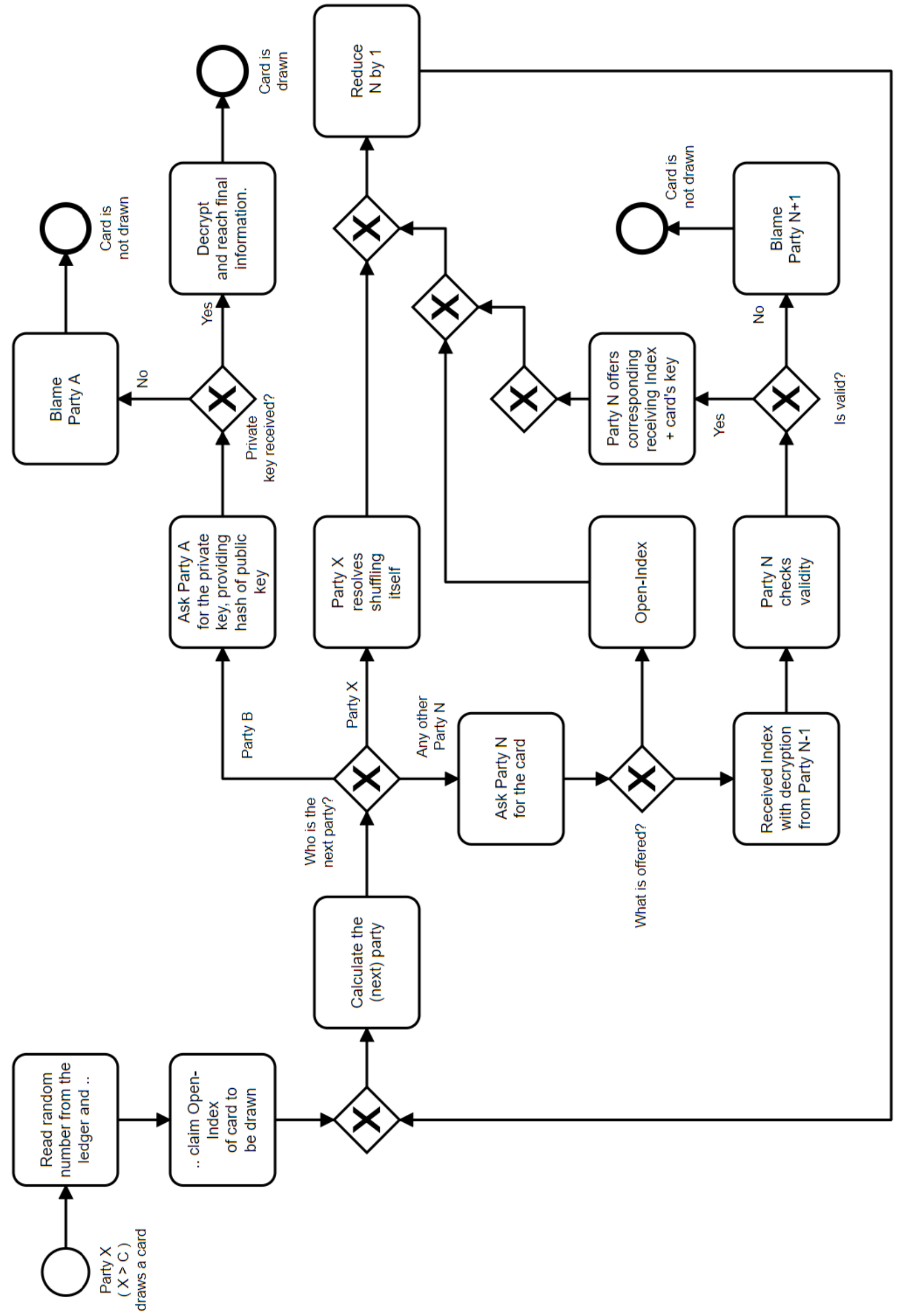}
		\caption{Decryption processes using many nodes ($\geq D$)}
		\label{fig:ShuffledCardsDecryptionMultiple}
	}
\end{figure}

The key issue of this type is that only the drawing player is allowed to receive the card's information.
Nevertheless, the other players are not allowed to draw the same card from the pile thereafter.
Hence, they need to know which card's placeholder (e.g. index) was taken.
Naturally, the index is not allowed to reveal the cards information/identity.
A mechanism has to be implemented, which hides the cards information, but prevents overlapping transactions (double spending) from the pile.
To untangle the challenge, the single card's information and the card's destination in the pile need to be separated to provide a suitable placeholder mechanism.
This can be done by shuffling and encrypting the cards.
Still, the 'shuffling and encrypting'-party has full knowledge about both the previous as well as the subsequent state of the pile.
Therefore several parties have to be involved. \\
During the process at least two rounds of 'shuffling and encryption' are needed and at least three parties have to participate.
Basically, with every additional party, another round of 'shuffling and encryption' has to be conducted.
The first round of encryption has to be executed using asymmetric encryption.
The asymmetric encryption has to be split on two (distinct) parties, one encrypting and another offering the decryption keys.
Last, a third party is needed to prevent data exposure.
The process is given in more details in Figure (\ref{fig:NodesShufflingCards}) and the steps and their possible pitfalls are described hereafter. \\
\noindent First \textbf{Party A} creates asymmetric encryption key pairs for each card.
\textit{Party A} is not allowed to encrypt the cards.
If \textit{Party A} would encrypt the cards, it would become the single gatekeeper and knew every drawn card.
Hence \textbf{Party B} shuffles the unencrypted pile and then encrypts each card using the (public) keys received from \textit{Party A}.
Without \textit{Party B}'s shuffling the encryption would be useless as the \textit{default pile} is known.
The resulting pile, consisting of encrypted cards complemented with their key's hash, is therefore (only) given to \textit{Party C}.
At this moment \textit{Party A} is not allowed to be involved, as it could decrypt all cards to receive the pile's change of order.
To prevent \textit{Party A} from decrypting, \textbf{Party C} shuffles to diminish \textit{Party B's} knowledge (Indexes) and encrypts each card again to prohibit \textit{Party A's} access.
Now none of the parties is capable to access information from the pile without consultation of all other parties. \\
More precisely, after the process: \textbf{Party A} does not have access to the pile unless \textit{Party C}
grants permission and hands over a card which is only encrypted with one of \textit{Party A}'s public keys.
\textbf{Party B} knows the first pile's change of order (from origin), but does neither know \textit{Party C's} shuffle nor \textit{Party A's} private keys.
Last, \textbf{Party C} neither has \textit{Party A's} private keys, nor \textit{Party B's} change of order to the source pile.
Consequently, the resulting \textbf{twice} \textit{shuffled and encrypted} pile can then be written to the \gls{BC}.
During the game the pile's cards are drawn numbers (Indexes).
Each party knows which card has been drawn from \textit{Party C}'s published pile as the drawn indexes are known.
This solves the 'double spending' challenge. \\
To the knowledge of the author, there is no possibility to decrease the number of parties below three.
Nevertheless, an infinite number of parties can be added between \textit{Party C} and the publication of the pile, represented in figure \ref{fig:NodesShufflingCards} by \textbf{Party D}.
Each added party reshuffles the pile and encrypts it with its own keys, just as \textit{Party D}.
Except the mandatory asymmetric encryption used by \textit{Party A} and \textit{Party B}, all other encryption is free to use either a symmetric or asymmetric encryption procedure.
Naturally, with each additional party, the reconstruction becomes more intricate.
Generally only the shuffling is needed until \textit{Party B}'s encrypted card can be combined with \textit{Party A}'s key to retrace the final information.
But in a hostile environment, each party insists on full safety/transparency.
Hence, on each step towards decryption, the drawing party insists on proof of correctness.
Thus, for not getting tricked, the key on every stage is claimed and validated. \\
If the key does not work properly (e.g. wrong data was transmitted),
the sending party has to be blamed directly, as it might try to betray (the network).
Additionally, the party which is asked to provide the card's information claims valid input data to ensure that
particularly this card is chosen and has to be revealed (especially for Party A's private keys).
Hence, the \textit{encrypted information} from the previous party,
which equals to one of the asked party's produced cards \textit{plus} its \textit{corresponding index} have to be delivered.
Only if \textbf{both values} are supplied, representing a valid claim to receive information, data is provided. \\
In figure \ref{fig:ShuffledCardsDecryptionAtoC} the decryption processes for the \textit{parties A, B and C} are shown.
Given the case that there are more than three parties, such as another \textit{Party D} and beyond, the adapted process is shown in figure \ref{fig:ShuffledCardsDecryptionMultiple}.

\begin{figure}
	\centering{
		\includegraphics[width=.95\linewidth,keepaspectratio=true]{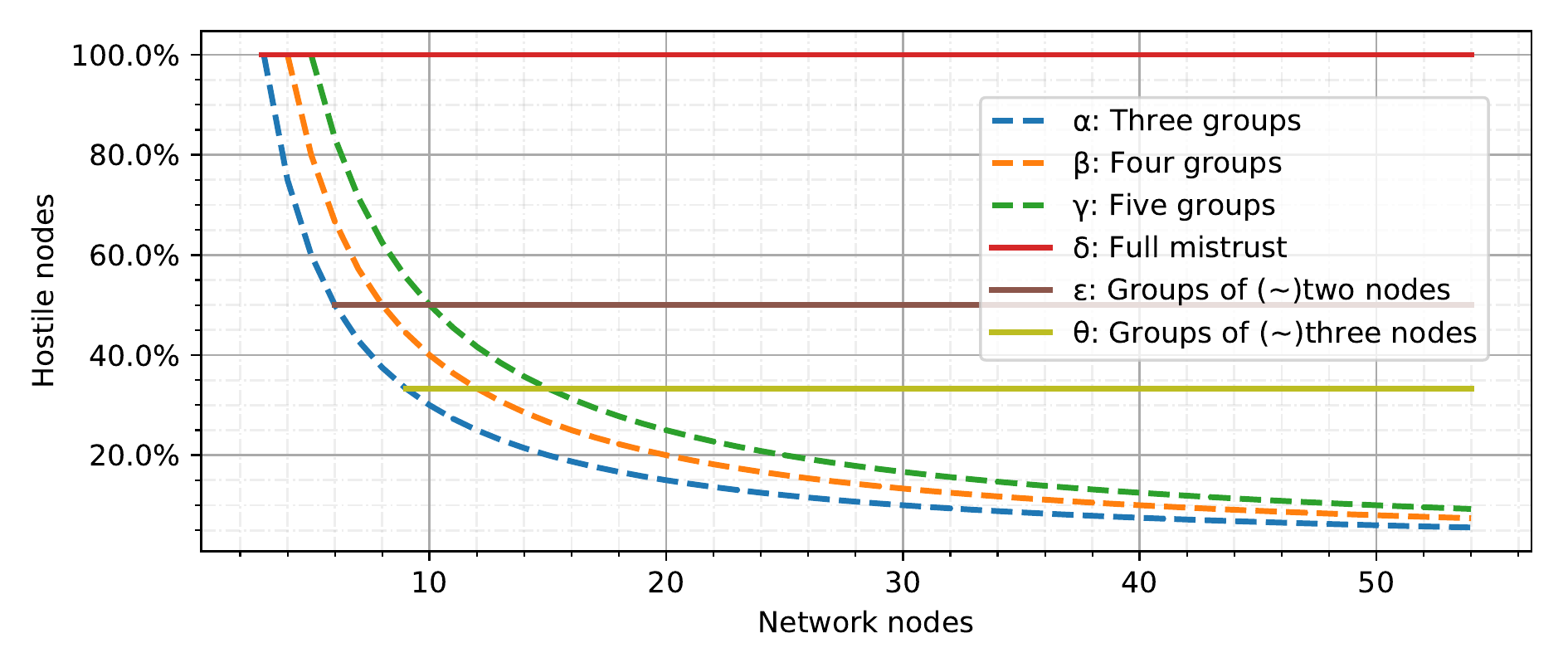}
		\caption{BFT shuffling cards}
		\label{fig:ShuffleCards}
	}
\end{figure}

\noindent Note that there are pitfalls if wrong information is transmitted or a party, which has to offer specific information, is not available.
This shows a major problem of this solution and stresses why recently it was not talked about \textit{network nodes}, but rather about \textbf{parties}.
The case of unavailability of any node, for whatever reason, freezes the network.
If the network freezes as described in the \textbf{Multichain} approach, there is no possibility to recover unless the affected node reconnects to the network.
Thus a game designer has to choose between a full mistrust scenario and cooperation of nodes.
In a three players game - as mentioned before the minimum amount of players - each node has to become one of the three parties.
But if there are more nodes participating in the process, a party may consist of multiple nodes as well.
The latter decreases trust and reduces security.
Still, if multiple nodes are capable to provide needed information responsiveness of the network rises.
This is especially desirable because the decryption as well as the shuffle trace back can only be performed in one specific order.
A review of the \hyperref[sec:ByzantineFaultTolerance]{BFT} problem to ensure fraud resilience has to be conducted.
\textit{Fraud resilience} follows the graphs shown in
figure \ref{fig:ShuffleCards}\footnote{\hspace{0.1cm}Figure \ref{fig:ShuffleCards} plot's Python code is given in the \hyperref[script:BFTshufflingCards]{appendix}.}.
Assuming \textbf{full mistrust} (Figure \ref{fig:ShuffleCards}: $\delta$), starting at a minimum of three nodes, all nodes would need to turn hostile to decrypt information.
Hence, no node can trick another or the whole network.
Nevertheless, it is the most vulnerable type in regards of a \textit{frozen network} state.
Increasing the number of nodes able to provide the information leads to the graphs $\epsilon$ and $\theta$.
Dividing all nodes into \textbf{groups of two} (Figure \ref{fig:ShuffleCards}: $\epsilon$) implies that each group needs to contain at least one hostile node to enable fraud.
Hence at least $50.0\%$ of nodes have to become hostile to break the system.
In the best case only one group with fully honest players is able to keep the system fraudless.
An even higher redundancy, dividing all nodes into \textbf{groups of three} (Figure \ref{fig:ShuffleCards}: $\theta$), leads to at least $33.3\%$ fraud resilience.
Raising redundancy even higher as of fixed \textbf{three, four or five groups}
containing many nodes (Figure \ref{fig:ShuffleCards}: $\alpha$, $\beta$ \& $\gamma$) reduces fraud resilience dramatically.
Depending on the desired resilience level, $\delta$, $\epsilon$ and $\theta$ are recommended.
Nevertheless, in particularly huge games with dropping out players a low number of groups might be superior.
Here it has to be reminded of \citet{Laneve.2019} who state that \gls{BCT} "[...] is reliant on the game’s popularity to keep it running, making it a risky choice for developers" (p. 27).
The final decision remains implementation dependent.

\FloatBarrier

\subsection{Fog of War}
\label{sec:FogOfWar}
\noindent Comparable \hyperref[sec:SmartContract]{SC}s might be needed, if a game includes
\textit{fog of war} elements, which deal with uncertainty of combat situations (\citet{Setear.1989}). \label{lbl:FogOfWar}
Again, there was no literature to be found which covered \textit{fog of war scenarios in \gls{BCT}}. \\
Four general uncertainties are assumed: '\textit{enemy's intentions}', '\textit{natural environment}',
behavior of '\textit{friendly forces}' and '\textit{underlying laws of war}' \cite[3-4]{Setear.1989}.
\begin{figure}
	\centering{
		\includegraphics[width=.995\linewidth,keepaspectratio=true]{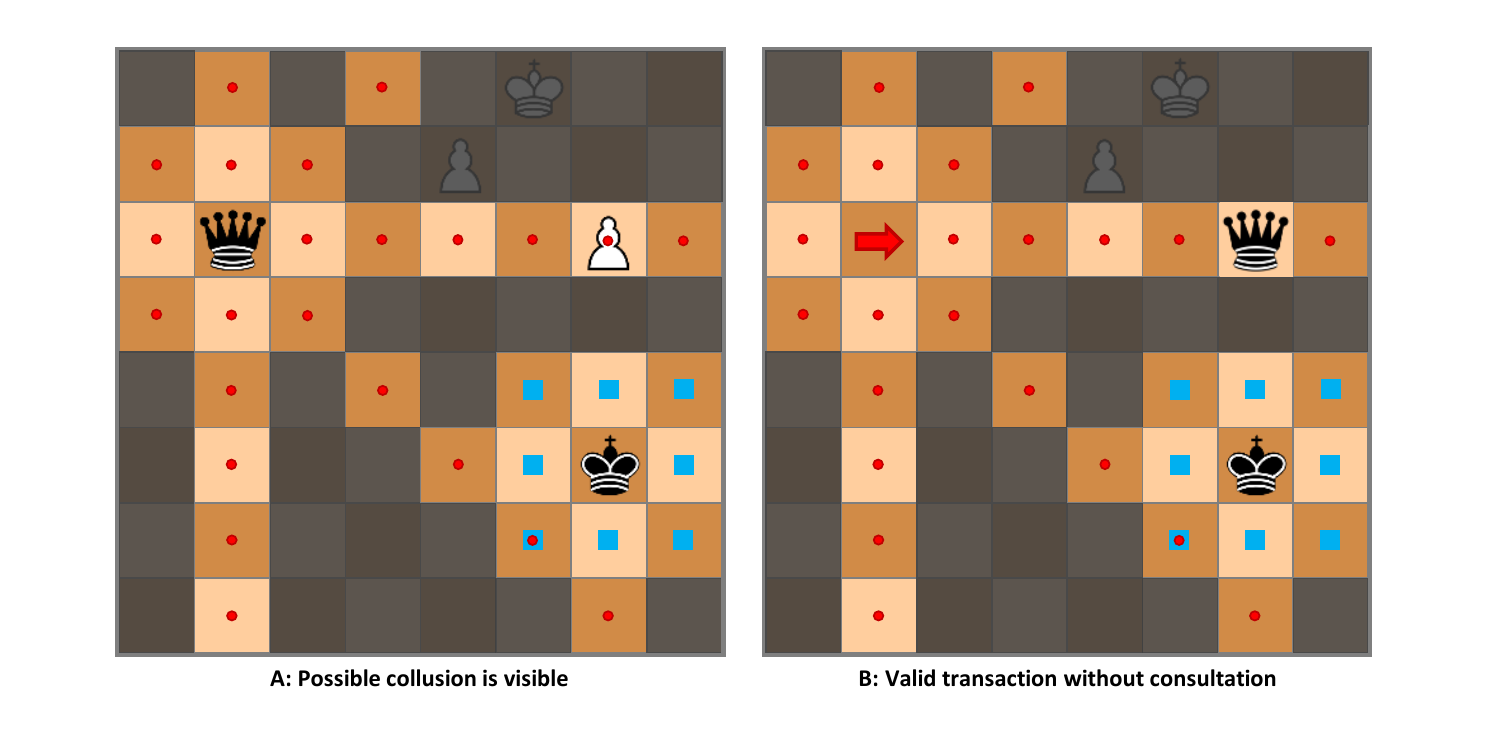}
		\caption{Fog of war in chess - Open moves (1)}
		\label{fig:FogOfWar_Chess_AB}
	}
\end{figure}
The last case does not apply in this scenario as it is diminished by defined \gls{SC}s and game rules.
Nevertheless, in many (realtime) strategy games especially the first two listed uncertainties are used.
In gaming environments with a central server, which is aware of every move, this does not offer special challenges as the server can always provide all needed values.
Back in a distributed environment, a challenge similar to the previous card draw scenario occurs.
\begin{figure}
	\centering{
		\includegraphics[width=.995\linewidth,keepaspectratio=true]{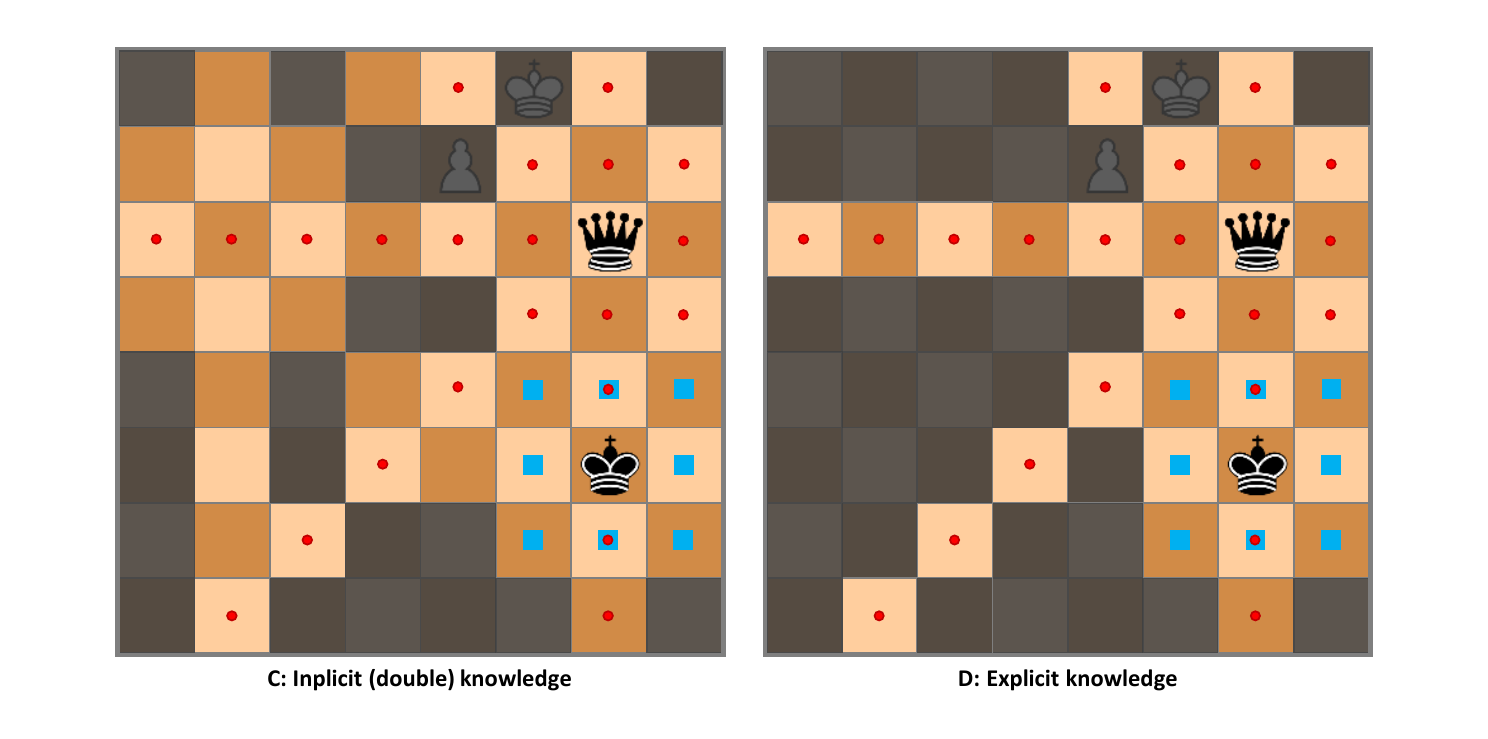}
		\caption{Fog of war in chess - Open moves (2)}
		\label{fig:FogOfWar_Chess_CD}
	}
\end{figure}

\noindent \textbf{First}, the easy case:
The game is designed in a way that every possible move can only be performed on fully revealed instances (Figure \ref{fig:FogOfWar_Chess_AB}, A).
Hence, each move can be calculated by the executing node (Figure \ref{fig:FogOfWar_Chess_AB}, B).
The information for the next move will be provided during the next network round.
All other nodes provide needed data (lifting the fog) for the next round on the affected fields (Figure \ref{fig:FogOfWar_Chess_CD}, C: Squares and dots).
On those fields, which are out of the new area of visibility, fog rises again (Figure \ref{fig:FogOfWar_Chess_CD}, D).
\begin{figure}
	\centering{
		\includegraphics[width=.995\linewidth,keepaspectratio=true]{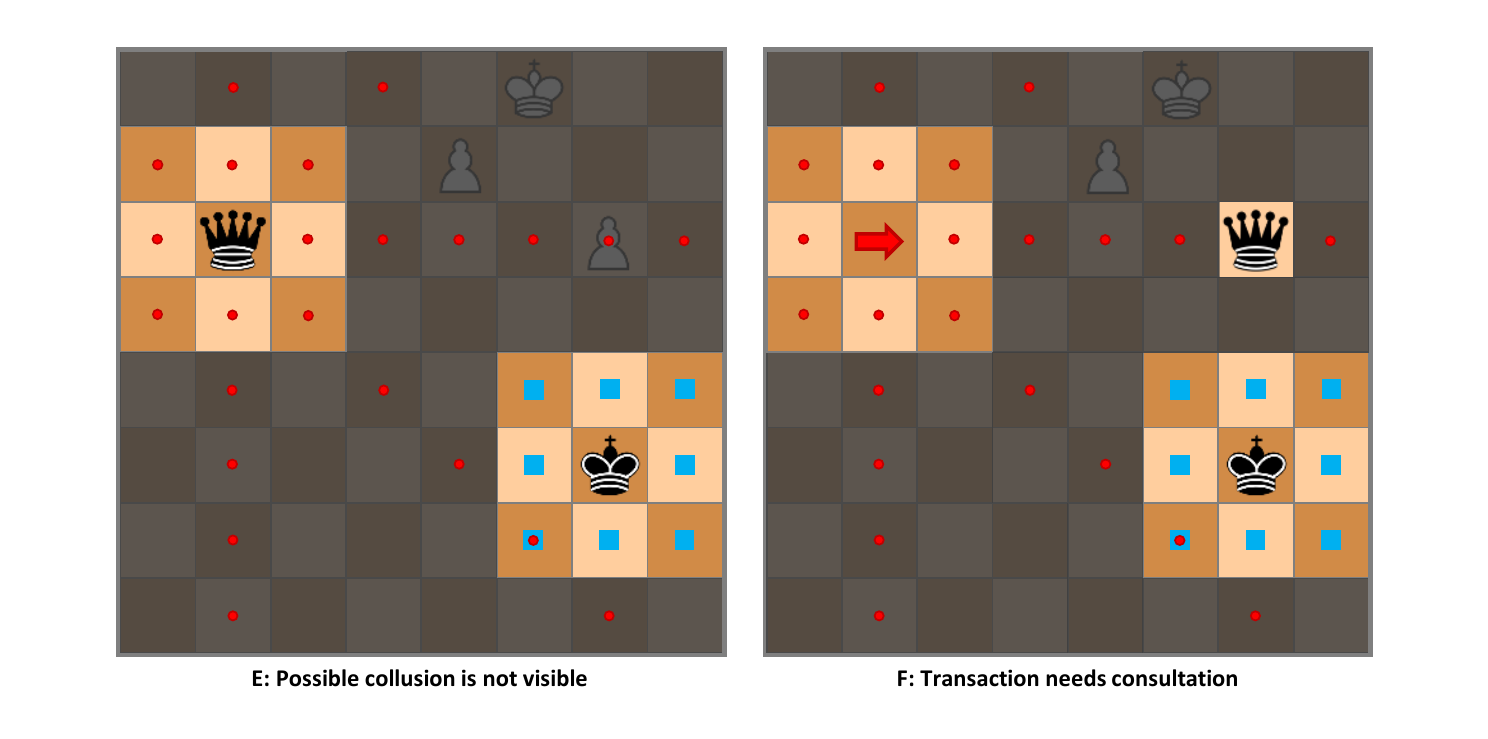}
		\caption{Fog of war in chess - Fogged moves (1)}
		\label{fig:FogOfWar_Chess_EF}
	}
\end{figure}

\noindent \textbf{Second}, the tricky case:
Just as games offer the possibility to inspect another player's hand of cards, in figure \ref{fig:FogOfWar_Chess_EF} (E),
all pawns in the game can only see the (directly) adjected tiles - regardless their movement possibilities.
Hence, performing the same move as seen in figure \ref{fig:FogOfWar_Chess_AB} to figure \ref{fig:FogOfWar_Chess_CD},
the player with black pawns does not know whether the queen will hit any enemy's pawn (Figure \ref{fig:FogOfWar_Chess_EF}, E to F).
The information has to be provided by the other players (see \hyperref[sec:RcTeDisputes]{Reveal Claims} below).
Consultation is needed to complete the move.
\begin{figure}
	\centering{
		\includegraphics[width=.995\linewidth,keepaspectratio=true]{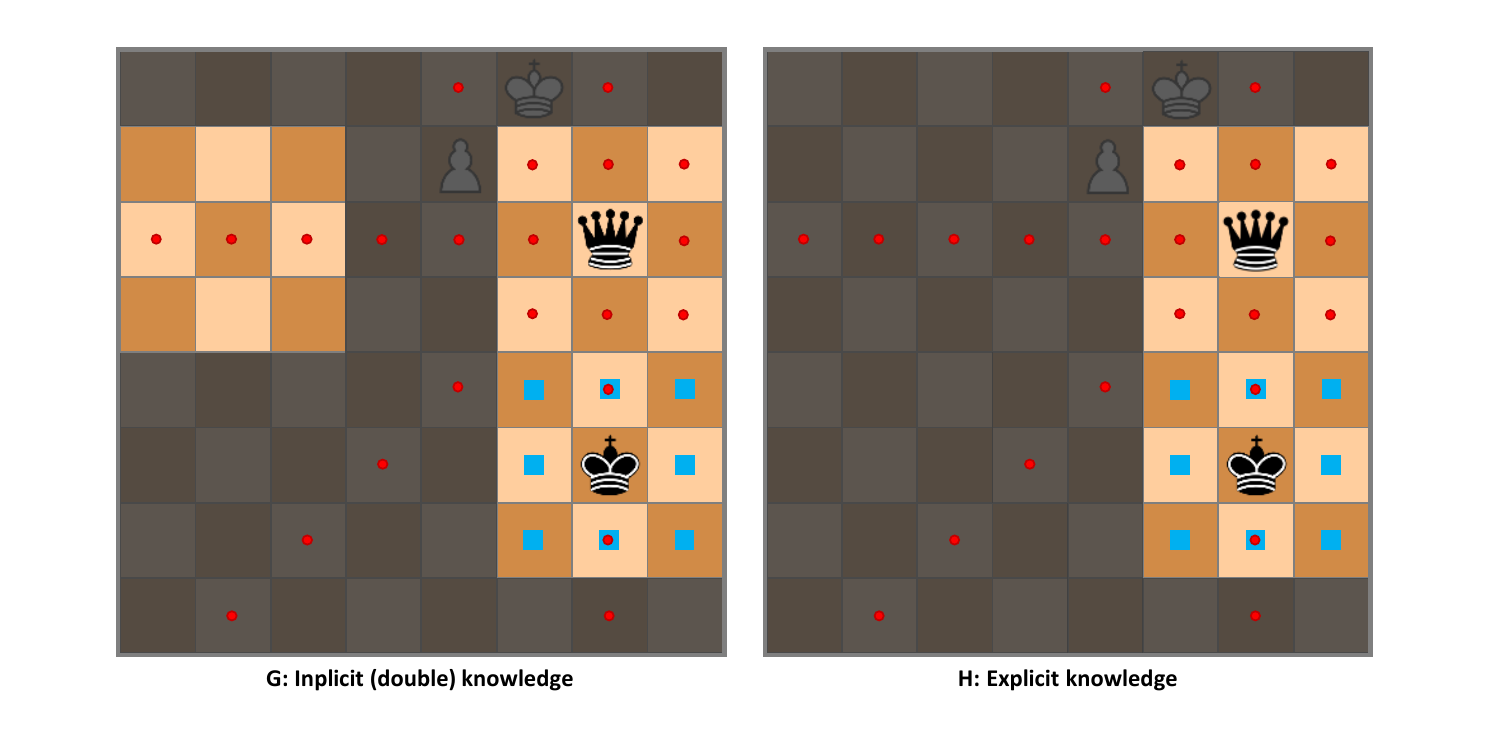}
		\caption{Fog of war in chess - Fogged moves (2)}
		\label{fig:FogOfWar_Chess_GH}
	}
\end{figure}
\noindent Implications:
\begin{enumerate}
	\item All this is generally possible using \textit{hidden transactions}.
	\item Due to the rules of chess \textbf{implicit knowledge} is granted - see figure \ref{fig:FogOfWar_Chess_CD}, C as well as figure \ref{fig:FogOfWar_Chess_GH}, G -
	which is not applicable to every game.
	\item The implication, that the player in black color has to offer all other players the (new) location of his queen, makes the whole 'fog of war'-situation \textbf{pointless} - at least for chess.
\end{enumerate}
\noindent Consequently, the underlying mechanisms two (\textit{implicit knowledge}) and three (\textit{pointlessness}) need to be considered throughout during a game's development process. \\
Point two and three implicate that fog of war is generally not possible in distributed environments conducting \gls{BCT}, but that is not completely true.
Chess has many constraints, such as only one move per turn can be performed, which diminishes the reappearance of fog.
Additionally, chess offers binary type information, such as the true/false answer to the question:
'Is the black queen located on field B4?'.\\
\textbf{Third}, the case in between:
Other games may offer to hide information better, especially if information is given in a numeric type and no \textit{exact amount} has to be delivered.
Less cryptically, assume a \gls{SC} exists in \hyperref[def:RftS]{RftS}, which says: \\
\textbf{A)} After his turn a player has to publish the number of ships (\textbf{X}) stored on each of his planets. \\
\textbf{B)} The real number (\textbf{X}) does not need to be published directly - it can be included in a \hyperref[HiddenTransactions]{hidden transaction}. \\
\textbf{C)} The publicly accessible number has to be within the range of +-\textbf{X}\%. \\
\textbf{D)} The deviation has to be calculated from a \hyperref[lbl:Randomization]{randomization} process.
The value extracted from the randomization may stay hidden until it has to be revealed.\\
Concluding, setting the number of ships to 100 \textit{constantly} and \textbf{X} to 30, the published values vary between 70 and 130 ships.
This could be recalculated long term using the mean.
Still, with extending variables such as the \textit{production of ships} (influenced by \hyperref[lbl:Randomization]{randomization}) and changing amounts due to \textit{incoming/outgoing fleets} etc., this \gls{SC}, based on numeric values, fits the guidelines of \textit{fog of war} \cite[3-4]{Setear.1989}.
Finally, the real number of ships can be fogged in distributed environments like \gls{BCT}.

\subsection{Reveal claims, trigger events \& Disputes}
\label{sec:RcTeDisputes}

\textbf{Reveal claims and trigger events} - Some games offer the possibility to inspect another player's hand of cards,
lift the fog on specific values of the game or in case of \hyperref[def:RftS]{RftS} claim a planet's (real) number of stored ships.
Living in the assumption of multiple players ($\geq3$) the information has to be provided for one player (only), a subgroup of players or to all players.
In the first two cases, the information is only encrypted to the \textit{legitimately claiming} node(s).
Still, the information transfer needs to be backed by a \hyperref[sec:SmartContract]{SC}.
Moreover, within a game spontaneous transactions/triggers could exist.
These may either be 'throw in transactions' or previously placed \hyperref[sec:SmartContract]{SC} which are just waiting to be triggered.
Anyway, these transactions follow sequential order and can therefore be conducted using \gls{BCT}.
Moreover this has certain implications on the \gls{CM} and will therefore be covered in chapter \hyperref[chap:PoT]{Proof-Of-Turn}. \\
\textbf{Disputes} - Although \gls{SC}s and the chosen \gls{BCT} \gls{CM} shall prevent disputes, sometimes they are inevitable.
Hence, a general pattern to resolve disputes is given:
\textbf{First}, the dispute may be solved within the network.
This could be conducted by a vote - the majority (>$50$\%) wins.
This is probably a good solution for disputes among a little group within the whole game network, as stakes in the game may influence the vote in an unfair manner.
\textbf{Second}, if possible, hand the dispute to any upper level instance as shown by \citet[92]{Kraft.2016}.
\textbf{Third}, the case which shall be prevented if possible: Stop the game(-channel) and prevent an unjustified win for any involved party.
The possibilities a game offers regarding the three cases determine the options after a dispute:
\begin{enumerate}
	\item Waiting / freezing the game until an absent player/node is answering.
	During this time the unavailable player may be kicked by vote
	(see section \hyperref[sec:PeerFluctuation]{Peer-Fluctuation \& adaptive turn time} in chapter \hyperref[chap:PoT]{Proof-of-Turn},).
	
	\item Impose any additional upper level penalty against the node e.g. by the publisher \cite[407]{Chatterjee.2019}.
	
	\item Stop the game generally.
	This is only recommended if players can be measured by points already.	
	
	\item Conduct any kind of recover strategy, which is implementation dependent.
	This might not be possible e.g. if cards had been dealt with the previous presented shuffling mechanism,
	as both the associated reverse shuffle order and encryption keys are lost.
\end{enumerate}

\FloatBarrier

\section{Data allocation improvements}
\label{sec:DataAllocationImprovements}
In the literature \textit{data allocation improvements} could only be found before
data was written to the \gls{BC} - e.g. game hash vs. encrypted data \cite[94]{Kraft.2016} but not altering the \gls{BC} thereafter.
The general idea of removing data from a \gls{BC} after blocks became \gls{CF} harms the characteristic of \textit{immutability of data} (\citet[17-18]{Butijn.2020}; \citet[57]{Dib.2018}; \citet[21]{Sharma.2020}).
Nevertheless, to meet the demand of reduced data consumption (See section \hyperref[sec:GamingMarket]{Gaming market}),
ways to reduce data allocation whilst keeping a shared consistency need to be evaluated. \\
If there was no central server from the publisher, not only the information in a transaction can be withhold (\textit{hidden transactions}).
Already the knowledge about a transaction can be a non-desirable factor.
To make this more tangible, in the aforementioned game, \hyperref[def:RftS]{RftS}, a player makes a move, sending a fleet to another planet.
This fleet is stored as an encrypted block.
Any opponent can now assume, that a potential fleet is on its way and is consequently able to
calculate round by round whether the fleet may reach the one or the other controlled planet.
As a player has to proof his turn later on with \textit{offset revealing}, publishing the move (somehow) to the blockchain is unavoidable. \\
Nevertheless, moves can be hidden in noise data using dummy data, here referred as a \gls{BT}. \label{def:BloatTransactions}
No matter if a player takes a move or not, an additional (random) number of \gls{BT}s is added.
\gls{BT}s can be be recognized as such right after their revelation.
All \gls{BT}s need to be revealed to claim the win (even if its the second place).
The enforced revelation prevents \textbf{double spending} of a planet's ships into several fleets.
More details on \textit{double spending} in the section \hyperref[sec:AttackVectors]{Attack Vectors}. \\
Additionally, game mechanics define the number of rounds until \gls{BT}s
have to be revealed (e.g. 10 rounds after the last non-\gls{BT}s is revealed).
Consequently, the number of transactions can not be calculated from the number of published blocks during a turn.
Therefore the number of moves can be hidden using \gls{BT}s.
Nevertheless, there is a \textbf{downside of \gls{BT}s} as well, \textbf{data storage}.
In long lasting games with many moves and players, the number of \gls{BT}s will probably derive to a problem. \\
The following, storage size calculation is highly assumptive and strongly dependent on the application's use case.
Nevertheless, general traits of different approaches shall be shown.
Therefore, presumptive values are given: \\
If most \gls{BT}s consumed $\sim0.01$ MB and in contrast most \textbf{relevant transactions} $\sim0.1$ MB,
\gls{BT}s could be distinguished from \textit{relevant transactions} easily by size.
Hence, \textit{relevant transactions} as well as \gls{BT}s need to allocate a comparable amount of storage space.
This increases data allocation by the factor '\gls{BT}s per \textit{relevant transaction}'.
Alternatively, \hyperref[def:GameHash]{game hashes} could be used decreasing each transaction down to ($\sim0.032$ KB).
This is always recommendable if the average transaction's size exceeds the \textit{game hashes} fixed amount. \\
During the following equations a \gls{BT}$(l)$ is a large blob of bloat data augmenting a \textit{hidden transactions} whilst a \gls{BT}$(s)$ is only augmenting a game hash.
If encryption is used, the data allocation can be calculated with \textit{hidden transactions} ($h$), and the 'factor of \gls{BT}s per hidden transaction' ($f($\gls{BT}$(l))$).
Additionally, revelations ($r$) of the encryption key ($k$) to different nodes and the network have to be taken into account.
\begin{center}
	Encryption: \hspace{0.2cm} $h + f($\gls{BT}$(l)) + r * k$
\end{center}
On the contrary, \textit{game hashes} ($H(h)$) can be used for \textit{hidden transactions}.
\begin{center}
	Game hashes (i): \hspace{0.2cm} $H(h) + f($\gls{BT}$(s)) + r * k$
\end{center}
If a \textit{hidden transactions} (game hash) has to be revealed to any other node r the whole network, it has to be written to the \gls{BC}.
\begin{center}
	Game hashes (ii): \hspace{0.2cm} $H(h) + f($\gls{BT}$(s)) + r * k + h$
\end{center}
Now, the two versions are compared:
\begin{center}
	$h + f($\gls{BT}$(l)) + r * k$ \hspace{0.2cm} \textbf{vs.} \hspace{0.2cm} $H(h) + f($\gls{BT}$(s)) + r * k + h$
\end{center}
For the sake of simplicity, '$h$' as well as '$r * k$' can be shorten.
Given the application's use case dependent values of \gls{BT}$(l)$ and \gls{BT}$(s)$,
the break even can now be calculated by equating the two formulas and solving for $f$.
\begin{center}
	$f($\gls{BT}$(l)) = H(h) + f($\gls{BT}$(s))$
\end{center}
This will show the use case's best suited solution.
Presumably, \textit{game hashes} are superior as they promise strong abstraction - especially of big files. \\

\noindent The following plot figures omit the number of published blocks by turn and/or round.
Instead, the types of transactions are relevant.
It is distinguished between \textit{relevant transactions} and \gls{BT}.
\textit{Hidden transactions} are a subcategory of \textit{relevant transactions}.
Hence, \textit{relevant transactions} can have the status \textbf{hidden}, \textbf{(recently) revealed} and \textbf{historic}.
The last category does not change the game state anymore and could be deleted (depending on the use case).
The plots assume either a long term or heavily written data due to \gls{BT}s.
A worst case scenario is assumed as a $50:50$ \textbf{ratio} between \textit{relevant transactions}
and \gls{BT}s (no matter if game hashes are used or not) are shown.
\begin{figure}
	\centering{
		\includegraphics[width=.95\linewidth,keepaspectratio=true]{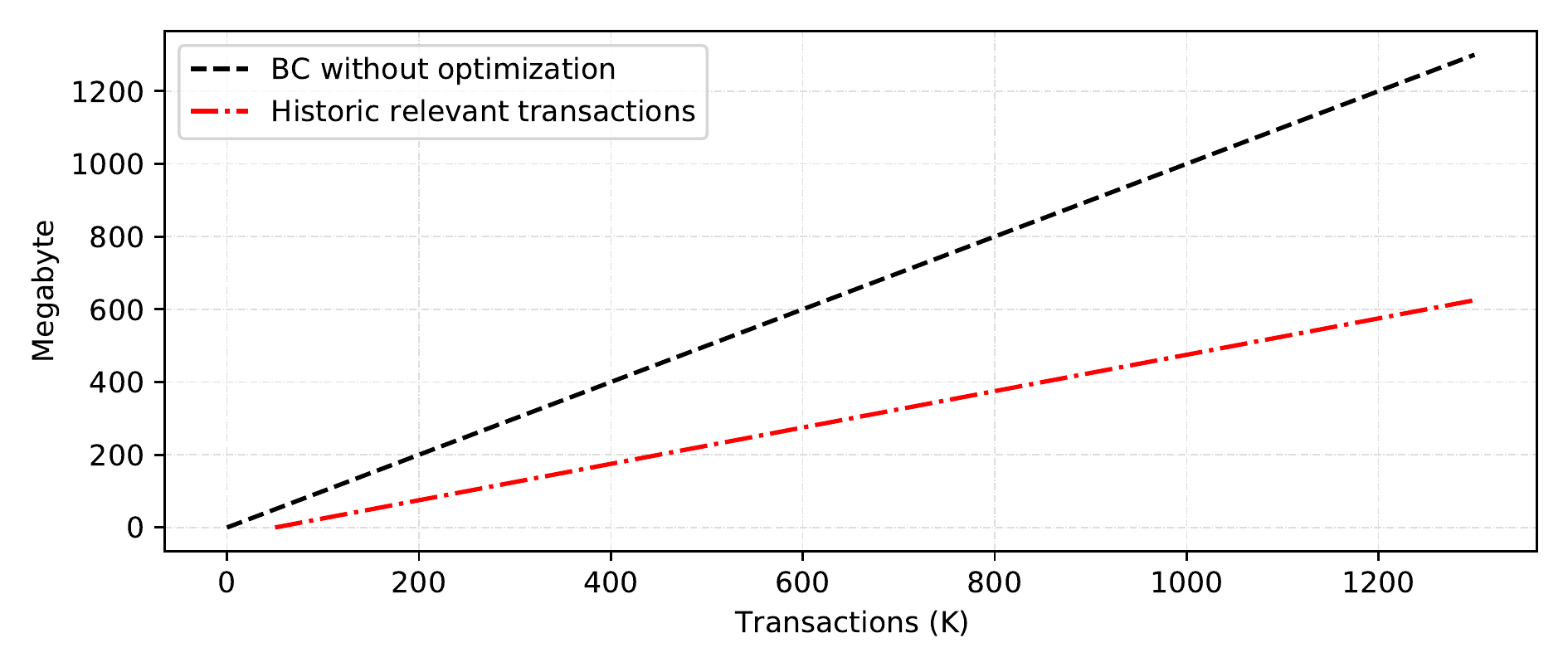}
		\caption{General \gls{BT} storage allocation chart}
		\label{fig:StorageAllocation_Base}
	}
\end{figure}

\noindent \textbf{First} of course each \gls{BC} starts with a \hyperref[def_GenesisBlock]{genesis block} at 'zero'.
The displaced start of the two graphs
(Figure \ref{fig:StorageAllocation_Base}\footnote{\hspace{0.1cm}Figure \ref{fig:StorageAllocation_Base} plot's Python code is given in the \hyperref[script:GeneralStorageAllocation]{appendix}.}) reflects the \textit{offset revelation},
as unrevealed transactions can not be distinguished upon being \textit{historically relevant} or \textit{deletable}.
Accordingly to the predefined \textit{ratio}, the upper $\sim50$\% of the \gls{BC}'s size (Figure \ref{fig:StorageAllocation_Base}, '\gls{BC} without optimization')
accounts for \gls{BT}s (Figure \ref{fig:StorageAllocation_Base}, Area between the lines).
Nevertheless, only the '\textit{Historic relevant Transactions}' are needed long term (Figure \ref{fig:StorageAllocation_Base}, Lower line). \\
Estimating an average block to be of $\sim0.001$ MB and an active \gls{BC} to consist of $\sim100,000$ transactions,
it would consume $\sim100$ MB of storage space (Table \ref{tbl:BloatblocksBlockchainSize}, Row 1).
Consequently, $600,000$ transactions $* 0.001$ MB $= 600$ MB (see Figure \ref{fig:StorageAllocation_Base}).
Considering that there is not only one game being played simultaneously, the needed storage space raises linearly (Table \ref{tbl:BloatblocksBlockchainSize}, Row 2). \\
Although a single block consumes little storage space, massively generated \gls{BT} raise the allocated storage space tremendously.
Importantly, each entire \gls{BC} is supposed to be stored on all network nodes.
Generally, \gls{BCT} is built upon networks with well equipped nodes regarding hardware.
In contrast, (casual) gamers might want to join using devices which do not offer much storage space, such as smartphones or handheld consoles.
Here a storage size consuming \gls{BC}s could be limited by (other) software restrictions.
In the worst case it may lead to a removal of the application/game by the gamer or even by any higher authority,
such as popular app stores\footnote{\hspace{0.1cm}E.g.: Apple 'App Store', Google 'Play Store', Epic Games Store and the Steam Store}. \\
Consequently, a mechanism to reduce long termed storage allocation is needed.
Still, the thought of orderly deleting data within \gls{BCT} is far from the usual approach,
such that \gls{BCT} is thought to prevent removal of any content in general.
\begin{table}
	\centering
	\begin{tabularx}{0.59\textwidth}{ l | c | c }
		& 1 Transaction & 100,000 Transactions \\ \hline
		1 \gls{BC} game & $0.001$ MB & $100$ MB \\ \hline
		10 \gls{BC} games & $0.01$ MB & $1$ GB \\
		\hline
	\end{tabularx}
	\caption{Blockchain Size}
	\label{tbl:BloatblocksBlockchainSize}
\end{table}
Nevertheless, some approaches exists which make "[...] it possible to re-write or compress the content of any number of blocks in decentralized services [...]" \cite[111]{Ateniese.2017}.
These approaches are not considered in the scope of this thesis as they (only) \textit{compress} but do not \textit{delete} content.
Additionally, most \gls{BC}s prohibit deleting single transactions passively, as they are encapsulated in one block with other still needed transactions.
But in this special case of utilizing \gls{BT}s, the following three mechanisms could be identified to reduce allocated storage space:
\hyperref[sec:PruneProcedure]{Prune bloat transactions}, \hyperref[sec:CcTransition]{Child-chain transition}
(alike \citet{Kraft.2016}'s game channels) and \hyperref[sec:MeatState]{Meta-State Blocks}.

 \subsection{Prune bloat transactions}
 \label{sec:PruneProcedure}	
\begin{figure}
 	\centering{
 		\includegraphics[width=.95\linewidth,keepaspectratio=true]{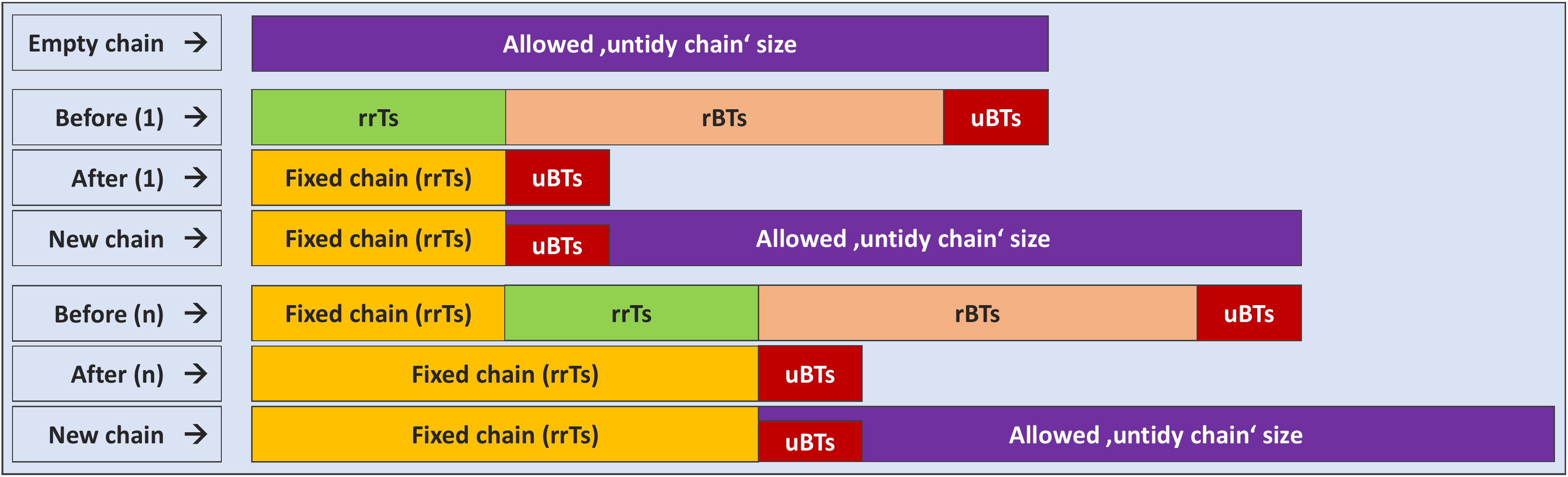}
 		\caption{Prune bloat transactions}
 		\label{fig:BloatBlocksPruneFromOrigin}
 	}
 \end{figure}

 Once a critical chain size is met, the chain could be recalculated using a \textbf{prune procedure}.
 As \textit{bloat transactions} could not be found in the literature, the deletion of these transactions is missing as well. \\
 The chain before the recalculation consisting of
 \gls{rrTs} (Figure \ref{fig:BloatBlocksPruneFromOrigin}),
 \textbf{revealed \gls{BT}s} (Figure \ref{fig:BloatBlocksPruneFromOrigin}, r\gls{BT}s),
 \textit{unrevealed \gls{BT}s} (Figure \ref{fig:BloatBlocksPruneFromOrigin}, u\gls{BT}s) as well as other negligible meta transactions
 is called the \textbf{untidy chain} (Figure \ref{fig:BloatBlocksPruneFromOrigin}, Untidy chain).\\
 The \gls{BC}'s network wants to get rid of meaningless allocated storage space and decides by vote to 'garbage collect' \textit{revealed \gls{BT}s} .
 One node has to conduct the \textit{prune procedure} - probably the node which broke a given \textit{untidy chain} size limit by adding transactions.
 The node calculating the prune is called the \gls{GCN}.

\noindent As \gls{rrTs}, \textit{revealed \gls{BT}s} and \textit{unrevealed \gls{BT}s}
are not separated or sorted within the \textit{untidy chain}, the \gls{GCN} generates a new chain from scratch except the \textbf{genesis block}.
During the recompilation, the \textit{untidy chain} is parsed by the \gls{GCN} and all \textit{revealed \gls{BT}s}
are deleted (Figure \ref{fig:BloatBlocksPruneFromOrigin}, Before(1) $\to$ After(1)).
This is only recommended in lightweight \gls{CM}s, as the \gls{GCN} needs to find suitable hashes for new blocks easily. 
Once the process is finished, another 'consensus vote' approves the new chain. \\
It has to be reminded that every network node exhibits the following behavior:
\begin{enumerate}
	\item Network nodes do not trust each other.
	The new chain, which was generated by the \gls{GCN} is checked/recalculated by every other network node.
	
	\item All network nodes have a desire to reduce their allocated storage space and look
	forward to a reduced chain - as long as the new chain complies with \textbf{one}.
\end{enumerate}
Consequently, if the \gls{GCN} fails to generate a valid new chain, the network tries to calculate a new chain delivered by another \gls{GCN}.
in this case, the first \gls{GCN} does not only have to calculate the first version, it also has to verify the second version.
Therefore every \gls{GCN} wants to avoid this (overhead workload) loose-loose situation which leads to honest recalculation and trust within the network. \\
The \gls{rrTs} are transitioned into a \textit{fixed chain} which will not be changed anymore (Figure \ref{fig:BloatBlocksPruneFromOrigin}, After (1): Fixed chain).
Transactions in the \textit{fixed chain} are unencrypted and do not store any additional bloat/coverage data.
Subsequent calculated \textit{prune procedures} will only focus on the 'new' \textit{untidy chain} (Figure \ref{fig:BloatBlocksPruneFromOrigin}, New chain: untidy chain).
Additionally, the \textit{untidy chain} size starts after the last \gls{rrTs} and directly contains all transactions which are not yet revealed. \\
All \gls{rrTs} which are transitioned into the fixed chain during this \textit{prune procedure}
may be written into a single block and signed by the \gls{GCN}, the transcribed encapsulated transactions are still signed by their emitting nodes.
Hence, the integrity of the chain is sustained.
Depending on the share of garbage data (\gls{BT}s/relevant transaction), the chain can be boiled down.
At the end of the \textit{prune procedure}, a marker is set which defines the start of the \textit{untidy chain}.
The next \textit{prune procedure} will start from that block (Figure \ref{fig:BloatBlocksPruneFromOrigin}, Before(n)). \\
Although the permanent growth of the \gls{BC} along '\textit{Historic relevant Blocks}'
(Figure \ref{fig:StorageAllocationPrune}\footnote{\hspace{0.1cm}Figure \ref{fig:StorageAllocationPrune}
	plot's Python code is given in the \hyperref[script:StorageAllocationPPSC]{appendix}.}, lower line)
can not be prevented using \textit{prune procedures} (Figure \ref{fig:StorageAllocationPrune}, Serrated line), the overall storage allocation can be dropped significantly. \\
If not triggered regularly, the prune approach is a computation heavy bulk operation.
It is computation heavy in the means of two factors:
\textbf{First}, generating the new chain as well as checking the integrity and \textbf{second}, in regards of network traffic to pass the whole new chain.
During a prune is calculated the game may be paused as described in section \hyperref[sec:PeerFluctuation]{Adaptive turn time} of the \hyperref[chap:PoT]{Proof-Of-Turn}.
Last, finding an optimal \textit{prune procedure} frequency (allocation size) is not part of this document.
For a more continuous approach to reduce allocated storage space, \textbf{Child-chain transitions} may be used.

\subsection{Child-chain transition}
\label{sec:CcTransition}
The idea of \textbf{child-chain transition} derives from \citet{Kraft.2016}'s \textit{game channels}.
Still, although \textit{game channels} conduct off-chain transactions using sidechains,
their intention is to speed up the main chain's throughput.
To the writers knowledge these \textit{game channels} are kept long term by the consulting parties.
Therefore, \textit{game channels} are not used for storage allocation improvements.
There was no further literature found regarding \textbf{child-chain transitions} for storage allocation improvements.
\begin{figure}[!b]
	\centering{
		\includegraphics[width=.95\linewidth,keepaspectratio=true]{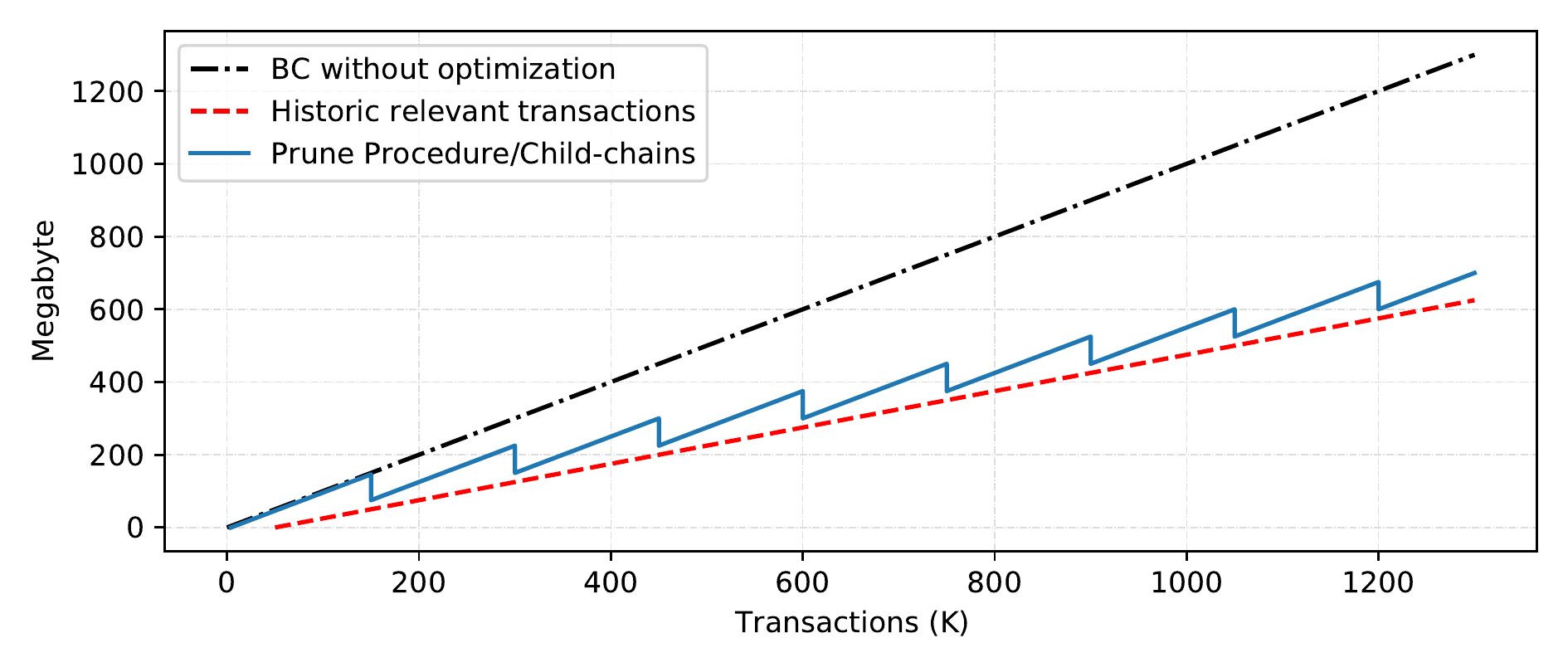}
		\caption{Prune procedure/Child-chain chart}
		\label{fig:StorageAllocationPrune}
	}
\end{figure}

\begin{figure}
	\centering{
		\includegraphics[width=.95\linewidth,keepaspectratio=true]{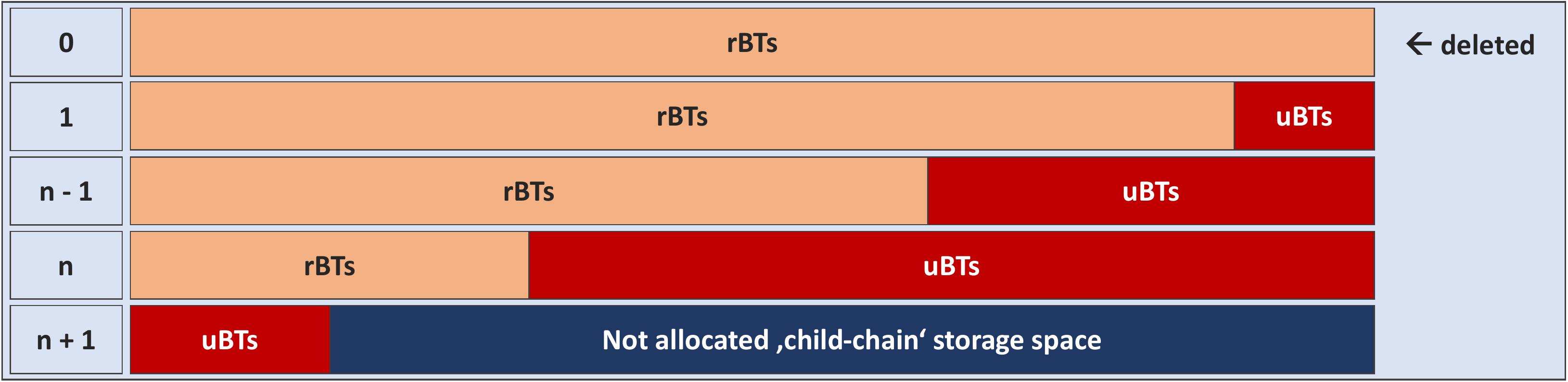}
		\caption{Child-chain storage allocation}
		\label{fig:BloatBlocksChildChain}
	}
\end{figure}

\noindent A downside of \textit{prune procedures} is the aforementioned peak of computation and network traffic on one single node.
A continuous \textbf{child-chain transition} from \textit{child-chains}
(see \hyperref[sec:PerformanceImprovements]{Performance Improvements})
to the \textit{main-chain} may be a reasonable answer to this challenge.
Here, a game uses an undefined number of \textit{child-chains} (Figure \ref{fig:BloatBlocksChildChain}),
each limited to e.g. $1$ MB ($\sim 1000$ transactions, following Table \ref{tbl:BloatblocksBlockchainSize}). \\
Whilst the \textbf{main-chain} stores all \gls{rrTs}, all unrevealed transactions are stored in \textit{child-chains}.
Once the storage size of a \textit{child-chain} is exceeded, all newly published blocks will be stored
on the next \textit{child-chain} (Figure \ref{fig:BloatBlocksChildChain}, \textit{n} $\to$ \textit{n+1}).
The newest \textit{child-chains} may therefore consist of \textit{unrevealed \gls{BT}s} only (Figure \ref{fig:BloatBlocksChildChain}, \textit{n+1}).
Still the filled \textit{child-chains} are temporarily kept (Figure \ref{fig:BloatBlocksChildChain}, \textit{1}, \textit{n-1} and \textit{n}). \\
By the time a transaction is revealed, the revelation key will be put on the dedicated \textit{child-chain}
- next to the unrevealed transaction - and the unencrypted residue is stored on the \textit{main-chain}.
Hence the \textit{main-chain} is unencrypted throughout.

\noindent By the time all transactions on a \textit{child-chain} are revealed or obsolete, the \textit{child-chain} can be deleted (Figure \ref{fig:BloatBlocksPruneFromOrigin}, 0).
The deletion is conducted by each node individually.
Hence a node - if desired - could keep the data as well.
Concluding, bloat/coverage data, which has initially been added to \gls{rrTs} becomes obsolete and
the key emitting (/revealing) node itself can sign the blocks on the \textit{main-chain} as well.
In this scenario \textit{child-chains} with small storage sizes offer faster deletion as it has to be waited until the last transaction is revealed.
Nevertheless, although small storage sizes lead to smooth overall storage allocation,
more filled \textit{child-chains} have to be stored until they can be deleted. \\
The overall storage size equals to the one from \textit{prune procedures} (Figure \ref{fig:StorageAllocationPrune}, Serrated line).
In this regards, frequent \textit{prune procedures} and small \textit{child-chains} equal each other as
seldom \textit{prune procedures} and big \textit{child-chains} do.
Whilst \textit{child-chains} offer an comparably equal distribution of computation and network traffic,
\textit{prune procedures} could be designed to be conducted on nodes offering strong computation power only. \\
Still, for \textit{child-chains} a lightweight \gls{CM} is needed throughout.
Finding an optimal \textit{child-chain} size is not part of this document.

\subsection{Meta-State Blocks}
\label{sec:MeatState}
\begin{figure}
	\centering{
		\includegraphics[width=.95\linewidth,keepaspectratio=true]{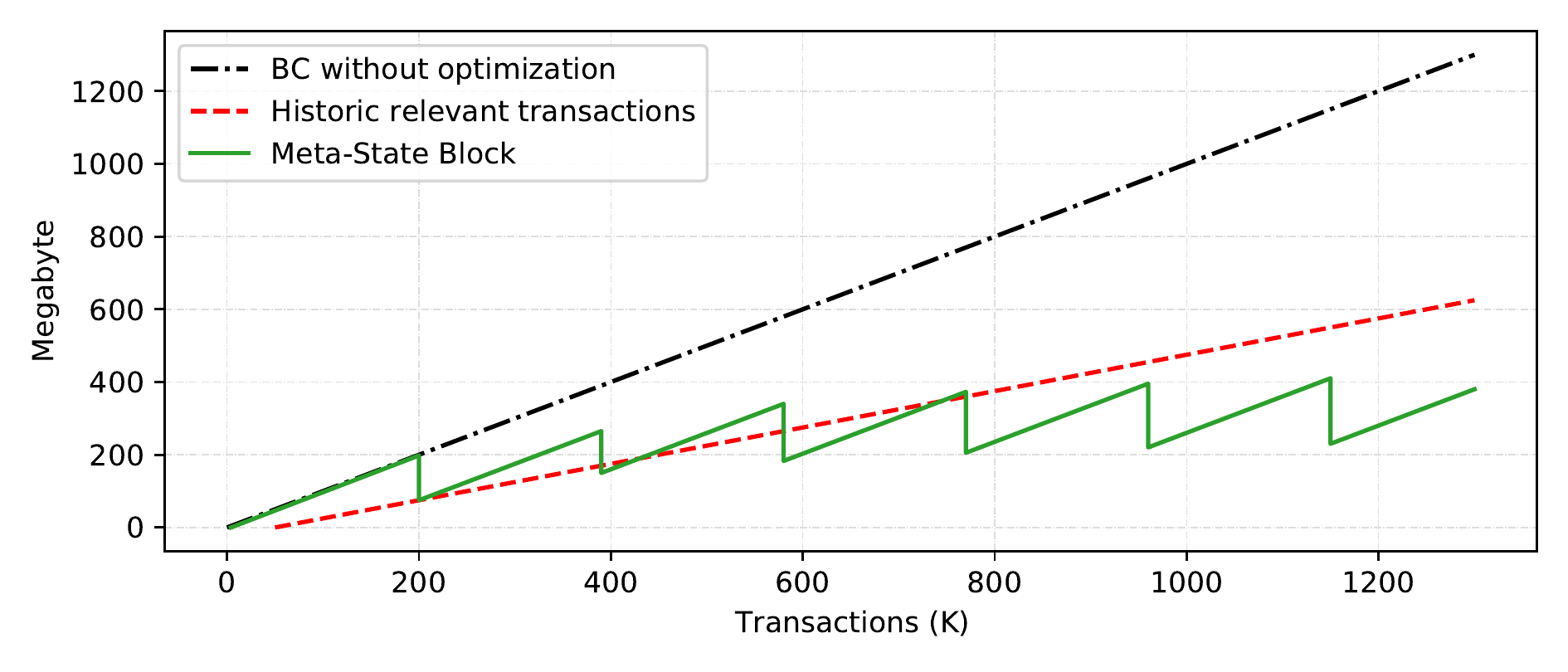}
		\caption{Meta-State Block chart}
		\label{fig:StorageAllocation_MetaState}
	}
\end{figure}

Both, the \textit{prune procedure}- as well as the \textit{child-chain}-approach offer a full (game) history.
Nevertheless, if storage space is limited, the network could decide to define any fully
\begin{figure}[!b]
	\centering{
		\includegraphics[width=.95\linewidth,keepaspectratio=true]{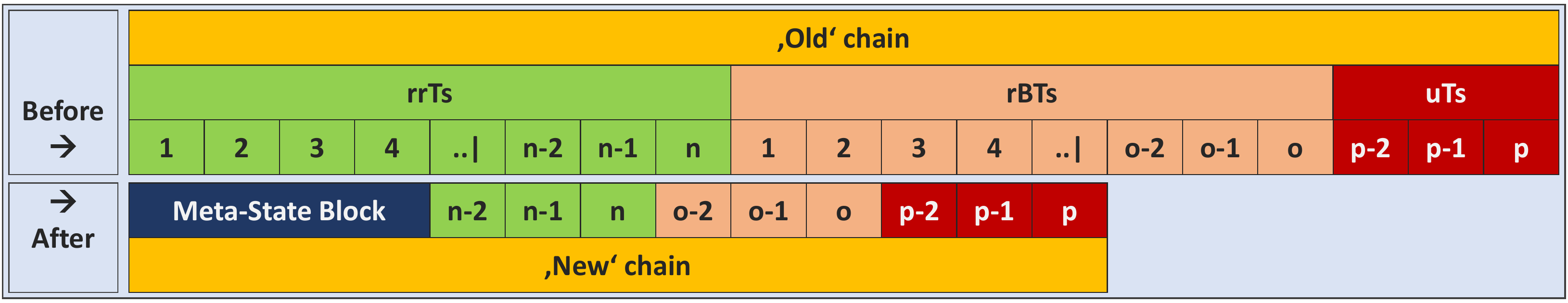}
		\caption{Meta-State Block process step}
		\label{fig:BloatBlocksMetaState}
	}
\end{figure}
revealed state - before the first \textit{unrevealed (bloat) transaction} - to be the new \textbf{genesis} (block).
Again there was no literature found which changes the genesis block to rewrite a \gls{BC}'s history for data allocation improvements. \\
A new \textbf{meta-state block}, at the beginning of the chain, represents the game's state at the \textit{genesis}
(Figure \ref{fig:BloatBlocksMetaState}, \textit{After}: Meta-State Block).
The \textit{meta-state block} summarizes all until then emitted transactions, defining a 'new default start'.
Old obsolete transactions are dropped, obfuscating 'historical relevant' data (Figure \ref{fig:StorageAllocation_MetaState}\footnote{\hspace{0.1cm}Figure \ref{fig:StorageAllocation_MetaState} plot's Python code is given in the \hyperref[script:StorageAllocationMetaState]{appendix}.}, Meta-State Block).
Except historically relevant values, the data in the \textit{meta-state block} does not differ from the network's knowledge 'at the starting point' (Figure \ref{fig:BloatBlocksMetaState}, \textit{Before}: '..|' $\to$ \textit{After}: Meta-State Block).
Therefore, it is encouraged not to use the last unrevealed (bloat) transaction, but to leave some historic data in between (Figure \ref{fig:BloatBlocksMetaState}, \textit{After}: 'n-2' $\to$ 'n', 'o-2' $\to$ 'o' and 'p-2' $\to$ 'p').

\noindent Although it is discouraged from, the equivalent in a cryptocurrency context would be
a \textit{meta-state block} at the beginning of (e.g.) the last elapsed
year - the wallets amounts remain the same, but transactions before
the \textit{meta-state block} can not be retrieved anymore.
Still, each node is free to either \textit{keep} or \textit{delete} the until then generated (old) chain.
Nevertheless, here it is assumed that the described nodes drop the old chain and receive
released allocated storage space in exchange (Figure \ref{fig:StorageAllocation_MetaState}, Meta-State Block).
\begin{figure}[!b]
	\centering{
		\includegraphics[width=.95\linewidth,keepaspectratio=true]{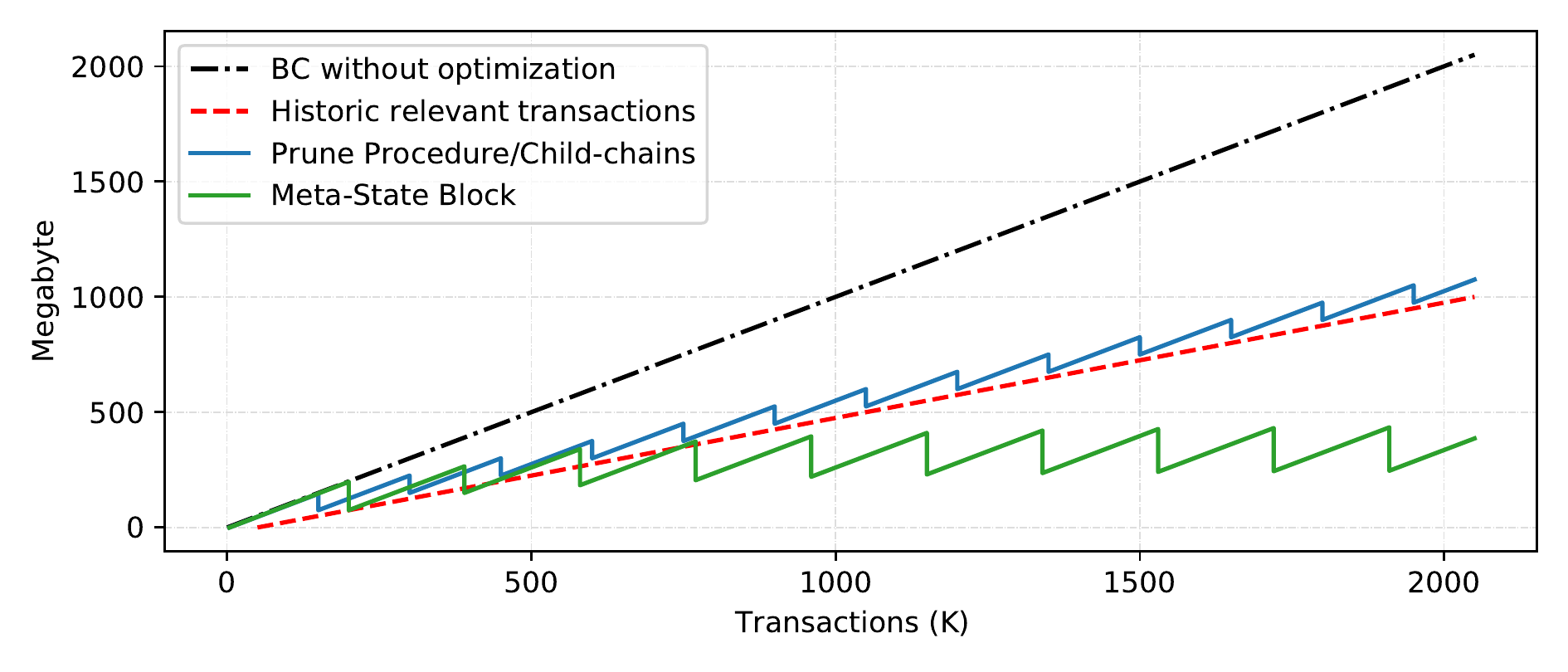}
		\caption{Storage allocation comparison}
		\label{fig:BC_StorageAllocation_All_Big}
	}
\end{figure}

\noindent Consequenly, in contrast to \textit{prune procedures} and \textit{child-chains}, all previous \textit{meta-state blocks}
containing movement information, randomization agreements and other transactions are deleted.
The \textit{meta-state block} is always the genesis.
Agreeing to the trade-off, the network looses its history before the genesis round, but retrieves storage space (Figure \ref{fig:StorageAllocation_MetaState}, Meta-State Block).
The baseline after the historic data was deleted was set to sample $\sim200$MB long term (Figure \ref{fig:StorageAllocation_MetaState}). \\
In figure \ref{fig:StorageAllocation_MetaState} it is assumed, that the application allocates an equal amount of storage in the long run, e.g. $\geq1.000k$ transactions.
Until this level is reached, the \textit{meta-state block}-approach is (here) not especially advantageous.
Given a (smaller) larger transactions size, this metric changes.
This method is well suited for applications, wherein history is considered obsolete in general.
Still all \gls{SC}s need to remain consistent for the upper level software.
Although the \textit{meta-state block}-approach is mentioned independently,
it can be combined with both, the \textit{prune procedure}- as well as the \textit{child-chain}-approach. \\
Finally, figure \ref{fig:BC_StorageAllocation_All_Big}\footnote{\hspace{0.1cm}Figure \ref{fig:BC_StorageAllocation_All_Big} plot's Python code is given in the \hyperref[script:StorageAllocationComp]{appendix}.}
shows the procedures together in one plot in larger scale ($2.000k$ transactions).
All procedures combine the characteristic that cheating is not possible, because only those blocks which are already verified by the network are deleted.
Of course, systems which contain 'above-average storage space' are able to keep historic data (longer).
Nevertheless, superior systems do not have any advantage regarding the quality of the stored information or benefits regarding upper level game play.

\section{Interim Summary - Blockchain Technology in Games}
\label{sec:InterimSummary-BCTinG}
Before presenting the \gls{PoT} \gls{CM}, a short summary of the so far identified results is given.
This chapter shows that \gls{BCT} is suitable for certain \hyperref[sec:GamingMarket]{games}
(e.g. \textit{gambling} and \textit{collectibles}) and is in some regards
already used to enable new types of games (here: \textit{collectibles}).
Still, publishers do not yet use \gls{BCT} to reduce network costs but rather
try to leverage cryptocurrencies and assets backed by \gls{NFTs} to boost their revenue.
Further, these applications proof that recent \gls{CM}s are already able to host slow games generally. \\
Additionally, a \hyperref[sec:ProblemSpace]{problem space} was defined to identify general, primarily board game related, \gls{SC}s.
Thereafter, \hyperref[sec:HiddenTransactionsPlusRandomization]{Hidden transactions \& Randomization},
\hyperref[sec:PileOfCards]{Piles of Cards}, \hyperref[sec:FogOfWar]{Fog of War}, as well as 
\hyperref[sec:RcTeDisputes]{Reveal claims, trigger events \& Disputes} were analyzed.
Last, possible \hyperref[sec:DataAllocationImprovements]{Data allocation improvements},
new to \gls{BCT}, were presented. \\
All these \gls{SC}s were given to answer the guiding question:
\begin{center}
	\textit{"Which kind of \gls{SC}s need to be established to cover typical in-game mechanics?"}
\end{center}
In this regards, one key factor has to be mentioned again:
A game can only be won by offering a full disclosure of all hidden content.
Without this final revelation it has to be assumed that fraud is being covered and
consequently the affected node is not allowed to win. \\

\noindent Nevertheless, the section covering the \hyperref[sec:GamingMarket]{Gaming market}
emphasizes the importance of well performing games on mobile devices such as smartphones.
Looking at the recent storage capacity of smartphones, \gls{BCT} games
developed for mobile devices probably demand the just described
\hyperref[sec:DataAllocationImprovements]{Data allocation improvements} to reach reasonable customer bases.
But the aforementioned \gls{CM}s do not offer the mandatory characteristic
to grant lightweight writing permission freely to specific nodes, whilst
keeping an adjustable \gls{BFT} - especially in the \textit{proof-of} category. \\
Therefore, a \gls{CM} offering these characteristics to enable games
on mobile operating systems, such as Google's Android and Apple's iOS is seen crucial
for \gls{BCT} to evolve further applications and use cases.
Additionally it would prevent the need for \textit{nerdy peer} setting up dedicated servers
and circumvent the \textit{shadow server} problem.
Last, the comparably easy option of providing updates to the gamer's devices via the app stores
reduces the publishers constraints to gamers not willing to update their dedicated servers to recent changes. \\
Consequently, next the \hyperref[chap:PoT]{Proof-of-Turn} algorithm is presented.


\chapter{Proof-of-Turn}
\label{chap:PoT}
\begin{figure}
	\centering{
		\includegraphics[width=.95\linewidth,keepaspectratio=true]{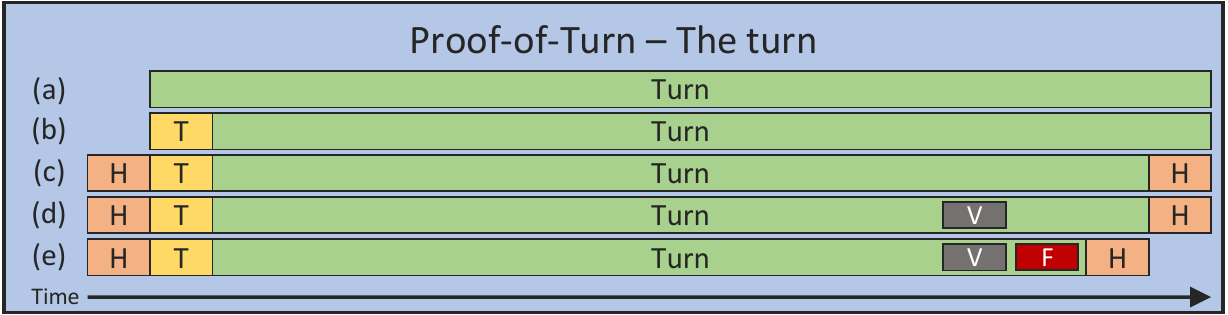}
		\caption{The turn}
		\label{fig:TheTurn}
	}
\end{figure}

In most of the mentioned \gls{CM}s extrinsic incentives, like tokens of cryptocurrencies
(\citet[4]{Butijn.2020}; \citet[3]{Khan.2020} \citet[85]{Kraft.2016}; \citet[4]{Nakamoto.2009}), are given.
The \gls{PoT} approach assumes intrinsic motivation instead, in this regard comparable to \gls{CM}s based on voting.
Still, the upper level software is assumed to offer enough incentives.
Nevertheless, mechanisms need to be established, which prevent nodes to behave hostile or malicious. \\
\gls{PoT} is guided by the business case of bringing \textit{turn based games} onto \gls{BC} infrastructure.
On a high level of abstraction, \gls{PoT} reminds of the round robin
nature of 'Token-Ring Local-Area Networks' (\citet{Bux.1989}; \citet{Ergen.2004}).
Although sticking to the phrase '\textit{turn}', this chapter abstracts from the gaming context.
Still, the \textit{turn}, as emphasized in its name, is the key factor of \gls{PoT} and therefore described first (Figure \ref{fig:TheTurn}).
In the most basic approach, each node awaits its turn, which is a predefined time slot in the round robin procedure (Figure \ref{fig:TheTurn}, (a): \textbf{Green turn}).
Derived from the gaming context, on the one hand, a turn may last only ten seconds.
On the other hand it is assumed that a turn will commonly last several hours or days.
By the time a node has its turn, it is called the \gls{LN}. \\
Moreover, the turns are assumed to enable \textbf{transition times} between the turns (Figure \ref{fig:TheTurn}, (b): \textbf{Yellow T}).
The transition times are supposed to be set to a constant
value of \textit{five} seconds, no matter how long the turn.
It is essential for transition times to last long enough to prevent network related race conditions \cite[75]{NetzerR.H.B..1992}.
Consequently the last \gls{LN} and its successor do not assume to be the \gls{LN} at the same time, preventing mutually exclusive forks. \\
In a more advanced case, the \gls{LN} and its successor both sign an arranged \textbf{handover-block} to seal the transition (Figure \ref{fig:TheTurn}, (c): \textbf{Orange H}).
This raises the security for the \gls{LN}'s successor to write to the latest branch(/fork) whilst ensuring the \gls{CF} for the \gls{LN}'s written data. \\
Additionally, once finished writing all data, the \gls{LN} may undertake a \textbf{network vote} (Figure \ref{fig:TheTurn}, (d): \textbf{Gray V}).
The vote asks upon the compliance of all in this turn published blocks.
Each node which received the vote call, raises the possibility for \gls{CF} as the possibility for a fork, leaving out the \gls{LN}'s data, is diminished.
Additionally, each positive reply increases the possibility for posted data to reach \textit{effective \gls{CF}}
(Section: \hyperref[sec:CFandBFT]{Consensus finality and Byzantine Fault Tolerance}) even before the \gls{LN}'s writing permission has ended.
If some data does not pass the vote, the \gls{LN} has the opportunity to revise the posted data until the turn's time slot is over.
A revision is supposed to only append blocks, setting things right.
Forks are prevented. \\
Once the \gls{LN} considers its turn to be finished, a \textbf{finalizing block} may be created,
which terminates the turn even before the designated time slot has passed (Figure \ref{fig:TheTurn}, (e): \textbf{Red F}).
Consequently, a \textit{finalizing block} is the 'soft version' of a 'hard and explicit' handover enforced by the end of the turn's time slot. \\
During its turn, the \gls{LN} is free to post all desired data.
In the first place the amount of data is only limited by the turn time and the \gls{LN}'s physical computation capacity.
Therefore, the turn's time slot has to be designed appropriately to ensure that all data can be written.
The latter is dependent on the upper level use case.
The bare minimum for data/blocks to be accepted is complying with \gls{BC}'s base structure
(Figure \ref{fig:BlockchainStructure}) combined with the \gls{LN}'s (valid) signature.
\begin{figure}
	\centering{
		\includegraphics[width=.95\linewidth,keepaspectratio=true]{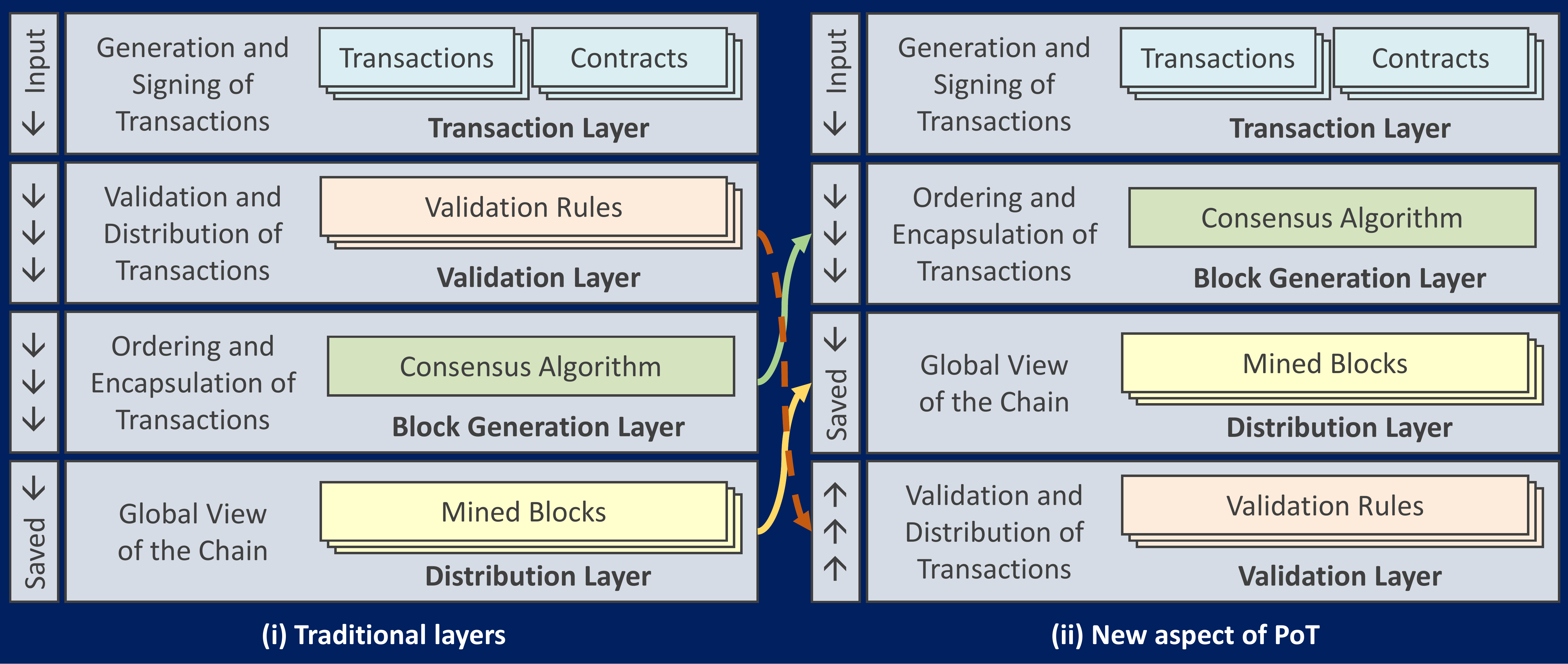}
		\caption{Layers in PoT (Adapted from  \citet{Oliveira.2019})}
		\label{fig:BlockchainLayersPoT}
	}
\end{figure}

\noindent At this point, the four layers of \gls{BCT} (Figure \ref{fig:LayersOfBlockchains}) are used in \gls{PoT} as well.
Depending on the attitude, on the one hand, one can argue that the validation is only conducted by the \gls{LN} and then directly pushed into the distribution layer.
The argument for this theory is the immediate \gls{CF} of each signed block.
The block is then part of the chain and immutable.
Following the definition, the \gls{CM} ends with the distribution (Figure \ref{fig:BlockchainLayersPoT}: (i)).
On the other hand, \gls{PoT} requires the data to be investigated after being distributed.
The data can not be deleted, but may be invalidated thereafter.
Consequently the validation layer is set last (Figure \ref{fig:BlockchainLayersPoT}: (ii)).
It is assumed that given a proper rule set, the invalidation will never occur.
Nevertheless, the nature of \textit{hidden transactions} may offer constraints to any previously published data (e.g. a node is offline during its turn (benign fault) and cannot reveal critical data) and
therefore raise the possibility of invalidation. \\
Regarding performance, nodes in other \gls{CM}s as \gls{PoW} have to try multiple times to mine a new block.
Comparably, proposing a new block in \gls{PoT} is cheap,
as the \gls{LN}'s fingerprint, derived from private-public key encryption,
on a block is sufficient for proposing the next block into the \gls{BC}.
Hence the runtime of each block's creation in \gls{PoT} is assumed to be $O(1)$. \\
As \gls{PoT} does not tie the \gls{LN}s block creation on mining success,
on-chain improvements like \hyperref[sec:PerformanceImprovements]{'Big block'} are obsolete.
In the first place, it is not distinguished between a \textbf{single block} containing
'all in the turn created data' from proposing \textbf{multiple blocks}.
Here, upper level rules may enforce to propose multiple blocks.
The latter, stems from randomization calls, which have to be finalized in the network,
before the \gls{LN} is allowed to continue its turn. \\
\textbf{\gls{PoT} is fully asynchronous.}
Therefore, in the most basic approach, \gls{PoT} does not allow other nodes to interfere a turn anyhow.
There is no repository for other node's transactions which have to be put into a block by the \gls{LN}.
Nevertheless, to enable randomization, disputes and likewise transactions, which demand consultation, either a fast single purpose round
(Section: \hyperref[sec:ClaimsAndTriggers]{Reveal claims \& trigger events}) has to be conducted
or any side-chain solution (Section: \hyperref[sec:Interoperability]{Interoperability}) may be used.\\
Of course, any time each node may send a direct message to the \gls{LN} to raise a dispute.
The \gls{LN} is supposed to follow \textit{legit disputes} to prevent double work,
invalidation, or (worst case) being sentenced with a handicap or any harder punishment by the whole network based on upper level rules.
Still, the \gls{LN} may decide to follow the validated (but not yet voted upon) dispute or ignore it.
The latter may lead to an invalidation of the affected data and a possible reset of multiple turns.
As the \gls{LN} has an intrinsic incentive for progress, it will most likely give in to a dispute call and raise a network vote upon the issue. \\
The \gls{PoT} concept is resembling to multiple, already existing approaches:
\begin{enumerate}
	\item \textbf{\gls{PoA}} assumes only a few nodes of the network to operate as writing/publishing nodes.
	The two clients \textit{Aura} and \textit{Clique} from the general \gls{PoA} are pretty similar to \gls{PoT}
	as they both elect one primary writing node, which is changing after a defined time slot \cite[2]{Angelis.2018}.
	Whilst \textit{Clique} allows multiple nodes to write simultaneously granting priority for the blocks written by the \gls{LN},
	\textit{Aura} and \gls{PoT} assume exactly one node to be allowed to write/publish data.
	But Aura votes for every emitted block directly,
	which leads to instant \gls{CF}, but slows down the system itself.
	Voting upon newly published blocks is optional in \gls{PoT} and generally tied to finalizing a turn.
	Additionally, \gls{PoT}'s \gls{LN} is generally not assumed to write data originating from other nodes.
	
	\item \textbf{\gls{PoP}} and \gls{PoT} are both primarily located in the gaming context.
	
	\item \textbf{\textit{Multichain}} uses a loose round-robin procedure, whilst \gls{PoT} enforces a strict round-robin sequence.
	\textit{Multichain} prevents nodes to write until a defined percentage of other nodes has written any block \cite[7-8]{Greenspan.2015}.
	If all the nodes allowed to propose a block remain silent, \textit{Multichain} freezes.
	The possibility freeze is not applicable in \gls{PoT} due to strict (maximum) time slots for each node's turn.
	Hence, the network has to wait until the next \gls{LN}'s successor becomes the \gls{LN}, but the network does not stall completely.
	
	\item \textbf{\gls{PoET}} uses exact timing to determine the next writing/publishing node.
	Although \gls{PoT} uses transition times to prevent the need for exact timing hardware,
	timing plays an important role passing the right for writing/publishing data to the \gls{BC}.

	\item \textbf{Vote based consensus} decides on each block or writing node separately.
	The common ground with \gls{PoT} arises in special occasions e.g. disputes
	which leaves all nodes to vote to either accept or invalidate a previous block or transaction.
\end{enumerate}
\noindent Due to its stiff progression of \gls{LN}s, \gls{PoT} is considered
to be used in \textit{private} or \textit{public} environments,
operated in permissioned or hybrid manner (Table: \ref{tbl:BlockchainNetworkTypes}).
If nodes joined and left the network in high fluctuation (e.g. in a permissionless environment),
\gls{PoT} would stall from absent nodes and may break its intended performance characteristics. \\
In the following, mechanisms and characteristics of \gls{PoT} are described:
\begin{enumerate}
	\item \gls{CF} and \gls{BFT} are discussed.
	\item General \hyperref[sec:Peering]{peering} of nodes is addressed.
	\item The classification into the \hyperref[sec:CAPtheorem]{CAP Theorem} is given.
	\item Passing turns and handling of possible forks are covered using \hyperref[sec:TransitionBlocks]{transition blocks}.
	\item \hyperref[sec:Interoperability]{Interoperability} with other \gls{BC} \gls{CM}s is described.
	\item Measures regarding \hyperref[sec:PeerFluctuation]{peer-fluctuation} are presented.
	\item Implications of \hyperref[sec:ClaimsAndTriggers]{trigger events} are shown.
	\item \hyperref[sec:FurtherCharacteristics]{Further characteristics} are given.
	\item Possible \hyperref[sec:AttackVectors]{attack vectors} are described.
	\item \hyperref[sec:Limitations]{Limitations} of \gls{PoT} are addressed.
\end{enumerate}

\FloatBarrier

\section{Consensus finality \& Byzantine Fault Tolerance}
\label{sec:CFandBFT}

Each \gls{LN} is supposed to validate its blocks following the given \gls{SC}s.
As stated before the blocks, written to the \gls{BC} by the \gls{LN}, meet \gls{CF} instantly - no consultation is needed.
The price to be paid for this fast-forward solution, is the possibility of a block to be invalidated thereafter.
Here it has to be differentiated between \textit{legal enhancements} (legally \textbf{added} to the \gls {BC})
and \textit{legal transactions} (legal \textbf{content} of the transactions).
In \gls{PoT} \textit{legal enhancements} of the blockchain structure (appending any block) meet \gls{CF} instantly.
The capsuled \textit{transactions}, may be invalidated sometime later.
The latter does not happen in the most \gls{BCT} algorithms like \gls{PoW}, as each transaction has to be open to everyone in the network to  pass the validation.
Unfortunately, the nature of game play sometimes prevents the direct detection of fraudulent (hidden) transactions.
In a worst case scenario the node itself - due to other hidden transactions - does not even know that its own transaction is invalid.
The latter has to be prevented by upper level rules.
Once a fraud \textbf{can be detected} by another node, a 'non action' being the \gls{LN}, equals a consent.
Hence, after data has passed $\geq 50\%$ of the nodes without intervention - no matter obviously fraudulent or not - the data can be seen as \textit{effectively consensus final}.
The $\geq 50\%$ refer to active nodes.
After all, this drawback - \gls{CF} vs. effective \gls{CF} - has to be accepted.
At the latest, when the writing node becomes the \gls{LN} again, it can be sure the published and revealed data is accepted without dispute.
Consequently, the scenario of a \textit{computational super power} recalculating the
chain - as possible in \gls{PoW} \cite[8]{Nakamoto.2009} - is not applicable in \gls{PoT}.
\gls{PoT} does not follow the \hyperref[LongestChainRule]{Longest Chain Rule}.\\
Additionally, voting to kick nodes out of the network which behave \textit{obviously hostile} is possible.
For both, voting for 'invalidation of any block' and 'kicking due to hostile behavior', a vote process has to be conducted.
The threshold, used in such a vote, is supposed to be set to $\geq 50\%$, but can be shifted as needed.
It is important that the following is known: \textbf{this threshold marks the \gls{BFT} for the whole algorithm}.
Consequently, the \gls{BFT} of \gls{PoT} remains implementation dependent as stated in table \ref{tbl:SumConsensusMechanisms_3}. \\
Game play related \gls{BFT}, as dependent on card draw mechanics, are explicitly not part of the \gls{PoT}-specific \gls{BFT} performance.
Last, to prevent any node flooding the \gls{BC}, it has to be considered whether a limit of written blocks/data is set for each turn.
Generally a limit is not recommended, as first bloat blocks may exceed the limits and
second disputes votes and following inflicted handicap upon hostile nodes can easily be conducted in a retrospection.

\FloatBarrier

\section{Peering}
\label{sec:Peering}
Data needs to be distributed through procedures and processes in the network.
Although the following section is about networks and peering, specific details of the TCP/IP- as well as the
OSI-model (\citet{Forouzan.2003}) are not discussed. \\
As the \gls{CM} manages new transactions decentralized, passing them trough the layers (Figure \ref{fig:BlockchainLayersPoT}), the whole network and its state is not (always) updated on every attached node until a block can be considered \gls{CF}.
Those transactions are previously, encapsulated in mined blocks (\citet{Butijn.2020}; \citet{Khan.2020}; \citet{Nakamoto.2009}), pushed towards the ledger from each node within the network.
In the most \gls{CM}s each node can always push a block into the network legally (\citet{Khan.2020}).
Hence, the (active) ledger is represented by many nodes simultaneously.
On the contrary, a pull(only)-mechanism to fetch updates would generate an enormous amount of network traffic as every node has to pull from every other node constantly.
Given the number of nodes, $n$, the number of pull requests for updates would scale by '$f(n) = n^2$'.
In a binary world, using a push-mechanism for new data is evidently the right choice
as the next block's origin node can be any network node.
Nevertheless, popular clients as Ethereum, Hyperledger Fabric and Corda mix both mechanisms \cite[6]{Daniel.2019}.
Still, literally, a node has to push hard to fight its data into the ledger. \\
In contrast, in \gls{PoT}, the (active) ledger is only represented by one node.
Only the \gls{LN} is allowed to push new transactions.
Thus, push- and pull-mechanisms are generally possible (as well) and described in detail.
Nevertheless, before digging deeper, it has to be distinguished between two scenarios, \textbf{consultationless} and \textbf{consultationfull}.
On the one hand, a \textbf{consultationless} network would be a shared storage wherein data is pushed only for others to be seen.
No node has a desire to change other node's data.
There is no reason for disputes or spontaneous interactions - such as triggers or throw in transactions.
On the other hand, a \textbf{consultationfull} network is assumed to utilize randomization calls, votes,
disputes as well as spontaneous interactions multiple times during a full (round robin) network round. \\
Moreover, the nodes are supposed to be loosely coupled, as players of games may quit (the game) or shut down their client.
Additionally, high mobility of nodes may hinder static network addresses and connectivity.
Hence, a broadcast in dedicated (static) sub networks is not applicable. \\
Last, permanent direct, linked connections are subsequently excluded due to possibly high numbers of nodes.
Therefore, more than five nodes are expected. \\
The literature describes peering mechanisms of \gls{BCT} only on seldom occasion referring to single transactions \cite[4]{Daniel.2019}.
Consequently push-mechanisms from an emitting node into the (whole) network are supposed to be common,
whilst pull-mechanisms exceeding \gls{SC}s could not be found in the literature.
Consequently the literature lacks an overall picture using 'only a pull-Mechanism' or 'only a push-mechanism'. \\
Given these assumptions, a sample pull-mechanism and a sample push-mechanism are described:

\subsection{Pull-Mechanism} \label{def:PullMechanism}
As stated, the decentralized nature of \gls{BCT} prevents frequently 'full-network' pull-mechanisms generally.
Nevertheless, in the special case of \gls{PoT}, the \gls{LN} circles,
alternating from node to node, due to the round robin procedure.
Hence, the source for the latest update is defined by time.
Consequently, nodes may pull from the latest \gls{LN} directly or from its predecessor to prevent interventions of the \gls{LN}'s turn.
\begin{figure}[!b]
	\centering{
		\includegraphics[width=.95\linewidth,keepaspectratio=true]{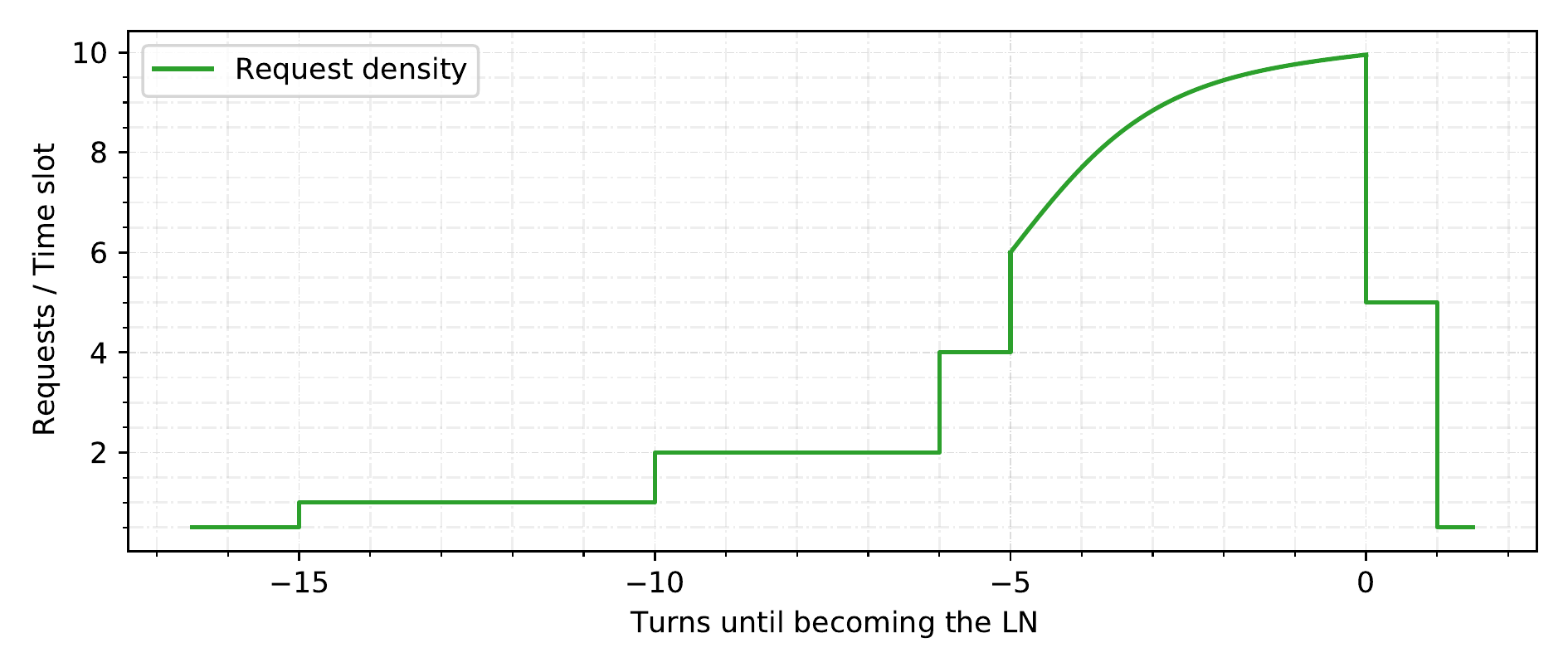}
		\caption{Sample 'peering by pulling'-strategy}
		\label{fig:PeeringByPulling}
	}
\end{figure}
Hence, the number of requests is tremendously reduced. \\
The following description is therefore set in a \textbf{consultationless} use case.
The nodes are lazy and try to pull seldom to \textit{reduce network traffic}.
Hence, each node is just waiting for its own turn.
A node missing its own turn is considered a worst case scenario.
Therefore, if a node expects to have its turn soon, it becomes more active.
A sample pull strategy is given in figure \ref{fig:PeeringByPulling} and different node's behaviors are described in the following. \\
On the y axis, the pull strategy in 'requests per time slot' is given.
For the sake of tangibility, first, 60 minutes is set as a turn's time slot ($t = 60 min$).
Second, the network consist of '$n=50$' nodes.
Third, the network is not stiff, as 'ending a turn prior to its maturity' is possible. \\	
This pull strategy is designed to be constraint to the node's distance to its next turn.
Along the round robin alternation, the node will have its turn again and again in the multiple of $n$.
But the number of turns until the next slot of being the \gls{LN} has to be within
'$x < n$'\footnote{\hspace{0.1cm}Depending on the base index, '$x \leq n$' is applicable as well}.
Predecessors of the \gls{LN} are not directly reset - hence '$x > 0$' is possible - and consequently a \textit{sample overflow} is set to ten ($n*1/5 = 10$).
Hence, most network nodes ($n*4/5 = 40$) assume to have a negative index.
The range of x can be described as:

\begin{center}
	$-n*(4/5) \leq x \leq n*1/5 = -40 \leq x \leq 10$.
\end{center}

\noindent The used index overflow is implementation dependent. \\
Effectively, assuming from a time measurement, that nine turns need to be conducted until becoming the \gls{LN}, offers the input '$x = -9$' on the x axis.
Following figure \ref{fig:PeeringByPulling}'s graph, '$x=-9$' leads to \textit{two} pull requests per turn (time slot) or likewise \textit{one} pull request ($p$) each 30 minutes.

\begin{center}
	$p = t/f(x) = 60min/f(-9) = 60min/2 = 30min$
\end{center}

\noindent Given these assumptions, the following cases (\textbf{A} to \textbf{E}) are distinguished:
\begin{enumerate}
	\item \textbf{Node A} is half way through the round robin
	
	\begin{center}
		$x = -(n/2) = -(50/2) = -25$
	\end{center}
	
	and looking forward to its next turn, which is more than 15 turns away (Figure \ref{fig:PeeringByPulling}: '$x < -15$').
	The sample pull strategy function returns '$y = 0.5$' for values below '$x = -15$'.
	One could say that node \textbf{A} is conducting in a 'network deep sleep', as it pulls only every second turn:
	
	\begin{center}
		$p = t/f(x) = 60min/f(-25) = 60min/0.5 = 120min$
	\end{center}

	Nodes do not ask node \textbf{A} for updates, as they prefer to pull from nodes, which are assumed to just have finished their turn (e.g.: $ 1 \leq x \leq 10$), likewise node \textbf{E}.
	Obviously, node \textbf{A} does not assume to have its next turn soon.
	Even if some peers finished their turns early, a seldom pull request is sufficient to not miss the time slot for becoming the \gls{LN}.
	Hence, an update is requested after \textit{two time slots have passed} (Figure \ref{fig:PeeringByPulling}: '$x < -15$').
	The request follows the logic of the \gls{BPMN} diagram (Figure \ref{fig:PeeringPullProcedure}).
	As turns can be ended prior to its maturity, node \textbf{A} only has a scale guess on the recent \gls{LN}.
	Consequently, node \textbf{A} tries to calculate the \gls{LN} heuristically, assuming no prior prematurely turnover.
	Hence node \textbf{A} requests its updates from the node, which is assumed to have its turn right now (here: node \textbf{D}).
	If node \textbf{D} is the \gls{LN}, it is free to either directly provide the update or refer to an already updated node (e.g. node \textbf{C} or node \textbf{E}).
	The latter shall prevent node \textbf{D} from being bothered during its turn. \\
	In the \textbf{best case}, although targeting the \gls{LN}, node \textbf{A} does not consult node \textbf{D}, but node \textbf{E} instead.
	Although node \textbf{E} is not the \gls{LN} anymore, it has just finished its turn and replies the request offering the latest version of the \gls{BC}.
	In the \textbf{worst case}, no answer is given at all, as node \textbf{D} is offline.
	Node \textbf{A} therefore asks node \textbf{D}'s predecessor (node \textbf{D+1}) and node \textbf{D}'s successor (node \textbf{D-1}), one after another.
	Except the \gls{LN} all nodes are assumed to be willing to provide updates immediately as
	shared data is predisposed to reveal flaws and advances progress.
	Node \textbf{A}'s requests are conducted in both directions until an appropriate answer is received.
	A pull request from itself is obviously skipped.
	In the \textbf{cases in between}, neither node \textbf{D} nor node \textbf{D-1} are targeted by the pull.
	Although data is received, the new information implies that the round robin procedure was (way) faster than expected.
	From the received update, node \textbf{A} can extract the timestamp from the last legal transition block.
	With this information it can recalculate the heuristic.
	Using this knowledge, node \textbf{A} may restart the process directly or wait until its next request trigger.
	
	\item \textbf{Node B} is approaching its turn and assumes to be eighth in line ($x = -8$).
	Consequently, the sample pull strategy function (Figure \ref{fig:PeeringByPulling}) causes node \textbf{B} to perform requests each 30 minutes:
	
	\begin{center}
		$p = t/f(x) = 60min/f(-8) = 60min/2 = 30min$
	\end{center}
	
	In regards of a full round, node \textbf{B} assumes to have its next turn somehow soon.
	Additionally, node \textbf{B} is not consulted by other nodes and utilizes the same procedure as node \textbf{A} (Figure \ref{fig:PeeringPullProcedure}).
	
	\item \textbf{Node C} is almost at its turn and knows to be first in line ($x = -1$).
	Whilst the pull requests increased during the last time slots continuously (Figure \ref{fig:PeeringByPulling}: See '$-5 \leq x < 0$'), node \textbf{C} tries to establish a stable, linked connection to the \gls{LN}.
	Once successful, node \textbf{C} does not need to repeatedly request for updates.
	Node \textbf{C} is only consulted by other nodes in exceptional cases - e.g. other nodes do not reach the \gls{LN} (node \textbf{D}) or its predecessor (node \textbf{E}).
	Node \textbf{C} awaits to become the the \gls{LN}.
	
	\begin{figure}
		\centering{
			\includegraphics[width=.95\linewidth,keepaspectratio=true]{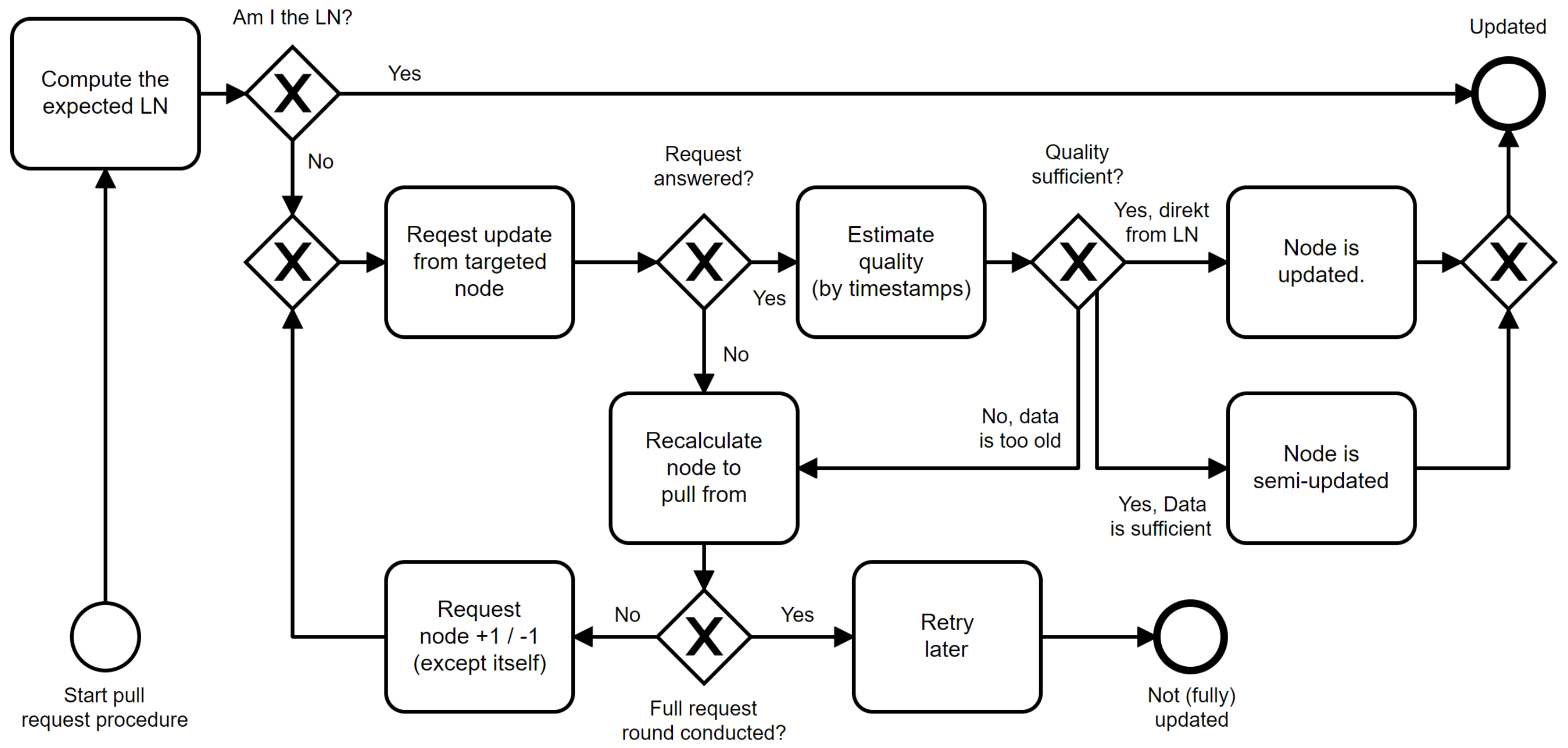}
			\caption{Pull request procedure}
			\label{fig:PeeringPullProcedure}
		}
	\end{figure}
	
	\item \textbf{Node D} is the \gls{LN} and in the middle of its turn's time slot ($x = 0$).
	During the transition, node \textbf{D} received the latest version of the \gls{BC} from its predecessor, node \textbf{E}.
	Consequently node \textbf{D} represents the latest version of the ledger and
	has no need to pull from any other node.
	Therefore the pull procedure is skipped by the first junction of the pull process (Figure \ref{fig:PeeringPullProcedure}).
	Still, a suitable value has to be found which balances between triggering
	the pull procedure too often and loosing connection to the network after the turn.
	All \textbf{four equations} (\textbf{I} to \textbf{IV}) represent the same intention: The \gls{LN} starts pulling from other nodes after its own turn. \\
	Whilst formula (\textbf{I}) does not allow to be calculated (Condition: '$f(0) = 0$'), resulting not triggering the pull process anymore,
	formula (\textbf{II.a}), using a value nearby zero ('$f(x) = 0.\overline{0}1$') increases the pull timer
	tremendously as shown formula (\textbf{II.b}).
	Until the next pull procedure is triggered after one year, many turns will have been missed already.
	Both, (\textbf{I}) and (\textbf{IIa} \& \textbf{II.b}) disconnect node \textbf{D} from the network permanently and are therefore dismissed.
	\begin{center}
		Not applicable: \\
		(\textbf{I}): $p = t/f(x) = 60min/f(0) = 60min/0 = NaN =>$ \textit{no pull} \\
		and \\
		(\textbf{II.a}): $p = t/f(x) = 60min/f(0) = 60min/0.\overline{0}1 = p \to +\infty =>$ \textit{no pull} \\
		- likewise - \\
		(\textbf{II.b}): $p = t/f(x) = 60min/f(0) = 60min/0.0001 = 60,000min >$ \textit{1 year} \\
	\end{center}
	Therefore '$f(0) = 5$' was used in figure \ref{fig:PeeringByPulling}, which can be seen in formula (\textbf{III}).
	Here, node \textbf{D} triggers the pull procedure each 12 minutes until the first junction
	of the \gls{BPMN} diagram (Figure \ref{fig:PeeringPullProcedure}) is answered 'No'.
	\begin{center}
		Applied: \\
		(\textbf{III}): $p = t/f(x) = 60min/f(5) = 60min/5 = 12min$ \\
	\end{center}
	\textbf{Instead}, a ratio could be used to send node \textbf{D} into a 'network deep sleep' as shown in formula (\textbf{IV}).
	The equation changes and uses a quarter round ($n*(1/4)$), constraint to the number of nodes.
	This leads node \textbf{D} to pull after $12.5h$:
	\begin{center}
		(\textbf{IV}): $p = t*(n*1/4) = 60min*(50*0.25) = 60min*12.5 = 12.5h$ \\
	\end{center}
	While node \textbf{D} is not pulling from other nodes, a stable, linked connection with the successor node \textbf{C} is established.
	Frequently node \textbf{D} is asked whether it is the recent \gls{LN}.
	If node \textbf{D} had ended its turn already, it would provide the update.
	The other way around, node \textbf{D} may reference to a likewise updated node, such as node \textbf{E} or node \textbf{C}.
	Once all data is pushed, node \textbf{D} may end its turn early.
	Of course node \textbf{D} is allowed to decide whether it wants to remain the \gls{LN} until the end of its turn's time slot or pass the turn to its successor before.
	An upper level application may incentive node \textbf{D} to pass the turn as fast as possible or prevent prior finalization of its turn.
	A reason against passing the turn ahead of schedule is that the \gls{LN}'s successor is not linked, yet.
	A variable turn time may allow the \gls{LN} to even prolongate its turn's time slot given certain conditions.
	The changing starting times with every prior-finished turn may allow to last a \gls{LN}'s turn until the stiff's end (Figure \ref{fig:TurnEndBeforeMaturity}).
	As the node might assumed to have the start of its turn later on.
	Either first, the node can not be blamed, second the rules stipulate a strict variable plan or third rules enforce online activity and aligned turn times.
	
	\item \textbf{Node E} has just ended its turn.
	The node \textbf{D} has taken the lead and node \textbf{E} will pull again when the trigger (formula \textbf{III} or \textbf{IV}) is activated.
	Node \textbf{E} will reset its index, once it reaches the \textit{sample overflow} ($x=10$).
	Consequently, for it's own updates, node \textbf{E} raises pull to the bare minimum and falls into a 'deep sleep', alike node \textbf{A}.
	Still, the next time slots will consist of many pull requests, which have to provide the latest data.
\end{enumerate}

\noindent Concluding, aiming to optimize network traffic, a pull strategy is supposed to be the most lightweight scenario as nodes are allowed to fall into a 'deep sleep'.
Still the pull strategy function has to be adjusted to the networks needs.
Here both, step function elements (Figure \ref{fig:PeeringByPulling}: '$-5 > x$' and '$x \geq 0$') and continuous function elements (Figure \ref{fig:PeeringByPulling}: '$-5 \geq x < 0$') may be used. \\
Obviously, the drawback of consultationless peering is the loss of short termed answers on behalves of the whole network.
Here, to fix the latter issue, a full network round until a decision is made, would be needed.
A special round for consultation, alike in the subsequent section \hyperref[sec:ClaimsAndTriggers]{Reveal claims \& trigger events}
would breaking the stiff time slots and is therefore not suitable in consultationless environments.
Implications of turns 'ending a turn prior to maturity' will be discussed shortly.
Anyhow, stability can be risen using more linked connections or with a higher density of pull requests. \\
Still, regarding the pull request density on the \gls{LN}, in larger networks this might lead unintentionally,
to several internal, flawed by design, \textit{network/transport-level flooding attacks} \cite[2046]{Zargar.2013}, also known as \gls{DDoS}.
Of course there might be possibilities to reduce the network load on the current \gls{LN},
but updates would still trickle slower throughout the network than push based mechanisms, due to the asynchronous nature of pull timings.	
An issue which has to be covered generally using 'ending a turn prior to maturity', is the handling of nodes which take their turn 'too late, but in time'.
Assuming nodes to stay in a permanent deep sleep until their next turn, only pulling once per round, may be sufficient in stiff environments (Figure \ref{fig:TurnEndBeforeMaturity}: A, Eight stiff turns, green).
But using early turnover (Figure \ref{fig:TurnEndBeforeMaturity}: B), the turns one to six might have been ended that fast, that turn seven did not receive the update in time.
Node six has to wait until node seven believes to have its turn again.
Here, the waiting time (Figure \ref{fig:TurnEndBeforeMaturity}: B, Red) would be long enough to conduct two full turns.
A upper level implementation would need to decide either to skip turn seven or to wait until the regular turn slot has passed.
The latter prevents the round robin to proceed faster.
Nevertheless, a fluid turn time demands a higher density of pull requests.
\begin{figure}
	\centering{
		\includegraphics[width=.95\linewidth,keepaspectratio=true]{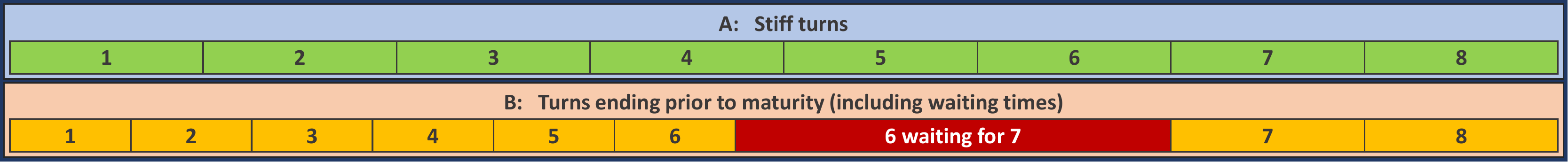}
		\caption{Ending turn before maturity}
		\label{fig:TurnEndBeforeMaturity}
	}
\end{figure}

\subsection{Push-Mechanism} \label{def:PushMechanism}
\begin{figure}[!b]
	\centering{
		\includegraphics[width=.95\linewidth,keepaspectratio=true]{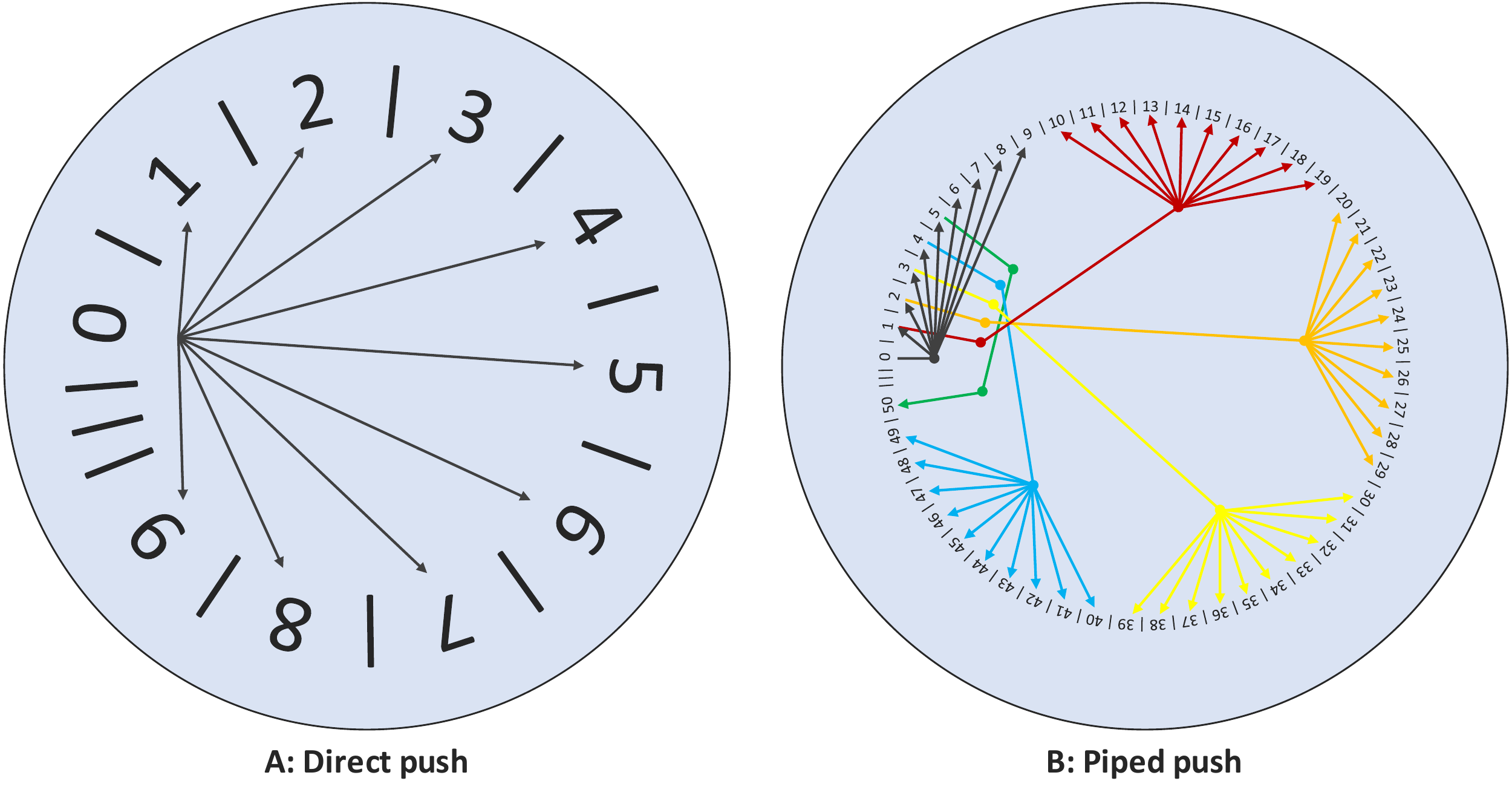}
		\caption{Update Mechanism}
		\label{fig:PeeringPushMechanism}
	}
\end{figure}

Still in a binary world, in contrast to pull-mechanisms, a push based procedure offers several benefits.
Again no literature regarding (full network) push-mechanism could be found.
Nevertheless, the loose nature of \textit{public permissionless} networks (see table \ref{tbl:BlockchainNetworkTypes})
, e.g. Bitcoin, does not fit the subsequently described, custom tailored push-mechanism for \gls{PoT}. \\
In \gls{PoT}, the \gls{LN} only needs to communicate whenever it is willing to do so.
Consequently, the \gls{LN} is not flooded with several update requests as seen the described pull-mechanism.
Depending on the network size ($n$), in small networks the \gls{LN} may send its data to all peers itself (Figure \ref{fig:PeeringPushMechanism}: A) , using a direct push.
In large networks (e.g.: $n>50$), a mechanism may be established to pipe the updates through \gls{DNs} to reduce the workload for the \gls{LN} (Figure \ref{fig:PeeringPushMechanism}: B).
In this regards, this example uses a '$10x$' scale.
Proceeding, instead of sending an update packet to every node,
only the next \textit{nine} nodes (here, indexes: '$0<x<10$') receive a direct update from the \gls{LN}.
The \gls{LN} is its own source and has \textit{nine} \gls{DNs}.
The \gls{DNs} build a partly-master structure and lead a sub net each (Figure \ref{fig:PeeringPushMechanism}: B, \gls{DNs} $1 \to $ nodes $10$ to $20 $, etc.).
An update may consist of any kind of new information (e.g. a new block).
Except the \gls{LN}, each node has one source-node and up to \textit{ten} \gls{DNs}.
The index in this mechanism is counting down towards the \gls{LN}.
The \gls{DNs} ($d$) of each node are calculated from the node's recent index ($x$).
The calculation follows the function:
\begin{center}
	$d = x*10+y$
\end{center}

\begin{figure}
	\centering{
		\includegraphics[width=.95\linewidth,keepaspectratio=true]{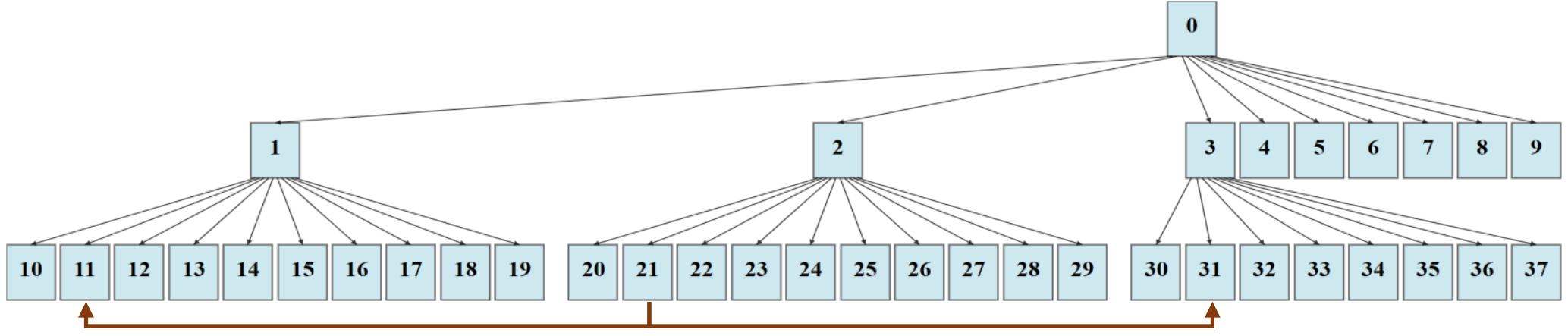}
		\caption{Pseudo layers of peering by pushing}
		\label{fig:PeeringPushLayered}
	}
\end{figure}

As each node has several \gls{DNs}, $y$ defines the number of \gls{DNs} (here: '$0 \leq y \leq 9$').
Hence, the node on index \textit{two} updates	
\begin{center}
	$d = x*10+y = 20+y => 20 \leq d \leq 29$
\end{center}	
the nodes from $20$ to $29$.
Consequently, the node on index \textit{four} updates the nodes $40$ to $49$ and
the node on index $55$ updates the nodes $550$ to $559$. \\
Even if a node has poor connectivity, this source-drain mechanism reduces the network traffic being the \gls{LN} tremendously from $n$ down to \textit{nine}.
A drawback of this \gls{DNs} variation are offline nodes, which have to serve other \gls{DNs}.
If e.g. node \textit{six} is offline, the nodes to be consulted by \textit{six}, $60$ to $69$, and their \gls{DNs} will miss the latest turn's update.
Nevertheless, the round-robin alternation leads to changing update paths with every turn. \\
During the next alternation, the offline node \textit{six} becomes the index \textit{five} and hence, now fails to update nodes $50$ to $59$, whilst $60$ to $69$ are consulted by the new node, which just moved on index \textit{six}.
Again, this stiff and predefined solution is not possible in traditional \gls{BC} solutions.
Changing peers and undefined source nodes prevent this kind of peering in other \gls{CM}s.\\
The so far explained push mechanism can be called vertical-push, as it establishes pseudo layers (Figure \ref{fig:PeeringPushLayered}).
Here the \gls{LN} represents the (\textit{1st}) source layer (Figure \ref{fig:PeeringPushLayered}: Node 0) and
it's \gls{DNs} are contextualized on the \textit{2nd} layer (Figure \ref{fig:PeeringPushLayered}: Nodes \textit{one} to \textit{nine}).
The push mechanism follows a traditional tree structure (Figure \ref{fig:PeeringPushLayered}: Black arrows) and offers the corresponding '$h = log10(x)$' to calculate the number of hops needed to transport an update from the \gls{LN} to any index $x$.
Additionally, further sample enhancements such as \textit{horizontal pipes}, \textit{Jump pipes} and \textit{full-push} are described.
\begin{enumerate}
	\item Horizonal pipes: \\
	A \textit{horizontal pipe} pushes information on the same layer from one node to another, e.g. from node $21$ to the nodes $11$ and $31$ (Figure \ref{fig:PeeringPushLayered}: Brown arrows).
	If the nodes $11$ and $21$ have not already received,
	the designated upper level node(s) may be asked to confirm the horizontal update.
	Alternatively, the neighboring nodes may be asked as well.
	
	\item Jump pipes: \\
	Jump pipes skipping a level may occur, if any source node does not reach a distribution node.
	Then, e.g. node \textit{zero} is not able to reach node \textit{two} and therefore pushes the update directly to the nodes $20$ to $29$.
	
	\item Full push: \\
	As stated before, pushing all updates to the whole network might be expensive for the \gls{LN}, especially in large networks.
	Nevertheless, in contrast to loosely coupled \textit{public permissionless} networks,
	in \gls{PoT} the \gls{LN} is expected to knows all the network's nodes.
	Therefore, some circumstances may benefit a 'full push'-solution (compare \textit{Aura} and \textit{Clique} in \citet{Angelis.2018}), such as transition blocks or any block which demands immediate consultation.
	Especially, the handover-/finalizing block is a crucial block for the \gls{LN}, as it marks the end of all in this turn written data.
	If this block is delivered, it sends all missed data to any receiving node.
	Hence, unattended and dropped behind nodes receive a fresh input.
	Additionally, once the \gls{LN} has finished its turn, it is not occupied anymore.
	Therefore the node is able to distribute the data without performance loss (during its turn) throughout the network.
\end{enumerate}
Further mechanisms to fix \textit{update prevention}, caused by offline nodes, remains implementation dependent.

\bigbreak

\noindent Concluding, the sample \textbf{pull-mechanism} offers to skip updates (e.g. pulling only every second turn) and therefore reduces network traffic throughout.
Obviously, if an increased pulling is not conducted, this comes to the cost of outdated data.
The latter is bad for spontaneous events, such as randomization and trigger events.
The other way around, much consultation in vain is happening as pulls are conducted without receiving new data.
An implementation dependent trade-off has to be found. \\
The key benefit of \textbf{push-mechanisms} is the ability to provide updates fast whilst reducing the traffic for requests.
Especially randomization and trigger events, as far as needed, demand for throughout high availability of every node
and therefore favor push based peering solutions.
Nevertheless, loosing the network is supposed to be one of the major threats for \gls{BCT} systems.
Therefore, pull mechanisms are an essential part of the network, especially if nodes loose connectivity.
Finally, a push-mechanism appears to be superior to a pull-mechanism, nevertheless it was shown
that \gls{PoT} can generally be operated by a pull-mechanism as well. \\
Additionally, the two mechanisms are not mutually exclusive and may be combined as needed.
Thus, a mix of both is generally recommended.
Hence, it is implementation dependent when the network is assumed to be lost e.g. if there was no update from the last three writing nodes.
Here it should be aware that \gls{BCT} solutions assume that nodes have an extrinsic (e.g. \gls{PoW}) or intrinsic (\gls{PoT}) desire to stay online.
Consequently, it is assumed that the peers next to the leading nodes behave in a way offering high availability.
Still, it has to be assumed that mobile clients only offer a reduced bandwidth.
This behavior is unlikely for fat clients, traditionally linked with mining in other \gls{CM}s.
Nevertheless, if nodes are represented by consumer systems, mobile clients are assumed to have an increased online availability
compared to desktop systems, because mobile clients are turned of or cut from the network (flight-mode) only on seldom occasion.
Smartphones, being usually switched on during mobility, strengthens this assumption.
In this regard, if not addressed by lower level peering, mobile clients might change their network address more often.
If this case is applicable, a higher interconnection may be needed.
After all, it is recommended to first design the upper level application to assemble the needed characteristics of \gls{PoT}.
With these characteristics the effective features regarding consultationfull and consultationless interconnection and consequent peering can be decided upon. \\
As the general information flow is supposed to be solved without a superior ledger,
other questions raises regarding \gls{CF}, Byzantine Fault Tolerance and so forth. \\

\FloatBarrier

\section{CAP Theorem measurement}
\label{sec:CAPtheorem}
\begin{figure}[!b]
	\centering{
		\includegraphics[width=.95\linewidth,keepaspectratio=true]{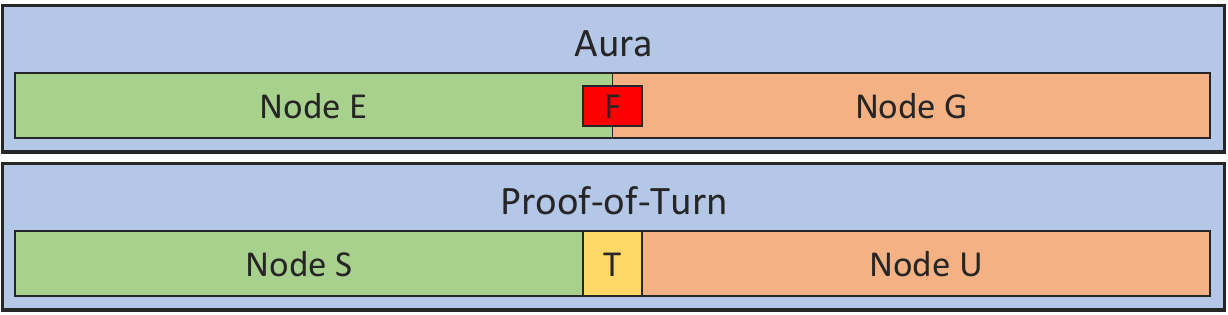}
		\caption{Permission Transition}
		\label{fig:PermissionTransition}
	}
\end{figure}
\noindent The CAP theorem (see: \citet{Brewer.2012}) is one of the most important models to describe distributed database systems.
"The CAP theorem asserts that any networked shared-data system can have only two of three desirable properties.
However, by explicitly \textit{handling partitions}, designers can optimize \textit{consistency} and \textit{availability}, thereby achieving some trade-off of all three" \cite[23]{Brewer.2012}.
\citet{Angelis.2018} compare in their paper two \gls{PoA} implementations, \textit{Aura} and \textit{Clique} with \gls{PBFT} regarding the CAP theorem. \\
They claim that \textbf{consistency} is achieved, when forks are avoided \cite[6]{Angelis.2018}.
Consequently, consistency in the context of \gls{BCT} is represented by instant consensus finality \cite[6]{Angelis.2018}.
\textit{Aura} fails in this matter as glitches might occur during the transition from one \gls{LN} to another, resulting in \textit{no consistency} \cite[7]{Angelis.2018}.
In detail, the time frames to write blocks in \textit{Aura} are shown by \textit{node E} and \textit{node G} (Figure \ref{fig:PermissionTransition}: Aura).
\textit{Node G} is the successor of \textit{node E}.
Due to exact timing, \textit{node E} could propose a block within the time slot \textit{F} (Figure \ref{fig:PermissionTransition}: Aura), wherein any node of the network may already assume \textit{node G} to be the \gls{LN}.
The \textit{Aura} client fails and the conflict is not solved.
\textit{Clique} does not meet \textit{consistency} as well, as several nodes are allowed to write simultaneously.
Solving this issue using the Ethereum GHOST protocol leads to \textit{eventual consistency} \cite[8]{Angelis.2018}.
In this regards, \gls{PoT}'s \textit{consistency} can be considered high because of the before mentioned \textit{transition times} (Figure \ref{fig:PermissionTransition}: Proof-of-Turn, T).
\begin{figure}
	\centering{
		\includegraphics[width=.95\linewidth,keepaspectratio=true]{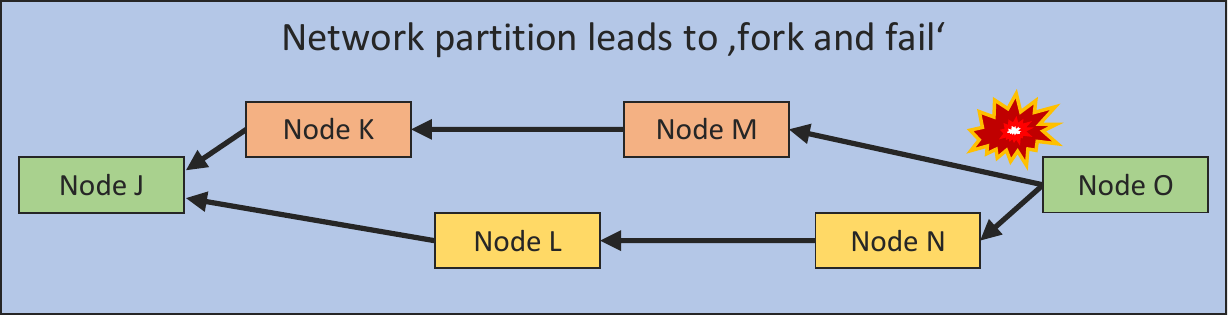}
		\caption{Network Partition in PoT}
		\label{fig:PoTfork}
	}
\end{figure}
First, a \textit{transition time} is only needed if \textit{node S} uses the full turn time.
\textit{Node U} has to assume that the writing permission will not be received before 'turn time plus transition time' is over.
Second, \textit{transition time} prevents the network from assuming different leaders due to \textit{time skews}.
Therefore the duration of each \textit{transition time} has to be chosen large enough to compensate inaccuracies of the client's time protocols.
After all, \gls{PoT} does not need as precise timing as \textit{Aura} or \gls{PoET}. \\
Viewing from another angle, a \gls{BC} "[...] is \textbf{available} if transactions submitted by clients are served and eventually committed, i.e. permanently added to the chain" \cite[7]{Angelis.2018}.
Following \citet{Angelis.2018}'s example of the \gls{PBFT} mechanism, \gls{PBFT} stalls sometimes giving up \textbf{availability} to achieve \textit{consistency} \cite[8]{Angelis.2018}.
The stall occurs, when the network can not agree on a \gls{LN} or a transaction proposed by any node is not added to the \gls{BC} for a long time.
Assuming \gls{PoT} as a 'networked shared-data system' it prevents writing by every client simultaneously, sacrificing \textbf{availability}.
If the \gls{LN}, by chance, has no network connection and is therefore not available, \gls{PoT} stalls (temporarily).
Data from the \gls{LN} can not be written to the \gls{BC} by other nodes, as other nodes can not sign the blocks accurately.
Arguing that the other nodes are not willing/allowed to write would break the 'networked shared-data system'-assumption and makes the CAP Theorem classification obsolete.
It would result in a circulating centralized server. \\
Last, when "[...] a \textbf{network partition} occurs, authorities are divided into disjoint groups in such a way that nodes in different groups cannot communicate each other" \cite[7]{Angelis.2018}.
This scenario is shown in figure \ref{fig:PoTfork}, where the network is split into two groups, orange (Node K \& node M) and yellow (Node L \& node N).
The two groups do not see each other and assume that the other side is 'just absent'.
Now, node O two distinct branches of the \gls{BC}.
If these branches represent different game states, they may be mutally exclusive.
As game states depend on each other, e.g. in \hyperref[def:RftS]{RftS}, node M unknowingly sends a fleet from a planet which has been conquered by node L just before.
The resolution of such forks can only be solved within an upper layer and might demand a reset of the \gls{BC} to any previous block/state.
In the worst case an upper level game is broken and has to be stopped.
The resolution of forks by the game layer remains implementation dependent.
Concluding, regarding the CAP Theorem, \gls{PoT} fails partition tolerance. \\
Along these thoughts, \citet[7-8]{Angelis.2018} categorize the \gls{PoA} implementations \textit{Aura} and \textit{Clique} as well as \gls{PBFT} into the CAP triangle.
\begin{figure}
	\centering{
		\includegraphics[width=.72\linewidth,keepaspectratio=true]{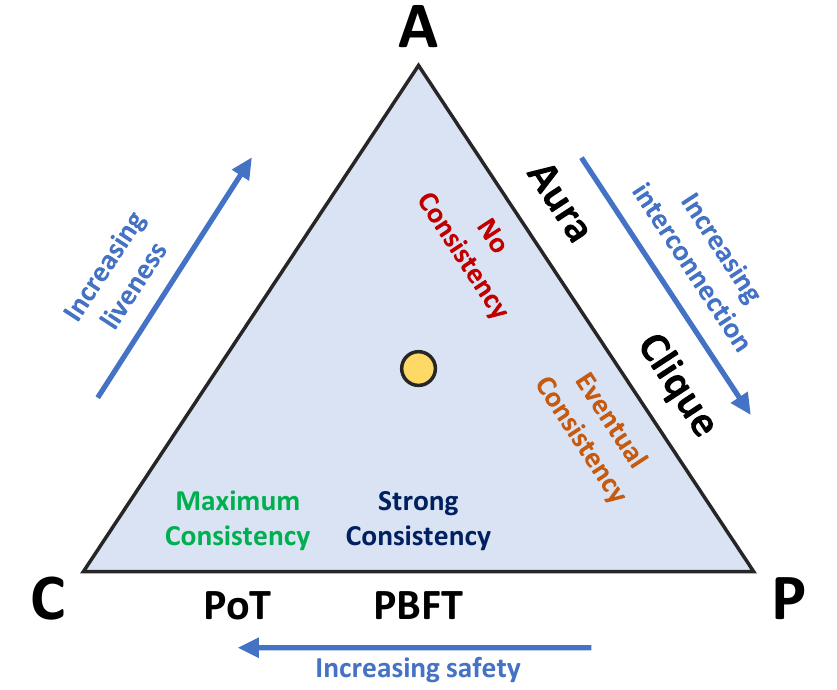}
		\caption{Mechanisms regarding the CAP Theorem (Adapted from \citet{Angelis.2018})}
		\label{fig:CAPTheorem}
	}
\end{figure}
If \gls{PoT} was categorized as a \textit{circulating centralized server}, as all nodes except the \gls{LN} are considered to be silent throughout,
it would be positioned in the middle of the triangle (Figure \ref{fig:CAPTheorem}: Yellow circle).
If \gls{PoT} was categorized with \textit{eventual consistency}, as a pull mechanism does not always provide the newest data,
it would be in the same area as \textit{Clique}.
If \gls{PoT} was categorized as handling \textit{network partition} well, as the upper level use case was only a dumb storage solution without constraints within pushed data,
it would be similar to \gls{PBFT}.
Nevertheless, following the argumentation, \gls{PoT} is added, as shown in figure \ref{fig:CAPTheorem}.

\FloatBarrier

\section{Forks \& transition blocks}
\label{sec:TransitionBlocks}

In any \gls{BCT} certain problems occur, such as forks (\citet[60]{Butijn.2020}; \citet[54]{Dib.2018}).
A fork takes place if there are two mutually exclusive continuations of the approved \gls{BC} \cite[22]{Yuen.2019}.
The network has to decide which trail is going to be followed (\citet{Courtois.2014}; \citet{Ewerhart.2020}) or has to start again from the last legitimate (non exclusive) block \cite[3-4]{FinlowBatesK..2017}.
It has to be mentioned that this section is not about \textit{soft-/hard forks}
on the \gls{BC}'s \gls{CM}'s protocoll \cite[4-5]{FinlowBatesK..2017}. \\
While most Proof-of-Mechanisms solve this issue using the \hyperref[LongestChainRule]{Longest Chain Rule} (\citet{Courtois.2014}; \cite{Nakamoto.2009}),
\gls{PoT} insists on following the rules regarding timing.
The blocks which can only be written by the \gls{LN} have to be pushed within the given time frame of the turn (see figure \ref{fig:PermissionTransition}: PoT).
Hence, the race condition is not dependent on computation power, but on a valid identity regarding time.
Having the turn grants authority to write.
At this point, \gls{PoT} could be classified as a special type of \gls{PoA}.
If a fork is detected, there are basically \textbf{four choices}: \\
\textbf{First}, the branches are soft exclusive, as the upper level software
allows to \textit{merge the branches} without dispute (\citet[7-8]{RanchalPedrosa.2020}; \citet[4-5]{Wang.2018}).
The data from one of the two branches is encapsulated by the recent \gls{LN}, signed and appended on the other branch.
Once merged, data from both branches are \gls{CF} on the main chain and the \gls{LN} continues with its own data.
\textbf{Second}, the branches are \textit{mutually exclusive} .
Nevertheless, encapsulation takes place to regain a consistent main chain.
Next, a dispute has to be raised to decide upon \gls{CF} and invalidation of the questionable parts.
This triggers a network vote on which (on-chain) fork has to be continued.
Herein might be decided on a reset, which resolves in a replay of some turns.
Alternatively, \textbf{third}, a node (or transition block) is \textit{chosen to be continued from},
terminating in a (soft) loss of some turns which will have to be conducted again.
This behavior equals \gls{PoW}'s \textit{Longest Chain Rule} \cite[2]{Ewerhart.2020}
as some transactions/blocks were computed in vain \cite[3-4]{FinlowBatesK..2017}.
\textbf{Fourth}, the worst case for the \gls{LN} originates from a failed vote - now
the \gls{LN} has to \textit{decide on its own} which branch to continue.
The easiest way for the \gls{LN} would be to just choose one branch (randomly, internally).
The drawback is that lacking consultation prevents consent and endangers the \gls{CF} of the \gls{LN}'s data.
Likewise, the decision can be made using a randomization call - which is the most interactive version, raising participation.
The other possibility is to choose the chain which implies the highest loyalty e.g. the most turns.
It is important to mention that the branch offering the most turns is not necessarily the one with the most blocks.
Consequently, the \textit{Longest Chain Rule} is not applicable here.
\begin{figure} [!b]
	\centering{
		\includegraphics[width=.95\linewidth,keepaspectratio=true]{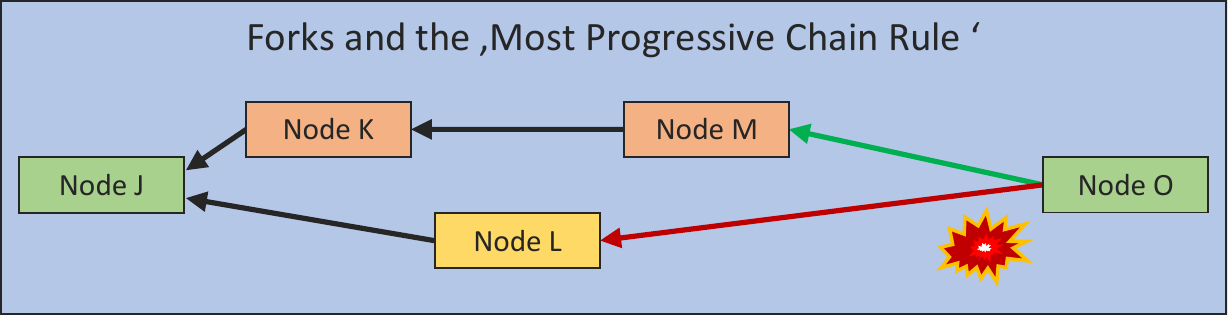}
		\caption{Most Progressive Chain Rule}
		\label{fig:PoTForkProgress}
	}
\end{figure}
Nevertheless, the chain which represents the most turns, is the one which offers the highest resilience against network votes (e.g. regarding invalidation).
The more stakeholders can be assumed for a branch, the higher the possibility for a positive network vote.
Concluding, the chain, continued by the \gls{LN}, is the one representing the most progress regarding turns (Figure \ref{fig:PoTForkProgress}).
After all, to give this scheme a name, \gls{PoT} uses a '\textit{Most Progressive Chain Rule}' instead of the \textit{Longest Chain Rule}. \\
Here, it is likely that a loyalty based approach is superior to a randomization based approach and randomization is only chosen, if the forks offer the same loyalty factor.
The node could also use the values from the failed network vote to influence its decision (slightly). \\
Last, here is a \textbf{fifth} negligible solution:
As creating new blocks in the \gls{PoT} \gls{CM} is cheap, a node could follow several branches.
This solution creates forks even if none has occurred in the first place.
Each node would fork from the last some transition blocks, resulting in turns which have to be calculated several times with changing preconditions.
Therefore the latter option is not considered as an option in the first place.
The effective arrangement is implementation dependent. \\
Of course, the validity of a transaction is checked by every other node.
If a block can neither be checked or rejected, as for \hyperref[HiddenTransactions]{hidden transactions}, it is accepted.
Still it does not change the game state, yet.
If one node does not want to accept an open block, voting determines the validity of the written block - \textit{silence infers consent}.
Nevertheless, the probability of a block to be accepted raises with every silent consent or approved check and is finally accepted if it is approved by a defined threshold of nodes.
Of course, silence could infer refusal as well, but first \gls{PoT} assumes online peers and second, if too many peers were offline, the network would stall constantly.
Hence a \textit{silence infers refusal}-approach is strongly discouraged from.
The threshold, for a vote to pass, can be defined as 'a few nodes' or 'the whole network'.
In table \ref{tbl:SumConsensusMechanisms_3} this was marked as \textit{implementation dependent}. \\
The chosen value affects the \gls{BFT}, the \gls{CF} as well as the smoothness of the \gls{BC} as shown in table \ref{tbl:ConsentSmoothness}.
\begin{table}
	\centering
	\begin{tabularx}{0.71\textwidth}{ c | c | c | c }
		\textbf{Consent} & \textbf{Round Robin} & \textbf{\gls{BFT}} & \textbf{Effective} \\
		\textbf{threshold} & \textbf{smoothness} &  & \textbf{\gls{CF}} \\ \hline
 		Low ($\leq 33.3$ \%) & High & Low & Fast \\ \hline
		Medium ($\sim 50$ \%)  & Medium & Medium & Medium \\ \hline
		High ($\geq 66.6$ \%) & Low & High & Slow \\ \hline
		\hline
	\end{tabularx}
	\caption{Consent smoothness}
	\label{tbl:ConsentSmoothness}
\end{table}
\noindent Luckily a high consent threshold does not lead to a freeze,
such as in \hyperref[def:MultiChain]{MultiChain} \gls{BC}s because of the \textit{silence infers consent}-principle.
Even if there is no interaction, the Round Robin procedure proceeds.
Consequently, latest after one full round a block reaches effective-\gls{CF}.
By then it can not be invalidated anymore.
The \gls{CF} is therefore both faster approached and can be realized with a higher threshold than e.g. \gls{PoW},
as \gls{PoW} suffers from a small residue probability to be forked by any computation superpower \cite[4]{Demi.2021}. \\
Blocks which are not supposed to be in discussion are transition blocks, those blocks which pass the turn from one \gls{LN} to the next.
These transition blocks may enable \textit{adaptive turn time} (See section \hyperref[sec:FurtherCharacteristics]{Further Characteristics}) and ensure consistency (figure \ref{fig:PermissionTransition}: PoT).
Therefore, transition blocks can be used for \textit{fast update verification}.
If a node stays offline for longer, it may ask for an update providing the last known block.
But due to forks or \hyperref[sec:DataAllocationImprovements]{Data allocation improvements} that block is not existing anymore.
Instead the requesting node could provide its last transition blocks, which is not allowed to be deleted.
The last known transition block from the requesting node will then become the start block of the delivered update.

\FloatBarrier

\section{Reveal claims \& trigger events}
\label{sec:ClaimsAndTriggers}

\begin{figure}
	\centering{
		\includegraphics[width=.995\linewidth,keepaspectratio=true]{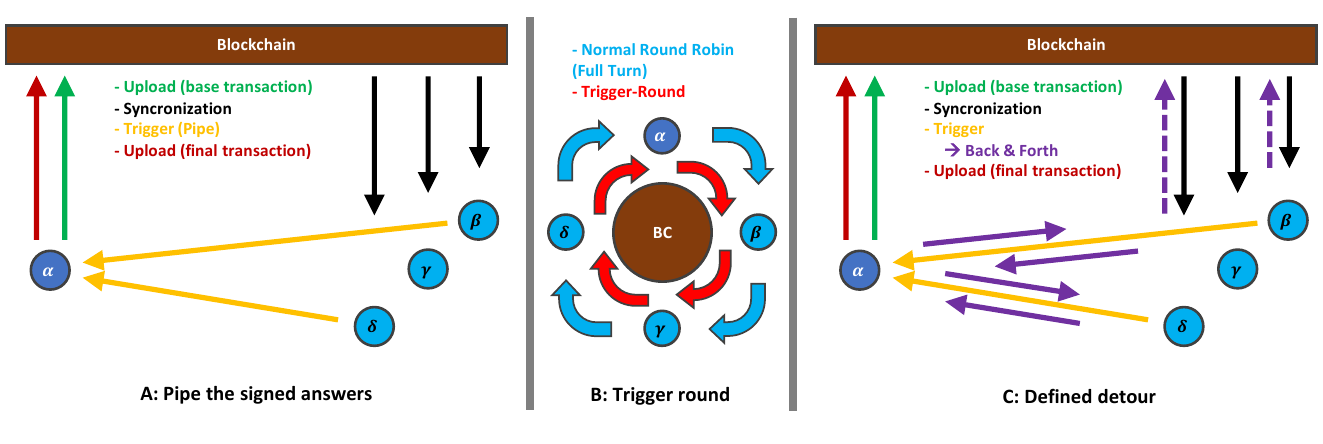}
		\caption{PoT: Trigger event mechanisms}
		\label{fig:TriggerEvents}
	}
\end{figure}

These spontaneous transactions were already mentioned at the end of chapter \hyperref[chap:BlockchainInGames]{Blockchain in Games}.
Although it was stated that general mechanism can be conducted using \gls{BCT}, certain examples were left out.
Therefore, three different approaches are presented hereafter (Figure \ref{fig:TriggerEvents}).
Following it is assumed that the turn is on node \textbf{$\alpha$}. \\
Claims and triggers are active transactions conducted by an intervening node (e.g. \textbf{$\beta$}), breaking the regular writing window, to change the outcome of player \textbf{$\alpha$}'s (sub-)move.
Due to the generally asynchronous nature of \hyperref[def:RftS]{RftS} this is not recommended, but still possible within the \hyperref[def:PoT]{PoT} mechanism.
Possible solutions are given in figure \ref{fig:TriggerEvents} (A-C) and range from 'piping a signed block' (A)
over conducting a 'trigger round' (B), up to 'defined detours' (C).
In \textit{case A} signed blocks are piped via node \textbf{$\alpha$} to the \gls{BC}.
\textit{Case B} describes a fast paced network round wherein only triggers can be pushed legally.
In \textit{case C} the writing permission is only actively given to claiming nodes, who write the triggers themselves before handing the writing permission back to node \textbf{$\alpha$}.
\noindent Whilst reveal claims are straight forward and any answer has to be given by the targeting node,
moves which \textbf{might} trigger, offer certain challenges regarding figure \ref{fig:TriggerEvents}:
\begin{enumerate}
	\item A trigger can only be performed by $\beta, \gamma$ and $\delta$ if both, \textbf{possibility} (any kind of resource) and \textbf{will} (to interfere) are given.
	
	\item In case A (figure \ref{fig:TriggerEvents}) nodes $\beta, \gamma$ and $\delta$ have to believe that node \textbf{$\alpha$} received the piped transaction in time and is not actively ignored.
	The same applies to case C. \\
	As the \gls{LN} is generally allowed to ignore other nodes claims, interoperability may fix this issue (Section: \hyperref[sec:Interoperability]{Interoperability}).
	
	\item Depending on the rules, triggers are not performed in line, players wait for others to execute triggers before they use their own resources.
	Hence, triggers are 'not in line'.
	For case B (figure \ref{fig:TriggerEvents}) this would lead to several rounds of consultation.
	Consequently, if not automatically performed, the manual interaction will slow down the game and/or annoy players due to repeated consultation.
	
	\item Additionally, depending on the numbers of players, case B lacks especially from a long freeze-like waiting time
	until all nodes have performed their answer regarding willingness to fulfill a trigger call.
	However, answering automatically, if the resource to pull a trigger is not available, could unwillingly offer vulnerabilities (e.g.: This node is not able to perform any defensive action).	
\end{enumerate}
Bullet point two could be handled by a child-chain storing raised triggers by $\beta, \gamma$ and $\delta$ including their timestamp.
Due to the random character of calls to pull a trigger, this chain would need to rotate fast with PoT or has to use another type of consensus mechanism, e.g. out of the \hyperref[sec:PbC]{proof based} \gls{CM}s.

\FloatBarrier

\section{Interoperability}
\label{sec:Interoperability}

\begin{figure}
	\centering{
		\includegraphics[width=.95\linewidth,keepaspectratio=true]{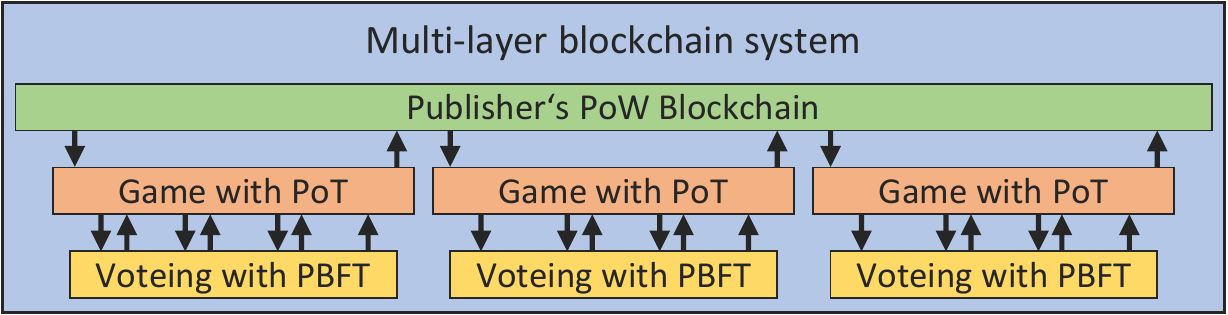}
		\caption{Multi layer BC system}
		\label{fig:MultiLayerBC}
	}
\end{figure}

Except from \citet{Besancon.2019} and \citet{Kraft.2016} there were no papers to be found regarding interoperability of \gls{BCT} for games.
Nevertheless, some implications are given:
Those in the previous chapter mentioned side-chains refer to the before mentioned on-chain, off-chain, child-chain, side-chain and inter-chain methods (\citet{Kim.2018}), see section \hyperref[sec:PerformanceImprovements]{Performance Improvements}.
These methods shall now be inspected for internal and external use regarding interoperability. \\
If not mentioned explicitly, there is no special use case.
Primarily it is focused on multiple layers next to the main-chain. \\
\textbf{First}, internally: If a game wants to augment direct immediate responses in an asynchronous network, preemptive actions have to be taken.
This would include some spare encryption keys or other conducted computation which include network consultation.
It has to be aware that allowing such measures might break the integrity of a game.
For example, computing a randomization in advance lets the player know the chances before actively deciding to take the randomly impacted decision.
Nevertheless, those pre-computed blocks which need consultation, but are not secure to be conducted, shall not be written to the main-chain - just as bloat blocks.
Not conducted turns can be invalidated with a revocation or left open (unrevealed) - just as bloat blocks.
Only those, which are finally used are written to the main chain.
Once all blocks in the side-chain are revoked or written to the main-chain, the side-chain can be deleted.
Moreover, an effective implementation of \gls{PoT} has to decide whether transactions stored in the main-chain only have to be \gls{CF} or at least effectively \gls{CF} (use case dependent).
The latter would call for a child-chain with \gls{CF} blocks/transactions, which is referred to by the main-chain.
The \gls{CF} blocks/transactions in the child-chain are referenced in the main-chain earliest after the defined \gls{BFT} threshold has been passed ($=>$ effective \gls{CF}).
As revoking of \gls{CF} blocks/transactions is supposed to be seldom, the transcription/reference can be done by the emitting node, after the round has passed.
If used, these mechanisms lead to an internal two-layer structure of \gls{PoT}.\\
\textbf{Second}, externally:
This part focuses on \gls{PoT} being an off-chain or side-chain to any other \gls{CM} or \gls{PoT} using any other \gls{CM}s as off-chain or side-chain solutions.
Hence it has to be distinguished between \gls{PoT} being below or above the other \gls{CM}.
For an improved understanding, figure \ref{fig:MultiLayerBC} shows \gls{PoT} in both situations - on
the one hand being a transaction channel child-chain for a \gls{PoW} parent-chain and on
the other hand being a parent-chain for a child chain with \gls{PBFT} functionality (e.g. for voting or trigger events etc.). \\
The below part was already shown by \cite{Kraft.2016}, who uses \gls{PoP} in Huntercoin for game-channels outside a main-chain using another proof-based \gls{CM}.
The other way around would be to fix the  issue of availability shown in by the CAP theorem.
Here, any child-chain offering a simultaneous proposing of blocks would be suitable.
Consequently, dispute resolution by votes and other messages which are of simultaneous nature may be covered with those child-chains, whilst the actual game play is maintained by \gls{PoT}.
A child-chain for votes and disputes could prevent \gls{LN}s from ignoring unwanted messages and enable simple trigger-event resolution
without circumlocutory logic (Section \hyperref[sec:ClaimsAndTriggers]{Reveal claims \& trigger events}).
Last, if needed, a publishers server may be established as a part of any of these chains - the publisher's incentive to do so is consciously left open.

\FloatBarrier

\section{Peer-Fluctuation \& adaptive turn time}
\label{sec:PeerFluctuation}

During the existence of a \gls{BC}, there are multiple reasons for nodes to leave or join the \gls{BC}'s network.
Whilst joining is generally considered an active move, leaving can both be an active operation as well as a passive coincidence. \\
\textbf{Join operations} can be maintained as any node within the network proposes a new node.
Either the network decides whether the proposed node will be added, or a trusted entity (e.g. a publisher's server) may add a node to the network ($n \to n+1$).
The network's (/game's) state predetermines where the node is added regarding the round robin state. 
The new node's index is chosen by the upper level software as it may be inserted at the end of the round or within the next some turns.
In a self organized network using the pull mechanism, without a publisher's server, it may take long to add a new node if no trigger event mechanism is used.
Nevertheless, to prevent two nodes from assuming to be the \gls{LN}, it is discouraged from replacing the recent \gls{LN}  as well as inserting the new node as a successor of the recent \gls{LN}. \\
\textbf{Leave operations} can be conducted by publishing all data the network needs to keep the upper level software running.
This is especially needed in a game's card draw-scenario and likewise situations.
The leaving node is cut out of the round robin and each round progresses faster ($n \to n-1$).
It has to be determined whether a node is allowed to rejoin.
Here, if private keys are shared during the leave operation, a rejoin is only possible using a new identity.
Both, join and leave operations are active operations. \\
Dropping out by (passive) \textbf{coincidence} is possible as well.
For nodes leaving without a trace like software/hardware errors or any longer network connection outage,
the upper level software has to be designed in a resilient way to cover such coincidence (e.g. shared keys for shuffling cards, figure \ref{fig:ShuffleCards}).
If it fails to cover these cases, the \gls{BC} could be hold hostage by any absent node.
As the Round-Robin procedure continues, there have to be mechanisms which lead both, the network and the node to regain the connection.
First, if all nodes are considered fluid parts of a larger network, a static server may help to keep trace as a lost node may ask the static server upon the other nodes of the network.
Alternatively, second, if the lost node is the \gls{LN}, the \gls{BC}'s network could sacrifice the static amount of time of the \gls{LN}'s turn in favor of a likely re-connection.
This action can be seen as the counterpart of 'ending a turn prior to its maturity'.
Both actions are seen as measures of \textbf{adaptive turn time}.\label{sec:AdaptiveTurntime}
The granted amount of time (from minutes to days) as well as punishments after the lost node's re-connection remains implementation dependent.
To find a suitable measure \citet[27]{Laneve.2019}'s claim of \gls{BCT} to be a "risky choice for developers" comes back to mind.
In a gaming context, the other players would have to wait longer during the prolongation.
Still, for in-game experience it is important to reduce waiting times as waiting "[...] is frustrating, demoralizing, agonizing, aggravating, annoying, time consuming and incredibly expensive." \citet{Buffa.1976} as cited in \cite[1]{Maister.1984}.
Consequently chopping down waiting time is considered a key feature.
To solve this deadlock, \gls{BCT} in general seems inappropriate. \\
Nevertheless, following the idea of adaptive turn time, measures such as a \textit{night switch} or a \textit{vacation break} could be implemented as well.
For Pause-by-Consensus votes, some players might not be able to answer manually, as they do not suppose to have their turn.
This is especially true for long termed games with multiple players.
Therefore opt-in, opt-out or predefined answers are recommended.
It has to be aware that a malicious node could pause the game repeatedly if too many nodes give automated positive answers for each prolongation.

\FloatBarrier

\section{Further Characteristics}
\label{sec:FurtherCharacteristics}
This chapter covers further mechanics which can/have to be used to enable certain game characteristics.

\begin{enumerate}
	\item \textbf{Helper Nodes} \label{sec:HelperNodes} \\
	Although most game mechanics can be guaranteed, using \gls{BCT} only, some mechanisms need intervention of external resources/nodes. \\
	In a basic scenario there are no such mechanisms which need strict timing to circumvent race conditions as transition times (figure \ref{fig:PermissionTransition}: Proof-of-Turn) prevent race conditions from delayed network messages.
	However, if transition times are diminished or left out to speed up, additional external resources (figure \ref{fig:MultiLayerBC}: Upper \gls{BC}) may be needed to solve race conditions.
	Still, each additional consultation results in a slower algorithm.
	It is encouraged to just set the transition time appropriately. \\
	Nevertheless, such helping nodes shortcut cumbersome solutions such as randomization and (hidden) card draw.
	A helping node might even be a publishers (central) server.
	Consequently, the publishers server does not host games entirely, but keeps being available for certain services.
	Especially for randomization this may help to meet tight time frames instead of waiting for slow (inactive) peers.
	In this case 'tight' means $\leq5$ seconds to maintain a reasonable user experience. \\	
	Additionally, if no lower level chain (figure \ref{fig:MultiLayerBC}: Lower \gls{BC}) for intervening messages is provided, a central server may be used as a referee in any dispute.
	Any \gls{LN} would therefore not be able to deny transactions/messages with for the \gls{LN} disagreeable content.
	However, this is not seen as a suitable solution, a structure as shown in figure \ref{fig:MultiLayerBC}) should be sufficient for this case and is therefore recommended.
	
	\item \textbf{Shared node resources and cooperative gaming} \\
	The distributed nature of \gls{PoT} enables each player to use several devices to act as one node (\citet{Harkanson.2020}).
	if multiple participants offer the same authorization and authentication parameters, they can be seen as one node (\citet{Karafiloski.2017}).
	Although this feature could be enabled, it has to be considered how the clients are supposed to behave.
	Here, contradicting data has to be avoided as, only one of the devices shall behave as a \gls{LN}.
	If its the player's turn, consequently only the used device shall be consulted to pull from.
	Nevertheless, \gls{BCT} enables cooperative gaming of
	several persons as one player in the game (\citet{Harkanson.2020}).
	The players have to be aware that once granted access rights (private keys) can not be reversed.
	Nevertheless, this feature could enable to play cross platform with many devices (e.g. Computer, Smartphone and Tablet).
	
	\item \textbf{Game session interaction} \label{sec:GameSessionInteraction} \\
	Imagining parallel games of \hyperref[def:RftS]{RftS} with likewise parameters this could allow (given certain parameters) to send a fleet from game A to another game B.
	Here, following figure \ref{fig:MultiLayerBC}, both games might be initiated on the upper level and played on distinct \gls{PoT} \gls{BC}s.
	The sending transaction would include to join the other game's network and become a node in both round robin circles.
	The games do not necessarily need to be in sync and the newly added node (/player) becomes a node following the peer-fluctuation rules of the specific implementation.
	Here, an upper level \gls{BC} as seen in figure \ref{fig:MultiLayerBC} or a helper node for the transition is believed to be necessary.
		
	\item \textbf{Reference Implementation} \label{sec:Reference Implementation} \\
	There are manifold implementation dependent characteristics for a network based on \gls{PoT}.
	\gls{PoT} has to fit the purpose of the business case and has to be set up accordingly.
	The described business case is not part of this document and hence a reference implementation does not fall into place.
	Therefore it is desisted from conducting a reference implementation.
	Nevertheless, table \ref{tbl:PoTparameters} lists parameters to be considered:
	
	\begin{table}[!b]
		\centering
		\begin{tabularx}{0.69\textwidth}{ l | c }
			\textbf{Parameter} & \textbf{Value} \\ \hline
			Nodes & Open ($1$ to $\infty$) \\ \hline
			Turn time & Open ($10$ seconds to several years) \\ \hline
			Maximum block size & $1$ to $\infty$ transactions \\ \hline
			Transition times & 1 to 10 seconds \\ \hline
			Finalizing blocks & Boolean \\ \hline
			Handover blocks & Boolean \\ \hline
			Peering & Push- \textbf{vs.} Pull- \textbf{vs.} Mixed-mechanism \\ \hline
			External resource & None \textbf{vs.} Server \textbf{vs.} Upper level \gls{CM} \\ \hline
			Sub-Chain & Boolean (+ chosen \gls{CM}) \\ \hline
			Votes (/\& Disuptes) & Boolean (+ define rules) \\ \hline
			Randomization & Boolean (+ define rules) \\ \hline
			Adding nodes & External resource \textbf{vs.} Self maintained \\ \hline
			New node's index & Start, Next, End, etc. \\ \hline
			Adaptive turn time & Boolean (+ define rules) \\ \hline
			Multiple devices & Boolean (+ define rules) \\ \hline
			\hline
		\end{tabularx}
		\caption{Parameters of \gls{PoT}}
		\label{tbl:PoTparameters}
	\end{table}

\end{enumerate}

\FloatBarrier

\section{Attack Vectors}
\label{sec:AttackVectors}

Now, some general attacks on \gls{BCT} are described.
Consequently, attacks against popular \gls{CM}s are described and the applicability on \gls{PoT} is given.
Additionally, some attacks special to \gls{PoT} are described as well. \\
An attack shall be considered any hostile behavior which might impact negatively on single nodes or the network itself.
It has to be distinguished between \textit{inside} and \textit{outside} attacks.
Here it is asked whether the saboteur is part of the network or not.
Moreover it has to be distinguished between tricking the upper level software and destroying the \gls{BC}'s network.
\bigbreak
\begin{enumerate}	
	\item \textbf{Byzantine Fault Tolerance} \\
	As this is one of the most discussed topics (\cite{Dib.2018}; \cite{Khan.2020}; \cite{Gramoli.2017})
	on the \gls{PoW} approach, it has to be mentioned here.
	As \gls{PoW} is designed to be used in a public permissionless (Table \ref{tbl:BlockchainNetworkTypes}) environment 
	wherein all nodes are able to enter and leave freely \cite[8]{Nakamoto.2009},
	an attacker with computational superpower (likewise the majority of CPU power)
	could take over the network (\citet[8]{Nakamoto.2009}; \citet[4]{Demi.2021}). \\	
	The nature of \gls{PoT} is completely different here, first \gls{PoT} is supposed to be used in private/public hybrid/permissioned networks.
	Stil,l proposing new nodes is possible.
	Second, instead of CPU power, the number of controlled nodes is the key for fraud in \gls{PoT}.
	The threshold which marks the \gls{BFT} for \gls{PoT} as described in section 'Consensus finality and Byzantine Fault Tolerance' marks the number of nodes needed to become the networks superpower.
	Once archived to control a share of network nodes above the threshold, lets the attacker vote (and kick) in any wished direction.
	The network has to maintain that such majorities are prevented.
	Still, a \gls{BC} network 'dies' when one all nodes have left.
	Any added node is a threat to the voting power of the other nodes.
	Hence, the upper level software has to define rules for joining nodes.
	
	\item \textbf{Benign faults} \\
	Next to the \gls{BFT}, benign faults are a source of inequality and injustice.
	Here "[...] leader misbehaviours can be caused by benign faults (e.g., network asynchrony, software crash) [...]" \cite[4]{Angelis.2018}.
	Although these benign faults are no attack in the traditional sense (\citet{CharronBost.2009}), they endanger the upper level software to stall (\citet[8]{Angelis.2018}; \citet[7]{Greenspan.2015}).
	As benign faults can barely be distinguished from e.g. 'unavailability on purpose', each implementation has to draw a line between the two cases.
	
	\item \textbf{Nothing at stake, long range and transparent forging} \\
	\cite{Lin.2017} describe two flaws of \gls{PoS}, the \textit{Nothing at stake} attack as well as the \textit{long range} attack.
	Both attacks are applied from inside the network.
	The \textit{Nothing at stake} attack is incentivized by cryptocurrency tokens from proposing a block even if there is no data to be written \cite[301-302]{Lin.2017}.
	\gls{PoT} does not suffer from this as block generation does not offer any cryptocurrency token rewards. \\
	The \textit{long range attack} derives from the \textit{Longest Chain Rule} and an attacker
	who once had a high amount of computing power being able to rewrite the chain later on \cite[302-303]{Lin.2017}.
	\gls{PoT} is resilient against this attack as well, as the data written by a node can only reflect the own turn's transactions.
	As long as the integrity of other nodes is not broken (e.g. published private keys), there is no way for an attacker to augment their moves.
	It has to be mentioned that \gls{PoT} does not rely on a \textit{Longest Chain Rule}, but rather on a \textit{Most Progressive Chain Rule}, which is an adaption of the \textit{Longest Chain Rule}. \\
	Last, some \gls{CM}s such as \gls{PoS} allow to propose the next \gls{LN}.
	If this is possible, a subgroup of nodes is able to pass the \gls{LN} within their peers, which is called \textit{transparent forging} \cite[302]{Lin.2017}.
	Due to the Round Robin procedure of \gls{PoT}, \textit{transparent forging} is not applicable.
	
	\item \textbf{Integrity of nodes} \\
	As details of encryption are seen out of scope, security of symmetric and asymmetric encryption are supposed.
	Still, the loss of the private encryption key(s) is critical to the integrity of any node.
	Once an attacker obtains access to the private key, the nodes identity can be augmented and used in hostile manner.
	Except the loss of the private encryption key(s), \gls{PoT} seems secure on outside attacks regarding integrity.
	Likewise, \citet[251]{Douceur.2002} describes the 'Sybil Attack' wherein large-scale "[...] peer-to-peer systems face security threats from faulty or hostile remote computing elements".
	Here, if "[...] a single faulty entity can present multiple identities, it undermining this can control a substantial fraction of the redundancy." \cite[251]{Douceur.2002}
	Hence, the integrity has to be kept throughout.
	Additionally, \citet{Douceur.2002}'s claim reflects the risk of \gls{BFT} being captured by an entity,
	which represents the majority in a vote through multiple controlled network nodes.
	
	\item \textbf{Timing} \\
	Revisiting the \gls{PoA} client \textit{Aura} from section \hyperref[sec:CFandBFT]{Consensus finality and Byzantine Fault Tolerance},
	malicious \gls{LN}s could try to post their blocks as near to the end of their writing permission as possible to break the algorithm.
	Of course this is considered an inside attack with the implication to break the network.
	This flaw is prevented in \gls{PoT} if \textit{transition times} are enabled.
	
	\item \textbf{Silencing nodes} \\
	The prevention of service of any node of the network can be accomplish by any (larger) entity.
	The idea can be abstracted from a single attacking device accomplishing \textit{Denial of Service}, therefore the term \gls{DDoS} is used.
	\gls{DDoS} can be both of internal and external nature (see \citet{Wu.2010}). \\
	\citet[72]{Johnson.2014} take a look at Bitcoin (\gls{PoW}) mining pools and see both potential and incentive to use \gls{DDoS} to gain competitive advantage over other mining pools.
	\gls{PoW} does not assume any node to provide a block, hence  the attack only harms the single node without hurting the remaining network.
	In \gls{PoT} the targeted node has to be distinguished between the \gls{LN} and any other node. \\
	Assuming that there is no data the network needs from the targeted node to continue, the \gls{DDoS} attack is only harms the state of the single node (update prevention).
	This is applicable if the targeted is not the \gls{LN} and does not have to deliver data (e.g. trigger/card draw/randomization). \\
	The other way around, if the targeted node is the \gls{LN}, the network is targeted entirely, as the progression stalls until the \gls{LN}'s turn is over.
	Still, information which has to be provided to the network by the \gls{LN}, can not be served. \\
	It can not be avoided, that the distributed and interconnected structure of \gls{BCT} has to expose the needed network address information, which is needed to conduct the attack.
	The general solution would be to guarded nodes from each other by a central server, abstracting the network addresses, but this breaks the \gls{BCT}'s paradigm.
	Moreover, the \gls{DDoS} could follow the round robin sequence and target one node after another, silencing the (whole) network.
	Nevertheless, if detected and once the attack is over, the round robin could be reset and continued from the last agreed state.
	
	\item \textbf{Double spending problem} \\
	One of the key problems \gls{PoW} solved is the ability to spend assets/tokens in distributed systems twice \cite[1]{Nakamoto.2009}.
	\gls{PoT} is able to guarantee this as well.
	If a node does not comly given \gls{SC}s, the block/transaction is invalidated by vote and does not reach \textit{effective} \gls{CF}.
	Consequently, the published information which uses \gls{PoT} has to deliver transparent data with every transaction.
	The double spending problem arises when hidden transactions become part of the upper level software.
	In this concern, \citet[56]{Dib.2018} state that "[...] it is always possible to add encrypted data using the recipients public key,
	but then the validators cannot verify the semantics of the transaction (e.g., double spending) [...]".
	Therefore, specific (upper level) implementation has to solve related issues.
	
	\item \textbf{Effective network split} \\
	This refers to a node, sending different data to different parts of the network, resulting in a forced fork.
	This could either happen, \textbf{first}, due to lost identity or team gaming with consequent simultaneous gaming or \textbf{second} hostile emitted blocks by a single node.
	Now "[...] some of the nodes of the blockchain recognize a node as the next block writer,
	while other nodes recognize another node as	the next block writer" \cite[22-23]{Yuen.2019}.
	All three scenarios can be detected by the network as long as the network rechecks the last (transition) block of each \gls{LN} to deliver equal fingerprints (Hashes) on each receiving node.
	Possible resolutions are described in section \hyperref[sec:TransitionBlocks]{Forks \& transition blocks}.
	Hence, consistency can be maintained by preventing network partition.
	
	\item \textbf{Data flooding} \\
	Assuming \gls{PoT} to be a lightweight algorithm for mobile games (e.g. on smartphones) and offering the possibility to write bloat blocks,
	enables to flood the network until some nodes have to drop out of the game due to allocated storage constraints.
	The mechanisms like pruning bloat blocks or meta-state blocks, wiping the history, are countermeasures but can not be seen as all-embracing solution.
	Depending on the final use case, a maximum write permission may be needed as well.
	It has to be mentioned that a limit (rule) may be broken if the network decides it in a vote (changes the rule).
	Nevertheless, single nodes may clean for themselves to keep the 'important' parts of the \gls{BC}.
	Thresholds for both the limit and the vote remain implementation dependent.
			
\end{enumerate}

\noindent Further, more technical, attacks on \gls{BCT} can be found in \cite{Min.2019b}.
As \citet{Min.2019b}'s attacks can not focus on \gls{PoT} yet, they are out of scope.

\section{Limitations}
\label{sec:Limitations}

First it has to be mentioned that this document is (only) based on theoretical background.
Many assumptions had to be made, as seen in table \ref{tbl:PoTparameters}.
Some of the assumptions, as a reminder, are given in the following:
\begin{enumerate}
	\item Many \gls{BCT} \gls{CM}s, due to their extrinsic incentives of cryptocurrency tokens, prevent offline peers.
	A software publisher has to decide whether intrinsic incentives are satisfactory to provide a network with \gls{PoT}.
	Taking the targeted platform with their consumers into consideration may lead to a decision. \\
	Here, strong tiers alike a company's servers, personal computers and gaming consoles tend to be switched off after a gaming session and might not stay online over a longer period of time.
	Consequently extrinsic incentives are needed. \\
	On the contrary mobile phones, likewise to WhatsApp usage, can be supposed stay online most of the time.
	Hence they do not need to be switched on on purpose.
	Nevertheless, offline nodes, stemming from flight mode during night times might stay a problem for \gls{PoT}.
	
	\item Especially if choosing mobile clients, disk size limit has to be evaluated critically.
	Although new devices may offer excessive resources,
	if the targeted consumers are throughout the whole range of devices, average free resources may drop dramatically.
	
	\item Although \gls{PoT} is considered a lightweight solution among other \gls{CM}s, it is still no solution for 'fast paced' upper level software.
	Transition times, needed to provide throughout consistency may be too long for immediate response.	
	This is especially true for large groups of nodes aiming for simultaneous interaction.
	\gls{PoT} is primarily designed for (long term) asynchronous data transfer.
	
	\item \gls{PoT} is not supposed and not able to reduce publishers fear of cloned replicas from open code due to \gls{SC}s or unencrypted shared data for re-engineering.
	The incentive to save maintenance costs on servers have to suffice to cover re-engineering risks.
	Additionally, not updated clients, assuming different parameters, may lead to incompatibilities.
	Changing rules (\gls{SC}s or parameters) after a \gls{PoT} \gls{BC} has been initialized is not recommended.
	Concluding: Once a software using \gls{PoT} is released, the advantage from \gls{PoT} are only the reduced network/server costs.
	
	\item Less about \gls{PoT} and more about games on \gls{BCT}, distributed systems without fallback option are critical for long term usage.
	This can be obtained both from precise timestamps without helper nodes and shared decks of card.
	For the latter, private draws without helping nodes are already cumbersome.
	Hidden draws are not even possible without external help.
	
\end{enumerate}

\section{Interim Summary - Proof-of-Turn}
\label{sec:InterimSummary-PoT}
The implications of \gls{PoT} are promising.
Chosen the right application's use case, \gls{PoT} offers an intelligible framework for fast lightweight and fair distributed gaming experience.
It outperforms \gls{PoW} in terms of speed, whilst still managing to grant equal rights among the peers in comparison to \gls{PoS} and \gls{PoA}.
There is no need for special Hardware as required for \gls{PoET} and no semi-fair additional measure for 'effort' has to be determined as in \gls{PoP}.
Compared to the approaches \textit{Multichain} and \textit{Aura}, freeze-states can be prevented due to granted time frames for turns and the transition time slot between two turns. \\
Still, \gls{PoT} meets a high \gls{BFT}, adjustable to each use case needs,
whilst offering fast \gls{CF} reducing forks and computation in vain.
Last, \gls{PoT} can be seen as open for interoperability as it can be used as a sub-chain as well as it can use other chains below itself, likewise to \citet{Kraft.2016}'s \textit{Game channels}.
Summing up, \gls{PoT} delivers many desirable features for its intended niche. \\
Therefore, after researching and evaluating different \gls{CM}s, the second leading question ..
\begin{center}
	\textit{"What is the best fitting \gls{BCT} \gls{CM} to cover an asynchronous game play scenario?"}
\end{center}
\noindent .. may be answered with "\gls{PoT}" - depending on the application's use case.
Additionally, by shifting the computation requirement from the publisher's central server
to the peers nodes, the research hypothesis .. 
\begin{center}
	\textit{"The \gls{PoT} \gls{CM} leads to a reduction of server running costs for game publishers."}
\end{center}
\noindent .. appears to be admissible. \\

\noindent \textbf{Finally}, after \textbf{first} analyzing recent \gls{BCT} \gls{CM}s,
\textbf{second} researching games, which use \gls{BC}s in their ecosystem
and \textbf{third} presenting a novel \gls{CM}, \gls{PoT},
to complement the existing solutions, a conclusion and outlook is given.


\chapter{Conclusion \& Outlook}
\label{chap:ConclusionAndOutLook}

To answer the primary research question ..
\begin{center}
	\textit{"Can \gls{BCT} be used to reduce publisher's server costs whilst \\
		providing (mobile) players a suitable gaming experience?"}
\end{center}
.. the two leading questions have been discussed and answered within chapter 
\hyperref[chap:BCT]{Blockchain Technology} and chapter \hyperref[chap:PoT]{Proof-of-Turn}.
Nevertheless, the primary research question cannot be answered within a straight boolean value spectrum.

\noindent From a purely algorithmic perspective, there is a solution to most challenges (e.g. \hyperref[sec:PileOfCards]{Card draw}).
But that there is a viable algorithm is only part of the answer.
This thesis brings to mind that every additional required algorithm is a trade-off:
What is gained in pure feasibility on one side, is payed in additional development effort, higher complexity (and lower maintainability), lower system performance (and higher transaction costs).
The risk of faulty technicality decreases, but other risks increase in return.
In the case of games, dropping out players due to limited chances to win would be only one of the problematic cases.
Especially the higher complexity should be evaluated throughout, before the need of cumbersome fixes arise.

\noindent Therefore, whilst \gls{CM}s may function unexpectedly good in a theoretical document,
the final '\textit{design of upper level (gaming) software}', '\textit{additional programming effort}' as well as
'\textit{cooperation and coordination of gamers}' (e.g. players do only regularly quit game sessions)
are major factors which have to be considered to answer the research question throughout.

\noindent This additional research would offer further insights to adjust and verify first the given working assumptions
and second the \hyperref[script:GraphPlots]{plots} containing (only) sample data.
Hence, sample games have to be invented, implemented and analyzed throughout.
Unluckily, the concomitant workload would have exceeded the scope of this document by far.

\noindent Nevertheless, the tight budgets in the gaming industry call for mechanisms to improve
their games backend and \gls{PoT} offers a solution for slow paced games promising low (networking) costs.
Thus, finding sample applications conducting \gls{PoT} is assumed not to be too far fetched,
as described in the \hyperref[sec:Outlook]{Outlook}.

\pagebreak

\section{Outlook}
\label{sec:Outlook}

The outlook focuses on the \gls{PoT} \gls{CM}, but abstracts from the use case of games.
Hence, research in other possible application's use cases with \gls{PoT}
and '\textit{Adaption in the Wild}' is aimed for:
\begin{enumerate}
	\item \textbf{A truly distributed chat} \\
	Given an initial interconnection, every pair of devices could establish a \gls{BC} for their chat protocol.
	To the writer's knowledge, every recent messenger relies on central server technology.
	Still the most messengers lack funding.
	If users had an especially high need for privacy wherein only their devices had encryption keys, 
	they could use a messenger build upon \gls{BCT} using a lightweight \gls{CM} as \gls{PoT}.
	Interconnection could still suffer speed as the other node peer might be offline,
	but speed up turns, down to $\sim$ten seconds seem possible.
	During inactivity, dynamic turn time may be used as well.
	Here, if turn time was set to long (sleeping chat),
	a direct message to the \gls{LN} could trigger its turn's end and decrease transaction delivery times.
	Last, it would be reasonable to sacrifice instantaneous delivery for an increased level of security,
	as non existing central servers can not loose their integrity.
	
	\item \textbf{Cryptocurrencies and parliamentary elections} \\
	The ability to rewrite the chain, grounded on any network vote,
	makes \gls{PoT} futile both for cryptocurrencies and parliamentary elections.
	Still, e.g. the possibility of flooding attacks and the setting in \textit{permissioned networks}
	limits the practicality in this regard.
	Therefore, it is discouraged to use \gls{PoT} in such critical infrastructures in general.
	
	\item \textbf{The citizens trust} \\
	Awarding of public administration contracts is mostly conducted using a public administration's central server.
	In this scenario, each company's offering is seen mutually exclusive to the others and all together create a race condition.
	Moreover, knowing the bids of other companies creates unfair advantage to the informed
	party - especially if the informed party still has to claim its price tag.
	Not only does the \textit{offset revelation} mechanism (independent from \gls{BCT})
	promise a fair and transparent match making in the market.
	\gls{PoT} offers here a lightweight (computation) and balanced (writing permission) distributed \gls{CM}.
	Hence, \gls{PoT} may help to lighten opaque administrative structures and raise the stakeholders' trust.
	
	\item \textbf{Responsible disclosure} \\
	Just found vulnerabilities in software systems, formally known as zero-day vulnerabilities
	as well as startling information gained from whistle-blowing activity endanger not only the targeted institution(s),
	but also the investigative journalist(s) and security researcher(s).
	The publishing party (person) is not dependent on any server to stay online or
	has to fear that the central server is taken down from any higher authority.
	Still \textit{offset revelation} can be used to inform affected parties first,
	formally known as 'responsible disclosure'.
	If implemented accordingly, the publishing party may stay anonymous and
	gains additionally the ability to publish the encrypted data from a different node than the affiliated key(s).
	Here, if nodes stay offline the \gls{PoT} algorithm may be adjusted to choose the next node according to online activity.
	Inactive nodes are - after a network vote - skipped.
	Hence, the publishing party is enabled to stay offline, until the right time to publish has come.
	This scenario, again, shows the adaptability of \gls{PoT}.
	Last, as \gls{PoT} assumes that nodes write their date themselves and
	nodes do not compete for tokens gained from mining, the possibility is
	high for the publishing party to release the data without intermediaries.
	Additionally, as some critical acts are sometimes discovered, not whilst conducting, but during the phaseout,
	transferring cryptocurrency tokens into traditional currencies was a risk for the publishing party.
	Therefore, the absence of cryptocurrency tokens in \gls{PoT} protects publishing parties by design.
			
\end{enumerate}

\noindent Next to \gls{PoT}, this thesis gives some hints on future research fields like
'finding an optimal child-chain size' for different use cases. \\
All these and other scenarios call for further research and '\textit{Adaption in the Wild}'.

	
	\appendix
	
\chapter{Appendix}
\label{chap:Appendix}
The appendix offers data used throughout the document in further detail.

\section{Data regarding gaming devices}

\begin{figure}
	\centering{
		\includegraphics[width=.90\linewidth,keepaspectratio=true]{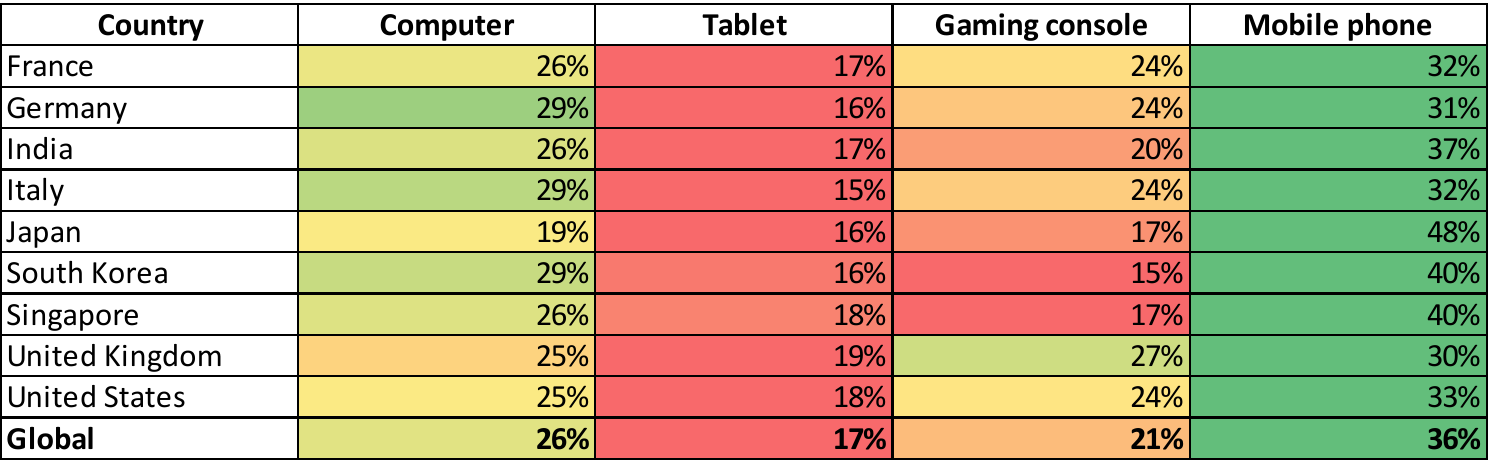}
		\caption{Share of devices by gaming time \cite[7]{LimelightNetworks.2020}}
		\label{fig:GameingDeviceDataTable}
	}
\end{figure}

\textbf{First}, the data relevant for used devices from \cite[7]{LimelightNetworks.2020} is shown in relative measures (Figure \ref{fig:GameingDeviceDataTable}).

\begin{figure}
	\centering{
		\includegraphics[width=.90\linewidth,keepaspectratio=true]{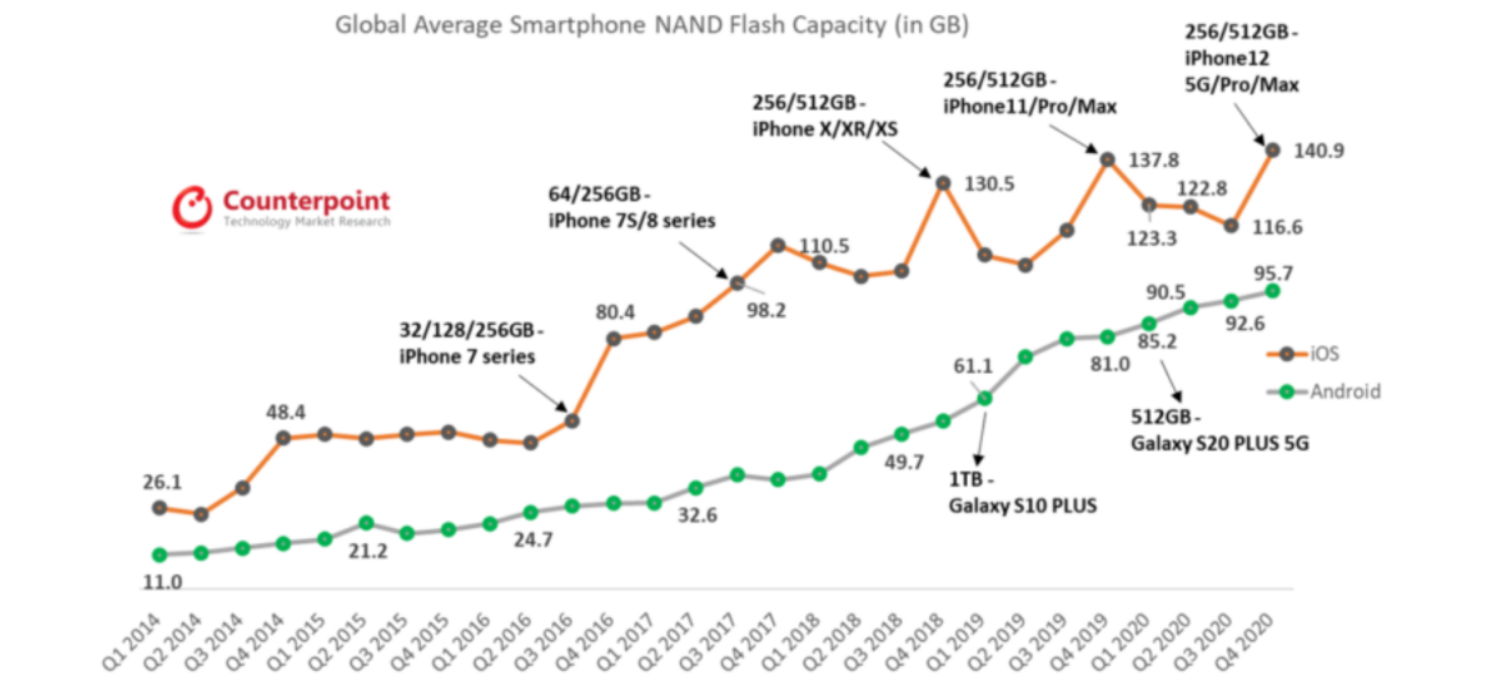}
		\caption{Average smartphone storage space (From \cite{Wang.2021})}
		\label{fig:GameingDeviceStorageSpace}
	}
\end{figure}

\noindent \textbf{Second}, a chart from 'Counterpoint Research' by \cite{Wang.2021} is given (Figure \ref{fig:GameingDeviceStorageSpace}),
which supports the claim that "Average Smartphone NAND Flash Capacity Crossed 100GB in 2020".

\begin{figure}
	\centering{
		\includegraphics[width=.90\linewidth,keepaspectratio=true]{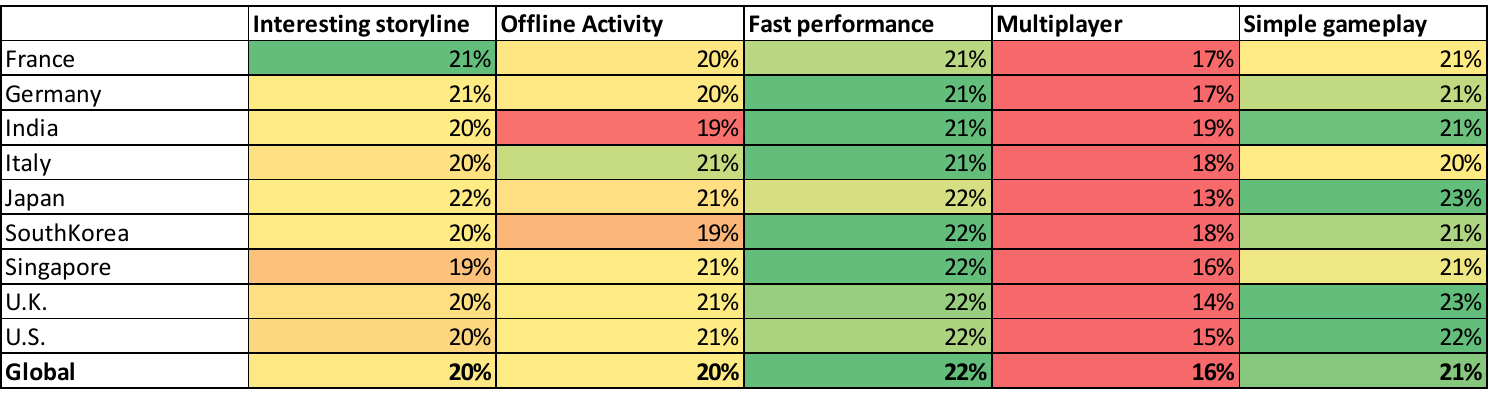}
		\caption{Game characteristics data \cite[18]{LimelightNetworks.2020}}
		\label{fig:GameCharacteristicsDataTable}
	}
\end{figure}

\noindent \textbf{Third}, the data relevant for game characteristics from \cite[18]{LimelightNetworks.2020} is shown in relative measures (Figure \ref{fig:GameCharacteristicsDataTable}).

\section{Scripts for graph's plots}
\label{script:GraphPlots}

\subsection{Plot: BFT shuffling cards}
\label{script:BFTshufflingCards}
Python code for figure \ref{fig:ShuffleCards}, '\textit{BFT shuffling cards}':
\pythonexternal{contents/graphsplots/ShufflePlot.py}
\pagebreak

\subsection{Plot: General storage allocation}
\label{script:GeneralStorageAllocation}
Python code for figure \ref{fig:StorageAllocation_Base}, '\textit{General \gls{BT} storage allocation chart}':
\pythonexternal{contents/graphsplots/BCstorageAllocationBase.py}
\pagebreak

\subsection{Plot: Storage allocation Prune procedure/Child-chain}
\label{script:StorageAllocationPPSC}
Python code for figure \ref{fig:StorageAllocationPrune}, '\textit{Prune procedure/Child-chain chart}':
\pythonexternal{contents/graphsplots/storageAllocationPruneSC.py}
\pagebreak

\subsection{Plot: Storage allocation Meta State Block}
\label{script:StorageAllocationMetaState}
Python code for figure \ref{fig:StorageAllocation_MetaState}, '\textit{Meta-State Block chart}':
\pythonexternal{contents/graphsplots/storageAllocationMetaState.py}
\pagebreak

\subsection{Plot: Storage allocation comparison}
\label{script:StorageAllocationComp}
Python code for figure \ref{fig:BC_StorageAllocation_All_Big}, '\textit{Storage allocation comparison}':
\pythonexternal{contents/graphsplots/BCstorageAllocationAllBig.py}

	\bibliography{BCinTBG} 
	
\end{document}